\documentclass[journal=jctcce, manuscript=article, layout=twocolumn]{achemso}

\usepackage[british]{babel}

\makeatletter
\let\l@addto@macro\relax
\makeatother

\usepackage[fontsize=10pt]{scrextend}
\usepackage[breaklinks=true,colorlinks=true,linkcolor=hyperlinkblue,urlcolor=hyperlinkblue,citecolor=hyperlinkblue]{hyperref}
\mciteErrorOnUnknownfalse
\usepackage{newfloat} % to change supp. info figure numbers
\usepackage{newfloat} % to change supp. info table numbers
\usepackage[version=4]{mhchem} % formula subscripts using \ce{}
\usepackage{siunitx}
\usepackage{chemmacros} % useful for chemical formulas
\usepackage{etoolbox} % needed for the patch

\definecolor{hyperlinkblue}{RGB}{34,31,150}
\DeclareFloatingEnvironment[name={Supplementary Figure}]{suppfigure}
\DeclareFloatingEnvironment[name={Supplementary Table}]{supptable}
\newcommand{\redpot}{E}
\newcommand{\redpotshift}{\delta E}
\newcommand{\fengdiff}{\Delta G}
\newcommand{\fengdiffhypo}{\Delta g}
\newcommand{\MDCB}{MD+CB}
\newcommand{\PBMC}{PB+MC}
\newcommand{\supp}{{Supporting Information (SI)}} 
\newcommand{\SIn}{{SI}} 
\newcommand{\figSone}{{S1}} % Fig.~\ref{fig:avg-struc}
\newcommand{\figStwo}{{S2}} % Fig.~\ref{fig:time-evolv}
\newcommand{\figSthree}{{S3}} % Fig.~\ref{fig:time-evolv-ox}
\newcommand{\figSfour}{{S4}} % Fig.~\ref{fig:time-evolv-red}
\newcommand{\figSfive}{{S5}} % Fig.~\ref{fig:PCA-natural}
\newcommand{\figSsix}{{S6}} % Fig.~\ref{fig:PCA-ox+red}
\newcommand{\figSseven}{{S7}} % Fig.~\ref{fig:Numb-residu}
\newcommand{\figSeight}{{S8}} % Fig.~\ref{fig:avg-Calpha}
\newcommand{\figSnine}{{S9}} % Fig.~\ref{fig:dist-34+92}
\newcommand{\figSten}{{S10}}  % Fig.~\ref{fig:dist-m4D2+19+DM}
\newcommand{\figSeleven}{{S11}} % Fig.~\ref{fig:PBMC}
\newcommand{\figStwelve}{{S12}} % Fig.~\ref{fig:raw-data}
\newcommand{\figSthirteen}{{S13}} % Fig.~\ref{fig:raw-histograms}
\newcommand{\figSfourteen}{{S14}} % Fig.~\ref{fig:posteriors}
\newcommand{\figSfifteen}{{S15}} % Fig.~\ref{fig:convergence}
\newcommand{\figSsixteen}{{S16}} % Fig.~\ref{fig:exp-fig}
\newcommand{\tabSone}{{S1}} % Tab.~\ref{table:abs}
\newcommand{\subsecsubcrooks}{{S.3}} % ~\ref{subsec:subcrooks}

\newcommand{\equal}{These authors contributed equally.\\[10pt]} % affiliations
\newcommand{\chemistrybristol}{Centre for Computational Chemistry, School of Chemistry, University of Bristol, Bristol BS8 1TS, UK}
\newcommand{\exeter}{Department of Physics and Astronomy, University of Exeter, Stocker Road, Exeter EX4 4QL, UK}
\newcommand{\potsdam}{Institute of Physics and Astronomy, University of Potsdam,  14476 Potsdam, Germany}
\newcommand{\biochemistrybristol}{School of Biochemistry, University of Bristol, Bristol BS8 1DT, UK}
\newcommand{\brissynbiobristol}{BrisSynBio Synthetic Biology Research Centre, University of Bristol, Bristol BS8 1TQ, UK}
\newcommand{\surrey}{School of Mathematics and Physics, University of Surrey, Guildford GU2 7XH, UK}

\makeatletter
\renewcommand*{\acs@author@fnsymbol@symbol}[1]{ % affiliations with numbers instead of symbols; * is for email
    \ifcase #1 *\or
    \dagger \or
    1\or
    2\or
    3\or
    4\or
    5\or
    6
    \fi
}
        
\renewcommand*\acs@contact@details{% addd * before  E-mail
    {\sffamily *\,E-mail: \acs@email@list }%
    \acs@number@list
}           

\patchcmd{\acs@address@list@auxii}% superscript for numbers in affiliations
{\acs@author@fnsymbol{\acs@affil@marker@cnt}}
{\textsuperscript{\acs@author@fnsymbol{\acs@affil@marker@cnt}}}
{}{}

\patchcmd{\acs@address@list@auxii}% superscript for numbers in affiliations
{{\acs@author@fnsymbol{\acs@affil@marker@cnt}\@nameuse{@altaffil@\@roman\@tempcnta}\par}}
{{\textsuperscript{\acs@author@fnsymbol{\acs@affil@marker@cnt}}\@nameuse{@altaffil@\@roman\@tempcnta}\par}}
{}{}        

\makeatother    
\renewcommand{\abstract}{} 

\author{A. S. F. Oliveira} % contributed equally
\affiliation{\equal}
\alsoaffiliation{\chemistrybristol} 
\alsoaffiliation{\biochemistrybristol}
\alsoaffiliation{\brissynbiobristol}

\author{J. Rubio} % contributed equally
\affiliation{\equal}
\alsoaffiliation{\surrey}
\alsoaffiliation{\exeter}

\author{C. E. M. Noble}
\affiliation{\biochemistrybristol} 
\alsoaffiliation{\brissynbiobristol}

\author{J. L. R. Anderson}
\affiliation{\biochemistrybristol} 
\alsoaffiliation{\brissynbiobristol}

\author{\\J. Anders} % corresponding
\email{Adrian.Mulholland@bristol.ac.uk} % emails inverted to make them appear in the right order
\affiliation{\exeter}
\alsoaffiliation{\potsdam}

\author{A. J. Mulholland} % corresponding
\email{janet@qipc.org}
\affiliation{\chemistrybristol} 

\title{Fluctuation relations to calculate protein redox potentials from molecular dynamics simulations}

\begin{document}

%%%%%%%%%%%%%%%%%%%%%%%%%%%%%%%%%%%%%%%%%%%%%%%%%%%%%%%%%%%%%%%%%%%%%
%% Abstract
%%%%%%%%%%%%%%%%%%%%%%%%%%%%%%%%%%%%%%%%%%%%%%%%%%%%%%%%%%%%%%%%%%%%%
\begin{abstract}

\section*{Abstract}

\textbf{
The tunable design of protein redox potentials promises to open a range of applications in biotechnology and catalysis. Here we introduce a method to calculate redox potential changes by combining fluctuation relations with molecular dynamics simulations. It involves the simulation of reduced and oxidized states, followed by the instantaneous conversion between them. Energy differences introduced by the perturbations are obtained using the Kubo-Onsager approach. Using a detailed fluctuation relation coupled with Bayesian inference, these are post-processed into estimates for the redox potentials in an efficient manner. This new method, denoted \MDCB, is tested on a {\it de novo} four-helix bundle heme protein (the m4D2 `maquette') and five designed mutants,
including some mutants characterized experimentally in this work. The \MDCB~approach is found to perform reliably, giving redox potential shifts with reasonably good correlation ($0.85$) to the experimental values for the mutants.
The \MDCB~approach also compares well with redox potential shift predictions using a continuum electrostatic method.
The estimation method employed within the \MDCB~approach is straightforwardly transferable to standard equilibrium MD simulations, and holds promise for redox protein engineering and design applications. 
}

\end{abstract}

%%%%%%%%%%%%%%%%%%%%%%%%%%%%%%%%%%%%%%%%%%%%%%%%%%%%%%%%%%%%%%%%%%%%%
%% Main text
%%%%%%%%%%%%%%%%%%%%%%%%%%%%%%%%%%%%%%%%%%%%%%%%%%%%%%%%%%%%%%%%%%%%%
\section{Introduction}

Electron transfer is fundamental in biological processes such as photosynthesis and respiration. Evolution has modulated the redox properties of proteins involved in redox processes to make electron transfer rates sufficient to sustain such processes.\cite{Liu2014,AyusoFernandez2019,hosseinzadeh2016,Page1999} Redox active metallocofactors, such as heme, enable many natural oxidoreductases to catalyse a wide range of reactions, including hydroxylation and oxygenation.\cite{poulos2014,Liu2014} Heme-containing proteins are ubiquitously found in nature and are involved in many biological electron transfer processes.\cite{Liu2014,yu2019} Their redox potentials play a role in determining their activities, such as oxygen binding, electron transfer and catalysis.\cite{Liu2014,yu2019} 

The redox potential $E$ of a heme center can be described as its tendency to acquire electrons and thus to become reduced. Thus, the higher the value of the redox potential, the more favorable the reduction of the group. The intrinsic properties of the heme macrocycle are critically modulated by the protein environment.\cite{Liu2014,yu2019} Many properties are important for this, including the axial residues directly coordinating the iron, the second coordination sphere interactions, such as hydrogen-bonds, and the electrostatic environment surrounding the center.\cite{Liu2014,yu2019} Given the multitude of factors involved in this ‘tuning’, the accurate prediction of the redox properties of a heme-containing protein remains a challenge. There is a need for computational methods capable of predicting redox potentials for  natural proteins (e.g., for analysing effects of mutations) and potentially in the engineering of proteins, both natural and \emph{de novo} with altered redox properties.

Engineering of existing redox proteins and construction of novel designs offers possibilities such as tuning enzymes towards alternative substrates and creating novel electron transfer systems \cite{Koebke2022,hosseinzadeh2016} for applications in biocatalysis, biosensing, biofuel generation and bioelectronics.\cite{prabhulkar2012,yu2019} Reliable prediction methods will assist functional protein design and complement directed evolution by identifying target sites for mutation. Calculations of redox properties of mutant proteins could usefully be incorporated into design protocols, e.g. to identify promising locations for mutations for synthesis, and narrowing the experimental search spaces for desired properties. 

Despite difficulties, examples of useful applications of ‘tuning’ redox potentials exist (see e.g.\cite{bhagi2014, bhagi2018}). However, such successes have been based generally on qualitative insight, and trial and error approaches. Alternative predictive methods for redox properties of proteins, whether designed \emph{de novo} or engineered natural proteins, could significantly accelerate applications in engineering biological systems at the molecular level. 

Biomolecular simulations, such as equilibrium molecular dynamics (MD) simulations, can contribute to such developments. Simulations are increasingly assisting the design of proteins, e.g. as catalysts.\cite{bunzel2021} Such approaches, which can be used qualitatively, to predict the stability of designs, are becoming increasingly capable of predicting thermodynamic properties. Nonetheless, reliably estimating redox properties of proteins using theoretical-computational approaches remains challenging.\cite{breuer2017, watanabe2017, cruzeiro2018, warshel2011, haiyun2016, formaneck2002, Blumberger2008, Blumberger2015, Blumberger2006} Challenges include for example a proper dynamic representation of the different redox states and their electrostatic interactions, and sampling the relevant conformational states (e.g. \cite{warshel2011, watanabe2017}). 

Different types of computational techniques have been used to study redox processes in proteins, including MD simulations and continuum electrostatics (CE) calculations.\cite{breuer2017, watanabe2017, cruzeiro2018, warshel2011} For example, in MD-based free energy simulations, the protein’s conformation changes are explicitly treated for a fixed reduction/protonation state. However, such calculations are computationally expensive. 

CE methods have also been widely used to predict changes in protonation and reduction in proteins (e.g. \cite{gunner2016,warshel2006,teixeira2002}). These methods are much faster than MD-based approaches as they sacrifice configuration aspects of the protein and/or solvent. 
Nonetheless, for most proteins, the lack of explicit dynamics can affect the accuracy of the predictions (e.g. \cite{warshel2011}). 
Hybrid approaches, combining MD and continuum electrostatics-based methods have also been developed to estimate protonation and reduction changes.\cite{cruzeiro2020,machuqueiro2009} 
Such methods require adequate sampling of conformational space, which is still a challenge for most proteins. \cite{gunner2016}  

Meanwhile, the emergence of stochastic thermodynamics has introduced  detailed and integral fluctuation relations that beautifully capture the properties of a wide variety of non-equilibrium processes.\cite{Evans1994,jarzynski1997,crooks1999}
They link the distributions, $p(\cdot)$, and averages, $\langle \cdot \rangle$, of stochastically fluctuating quantities, such as entropy $S$, work $W$ or heat $Q$, for a particular process $\Lambda$ with those of the time-reversed process $\tilde{\Lambda}$. Fluctuation relations establish limits on the microscopic fluctuations of small systems that are much more detailed than the macroscopic laws of thermodynamics.\cite{Evans2002,seifert2012}
Practically, they are used to infer free energy differences $\fengdiff$ based on data from highly non-equilibrium experiments. 
For example, a range of optical tweezer experiments have been conducted that mechanically stretch single molecules under a variety of conditions and collect the non-equilibrium statistics.\cite{liphardt2002,Collin2005,camunas2017}
Combined with fluctuation relations, these experiments have been been used, for example, to  determine ligand binding energies, as well as characterise selectivity and allosteric effects of nucleic acids and peptides.\cite{camunas2017}

Here we use a detailed fluctuation relation, the Crooks relation,\cite{crooks1999} with data from MD simulations to calculate protein redox potential changes. To our knowledge, this is the first application of these relations in this context.  We use this method to predict the redox potential $\redpot$ of a \emph{de novo} designed protein, the m4D2 ‘maquette’ (Fig. 1), and several mutants (single and double). m4D2 is a well-characterized soluble four-helix bundle mono-heme binding protein. Redox potentials have been experimentally determined for m4D2 and several mutants by optically-transparent thin-layer electrochemistry.\cite{huggins2019} m4D2 is quite small (about 110 residues long), making it an amenable target for MD simulations.\cite{huggins2019} It also lacks some complexities of natural proteins, such as allosteric regulation.\cite{huang2003,koder2006,huggins2019}  

\begin{figure}[t!]
\centering
\includegraphics[trim={1cm 2.5cm 0cm 1cm},clip,width=4.5cm]{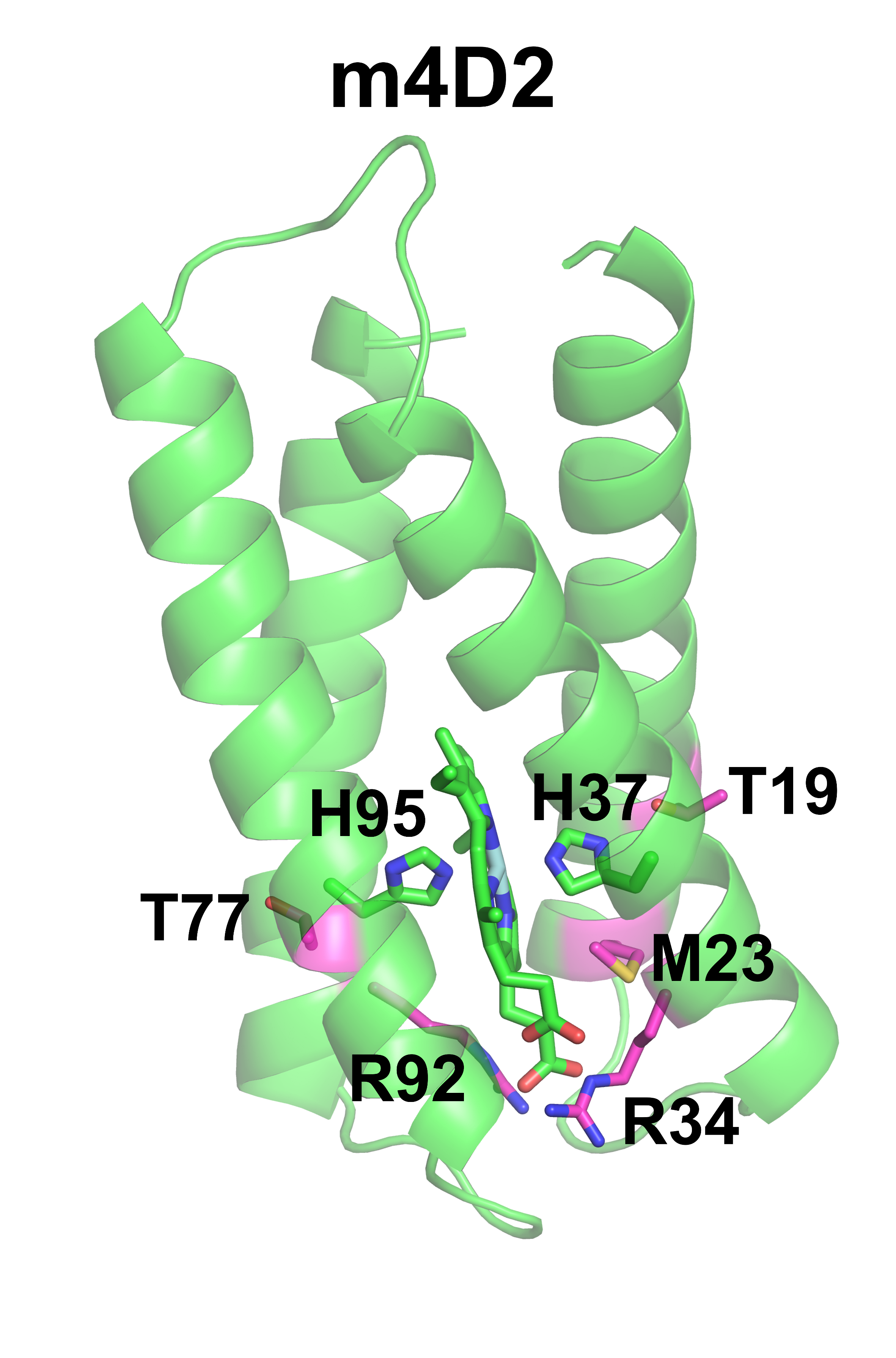}
	\caption{Predicted structure of the \emph{de novo} monoheme binding maquette protein m4D2. The structure shown  was built using Rosetta\cite{Fleishman2011} (for more details, see \cite{Hutchins2023}).
	m4D2 is a designed, {\it de novo} 4-helical bundle protein that binds heme B. Heme B and the histidine residues axially coordinating the Fe atom are shown with green sticks, whereas the sites of the mutations studied here are shown with magenta sticks.
 } 
\label{fig:m4D2}
\end{figure}
\section{Materials and methods} \label{sec:materials}

\subsection{m4D2 structure} \label{subsec:m4d2}

The computational design of the monoheme m4D2 structure was performed using Rosetta\cite{Fleishman2011} as described in detail in Ref.\cite{Hutchins2023} In our design (Fig.~\ref{fig:m4D2}), the heme group is coordinated by two histidines situated on diagonally opposed helices, and is positioned with its propionate groups pointing towards the end of the bundle. 
In m4D2, there are two threonine residues (threonines 19 and 77), which are directly involved in key hydrogen bonding interactions with the heme-coordinating histidines. 
Experimental mutation of threonine 19 to an aspartate \cite{Hutchins2023} changes the redox potential by $-28$~\si{mV} relative to m4D2 (T19D in Tab.~\ref{table:results}).
Replacing both threonine 19 and 77 with aspartate \cite{Hutchins2023} has a substantial effect on the heme redox potential, decreasing it by $-56$~\si{mV} relative to m4D2 (see double mutant, DM, in Tab.~\ref{table:results}). These residues were originally selected for mutation as there is an aspartate in an equivalent histidine-interacting position in horseradish peroxidase, which is believed to play a critical role in the enzyme's catalytic triad by increasing the imidazolate character of the proximal histidine. \cite{goodin1993,ortmayer2020}
This local increase of negative charge was expected to lower the redox potential of the heme.

Arginines 34 and 92 are located on the second and fourth helices of m4D2, respectively, and likely interact directly with the heme propionates through ion pairing and hydrogen bonding, as observed in the 4D2 crystal structure. \cite{giovanna2004, Hutchins2023} The distance between these arginines and the propionate groups was monitored in our trajectories (Fig.~\figSnine~ in the \supp), and as predicted, these residues can form direct interactions with the negatively-charged propionates. Mutation of either of these arginines is likely to alter the network of interactions involving the propionate groups. Experimentally, replacement of either arginines by glutamine (so removing the positive charge) is here found to change the redox potential by approximately $-30$~\si{mV}  (see R34Q and R92Q in Tab.~\ref{table:results}).

Finally, methionine 23, which is located on helix 1, is also close to the heme. Nonetheless and despite its proximity to the heme, experimentally mutating this methionine to asparagine (see M23N in Tab.~\ref{table:results}) has little effect on the redox potential of the heme (with a $+1$~\si{mV} change relative to m4D2). \cite{Hutchins2023}

\subsection{MD simulations} \label{subsec:eqMDsim}

MD simulations were used to generate ensembles of conformations of  m4D2 \cite{giovanna2004}, four single mutants (T19D, M23N, R34Q, R92Q) and a double mutant, namely T19D-T77D (hereafter labelled DM). 
The m4D2 model\cite{Hutchins2023} produced by Rosetta described above in section 2.1 was used as the starting point for the m4D2 simulations. Starting structures for simulations of the mutants were created using the mutagenesis tool in PyMOL. \cite{pymol2015}. For each protein, MD simulations were performed for the reduced and oxidized form. 

MD simulations were performed using GROMACS\cite{berendsen1995, vanderspoel2005, pall2015, abraham2015} on the University of Bristol’s compute cluster, BluePebble. The GROMOS 54A7 force field\cite{schmid2011} was used for protein, and parameters for the oxidized and reduced redox center were taken from our previous work. \cite{Hutchins2023}
Protein models were solvated in dodecahedral boxes, with a minimum distance of $2\,\si{nm}$ between the protein and the box limits. The simple point charge (SPC) water  model\cite{hermans1984} was used. 
The total net charge of m4D2 and M23N in the oxidized and reduced states is -2 and -3, respectively. For the T19D, R34Q and R91Q mutants, the overall charge of the proteins in the oxidized and reduced states is -3 and -4. Finally, the total charge for the double mutant (in which aspartate residues replaced both T19 and T77) is -4 and -5 for the oxidized and reduced state, respectively. 
The overall net charge in all systems was neutralized by adding the exact number of positively charged ions to offset the net charge on the proteins. Overall, 2 and 3 sodium (Na+) ions were added in the oxidized/reduced m4D2 and M23N systems; 3 and 4 Na+ ions were included in the T19D, R33Q and R91Q systems; and 4 and 5 Na+ ions were added to the double mutant system. In addition to the ions needed to neutralize the systems (i.e. to make the total net charge of the proteins equal to 0), an ionic concentration of 0.05 M sodium chloride was also included in the simulation boxes to mimic the experimental conditions.\

Simulations were performed at constant temperature and pressure, using the velocity rescaling thermostat\cite{bussi2007} at $T=298\,\si{K}$ and the Parrinello-Rahman barostat\cite{parrinell01981, nose1983} to maintain the pressure at $1\,\si{bar}$. A time step of $2\,\si{fs}$ was used for integrating the equations of motion. 
The LINCS algorithm\cite{hess1997} was used to constrain bonds in the protein and the SETTLE algorithm\cite{miyamoto1992}  was used to keep water molecules rigid. 
Long-range electrostatic interactions were calculated using a particle mesh Ewald method,\cite{essmann1995} with a Fourier grid spacing of $0.12\,\si{nm}$ and a $1.4\,\si{nm}$ cut-off for direct contributions.
$1000$ steps of energy minimization with the steepest descent method with harmonic restraints applied to heavy atoms, followed by a further $1000$ steps restraining the C$\alpha$ atoms only, then $1000$ steps with no restraints, were performed prior to MD simulation. 
Then, a $3\,\si{ns}$ restrained MD relaxation was performed to relax the system, prior to unrestrained MD simulations.

Multiple MD simulations were performed for each system and redox state: ten  $500\,\si{ns}$ unrestrained MD simulations were performed for the reduced and oxidized states of m4D2 and for the single mutants (T19D, M23N, R34Q and R92Q). Twenty $500\,\si{ns}$ unrestrained MD simulations were performed for the T19D-T77D double mutant (DM) in the reduced and oxidized states. 
The replicas were initiated with different sets of random velocities. 
In total, this amounts to $70\,\si{\mu s}$ of simulation.
All analyses were performed using the GROMACS package\cite{berendsen1995, vanderspoel2005, pall2015, abraham2015} and in-house tools. 
PyMOL\cite{pymol2015} was utilised for molecular representation.

All proteins were stable over the simulation time. The simulations showed that the structures of all of the mutants are overall similar to m4D2, as expected. The proteins all appeared to be equilibrated after $100\,\si{ns}$ (see Figs.~\figStwo-\figSseven~in \SIn). The first $100\ \si{ns}$ were taken as equilibration, and only the last $400\ \si{ns}$ was analysed (Figs.~\figSone-\figSeight~in \SIn), unless stated otherwise. Principal component analysis was used to evaluate the sampling of the conformational space by the replicates (see Figs.~\figSfive-\figSsix). 
To analyse dynamical changes caused by the mutations, root mean square fluctuation (RMSF) profiles of the C\(\alpha\) atoms were calculated (Fig.~\figSeight). The RMSF plots show that the effects of the mutations are localised to the regions surrounding the mutation site. Some decrease local fluctuations (e.g., T19D) while others increase them (e.g., R92Q), relative to m4D2. The RMSF profiles also show that the unstructured loop regions of the proteins are very mobile, representing some of the largest peaks observed Fig.~\figSeight. Such dynamic behaviour is also contributing to the high RMSD values observed in Fig.~\figStwo~in the \SIn. As can be seen in Figs.~\figSthree-\figSfour, the RMSD profiles for the structured parts of the protein (i.e. excluding the loop regions) show lower C\(\alpha\) deviations from the structures used as starting points for the simulations.
These analyses altogether indicate that the MD simulations provide a reasonable conformational sample for calculations of redox potentials.

\subsection{Nonequilibrium Perturbations} \label{subsec:DNEMD}

To determine the energy cost of reducing and oxidizing the heme group, conformations were extracted every nanosecond from the equilibrated trajectories ($400$ conformations per replicate) and used as starting points for the reduction/oxidation events (in a total of $4000$ conformations per system for m4D2, T19D, M23N, R34Q and R92Q, and $8000$ conformations for the DM). In each of these extracted conformations, the redox state of the heme was (instantaneously) changed.

For each conformation, the energy change associated with the oxidation/reduction of the protein was calculated as the difference in energy between the states in that conformation. Note that these energy differences, which were obtained using a molecular mechanics force field, do not take into account the electronic effects associated with adding/removing an electron from the heme group, i.e. the intrinsic energy for reduction/oxidation of the heme cofactor (the intramolecular contribution to the energy).

The energy difference ($\Delta \epsilon$) between oxidation states in the protein was determined using the Kubo-Onsager approach:\cite{ciccotti1975, ciccotti1979, ciccotti2016, oliveira2021c} specifically, by calculating the difference in the potential energy of the protein between every pair of reduced/oxidized conformations extracted from the simulations. 
The large number (thousands) of replicates allows for convergence of the calculated energetics associated with heme reduction and oxidation. 

Note that in this work, the oxidation/reduction reorganization energies are obtained by determining the instantaneous energy difference between redox states. This is a simple way to determine the work values $W$ (for more details see section 2.4). An interesting future extension would be to use dynamical nonequilibrium molecular dynamics (D-NEMD) simulations~\cite{ciccotti1975,ciccotti1979, oliveira2021c} to estimate such energies. Additional factors (e.g. barostat effects) need to be considered for this. 
Note also that for the energy difference between the two redox states extracted from the MD simulation we will include only the contribution from the protein. It is clear that the solvent can make a non-trivial contribution. However, when taking the energy differences between states of the protein plus the full solvent, we found that the fluctuations are then too large in comparison to the redox potential we want to extract. A solution to this problem would be to only include a portion of the solvent. But this just shifts the problem of where to make the cut. Future  theoretical development is needed that addresses how to adequately include the solvent effects.

It should be noted that the experimental redox potential shifts, while they might appear large, correspond, in fact, to very small free energies relative to the systematic and statistical errors associated with a typical computational calculation. Indeed, this is why predicting redox potential shifts is such a challenging task. Besides the contribution of the solvent, there are other factors that are not considered and that can affect the energy differences. These include quantum effects which are not captured by molecular mechanics approaches, such as changes in the heme's polarization, and density and ionization energy due to different environments. Sampling problems, imprecision in the model produced by Rosetta, biases introduced when building the models for the mutants, force field limitations (e.g. the lack of polarization), uncertainties in the protonations states of the titrable residues are all examples of factors that can affect the energy differences, and thus, computational predictions.

\subsection{Fluctuation relations} \label{subsec:FR}

The goal of this investigation is to predict redox potentials $\redpot$ using fluctuation relations applied to the data from these perturbations. We  employ a Bayesian generalisation of the procedure to estimate free energy differences via the Crooks fluctuation relation.\cite{crooks1999,maragakis2008}

First, we introduce the Crooks relation. This detailed fluctuation relation is formulated for a generic system, such as a harmonic oscillator or a molecule, with at least one externally controlled parameter $\lambda$ with two settings $A$ and $B$. For example, the harmonic  potential's frequency can be either $\lambda_A$ or $\lambda_B$, and in a molecule a charge can be absent ($\lambda_A$) or present ($\lambda_B$). Starting at setting $\lambda_A$, with an equilibrium state at inverse temperature $\beta = 1/k_B T$ where $k_B$ is the Boltzmann constant, the system  is pushed (arbitrarily far) out of equilibrium by varying the parameter $\lambda$ with some protocol $\Lambda$. This can either be a smooth variation in time, $\lambda(t)$, or an instantaneous change, e.g., $\lambda_A \to \lambda_B$. 

Work $W$ is received by the system during the action of the non-equilibrium protocol.  Crooks' relation~\cite{crooks1999}  quantifies the likelihood $p$ of a specific work value $W$ being required given the forward protocol $\Lambda$ in comparison to the likelihood of the corresponding negative value $-W$ being required given the backward protocol $\tilde{\Lambda}$,  i.e.,
\begin{equation}
    \frac{p(+W| \Lambda)}{p(- W|\tilde{\Lambda})} = \mathrm{e}^{\beta\left(W - \fengdiff \right)}.
    \label{eq:crooks-rel}
\end{equation}
While the left-hand-side contains non-equilibrium work distributions, the right hand side contains equilibrium properties of the system at settings $\lambda_A$ and $\lambda_B$. Specifically,  $\fengdiff = G_B - G_A$ is the free energy difference between the two equilibrium states at inverse temperature $\beta$. Obviously, in a quasistatic protocol $\Lambda_{\rm qs}$ where the system is always in equilibrium, one would have $p(+W| \Lambda_{\rm qs})=p(-W| \tilde{\Lambda}_{\rm qs})$ and $W = \fengdiff$ in every run, as expected from macroscopic thermodynamics. The power of relation \eqref{eq:crooks-rel} is that it is valid for protocols that drive the system far from equilibrium. 

In the context of our heme-containing proteins, the two settings are that an electron is absent (setting $\lambda_A$, oxidized) or present (setting $\lambda_B$, reduced). The free energy difference for reduction is then $\fengdiff = G_B - G_A = - n F E$,\cite{atkins2022} where $\redpot$ is the redox potential of the heme, $F$ is Faraday's constant ($96485.3\,\rm{C\,mol^{-1}}$), and $n$ is the number of electrons being transferred. 
In our case, $n = 1$, so that $\fengdiff = - F \redpot$.  

The reduction of the heme group is thus identified with the forward protocol $\Lambda$, while the oxidation process is the backward protocol~$\tilde{\Lambda}$.
To get the work values for the reduction (oxidation) protocols, we first note that there is no heat contribution since the simulations implement an {\it instantaneous} appearance (disappearance) of the electron in the heme, leaving no time for heat to be exchanged.\cite{jarzynski1997,seifert2012} 
We define the statistical work value $W_i$, received  by the heme in the $i$-th reduction process $\Lambda$ as the statistical energy difference
$W_i = \epsilon_{\rm{red}}^i - \epsilon_{\rm{ox}}^i =: \Delta \epsilon_i$ for $i=1, ..., \mu$.
I.e., in each run, an equilibrium simulation gives the initial statistical energy value of the oxidised protein, $\epsilon_{\rm{ox}}^i$, and a subsequent non-equilibrium perturbation, where the electron has been removed, gives the final statistical energy value,  $\epsilon_{\rm{red}}^i$, for the reduced protein.
Similarly, for the backward protocol $\tilde{\Lambda}$, 
the statistical work is
$W_i = \epsilon_{\rm{ox}}^i - \epsilon_{\rm{red}}^i = \Delta \epsilon_i$ for $i=\mu+1, ..., 2\mu$.
For an ensemble of non-equilibrium processes, for a given m4D2-mutant, we obtain a set of work values $\boldsymbol{W} = (W_1, \dots, W_{2\mu})$, where entries $1, ..., \mu$ correspond to reduction and entries $\mu+1, ..., 2\mu$ to oxidation.

\begin{figure*}[t!]
\centering
\includegraphics[trim={0cm 0cm 0cm 0cm},clip,width=\textwidth]{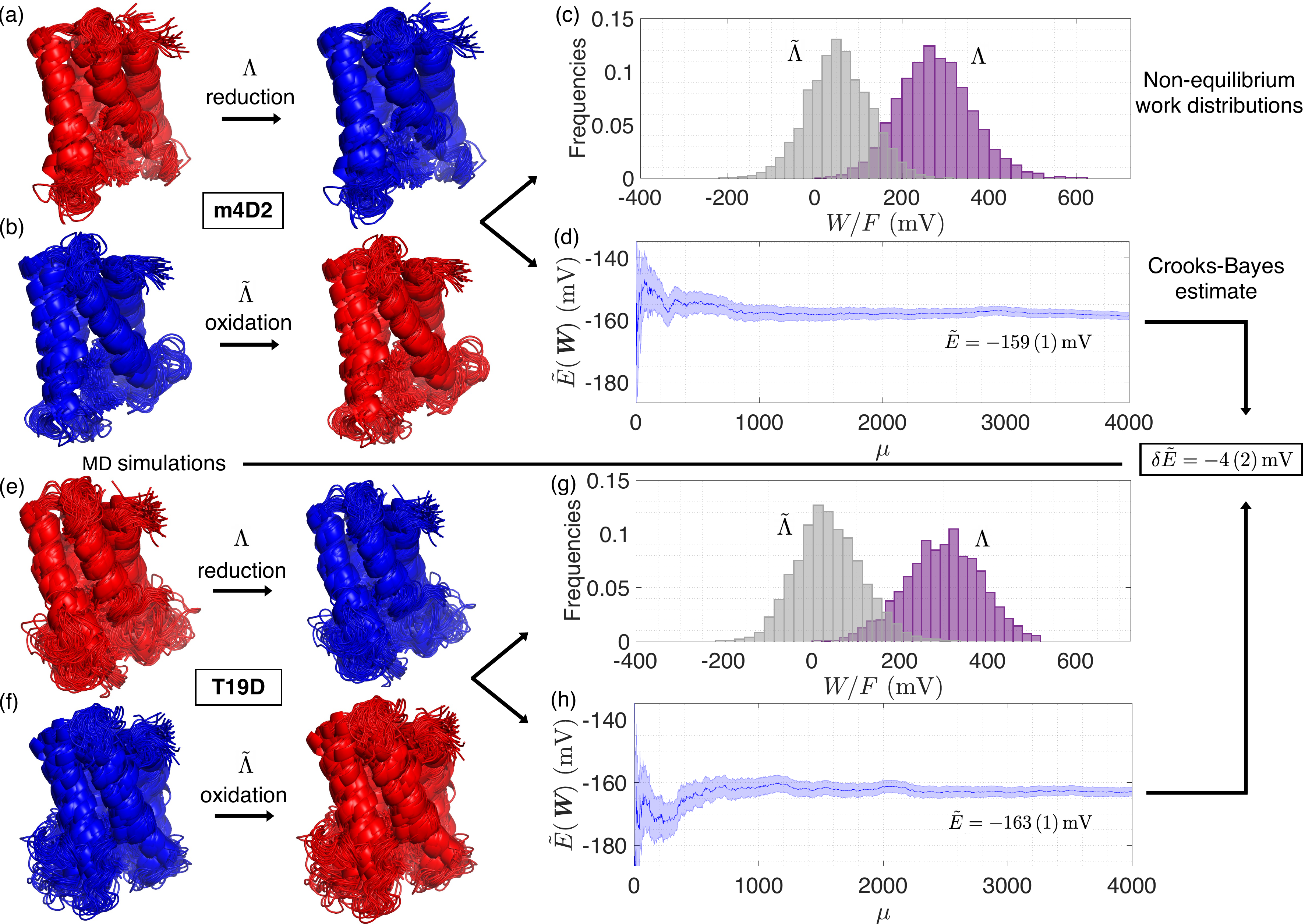}
	\caption{Schematic of the \MDCB~method used to calculate the redox potential shift ($\delta E$) of T19D relative to m4D2. 
	MD simulations and a nonequilibrium perturbation were used to determine the energy cost of reduction (panels {\bf (a)} and {\bf (e)}) and oxidation (panels {\bf (b)} and {\bf (f)}) for m4D2 and T19D. 
	Reduction and oxidation are identified as the forward ($\Lambda$) and backward ($\tilde{\Lambda}$) protocols, respectively, which are needed as input for the detailed Crooks fluctuation relation.
    The statistical work, $W$, was determined as 
    $W = \epsilon_{\rm{fin}} - \epsilon_{\rm{ini}}$, where for the reduction process, the final energy is the energy for the reduced protein, $\epsilon_{\rm{fin}}= \epsilon_{\rm{red}}$, and the initial energy is the energy for the  oxidized protein, $\epsilon_{\rm{ini}}=\epsilon_{\rm{ox}}$.
    For the oxidation process, it is the other way around, i.e. $\epsilon_{\rm{fin}}= \epsilon_{\rm{ox}}$ and $\epsilon_{\rm{ini}}=\epsilon_{\rm{red}}$.
    The resulting work histograms, $p(W|\Lambda)$ for the forward/reduction (purple) process and $p(-W|\tilde{\Lambda})$ for the backward/oxidation (grey) process are shown in panels {\bf (c)} for m4D2 and {\bf (g)} for T19D.
    These histograms are shown for illustrative purposes, but they are not employed to perform the estimation of redox potentials due to the finite-sample caveats discussed in the main text.
    Instead, we use the more accurate Crooks-Bayes estimate\cite{maragakis2008} \eqref{eq:crooks-bayes}, with error \eqref{eq:est-err}. 
    For m4D2 and T19D, respectively, panels {\bf (d)} and {\bf (h)} show the convergence of these estimates as the number of iterations $\mu$ increases. 
    The estimate $\delta \Tilde{E}$ for the shift $\delta E$ was calculated by subtracting the $E$-estimate ($\tilde{E}$) for m4D2 from that for T19D, and the errors were propagated.  
    The same procedure was applied to all of the mutants, with results given in Tab.~\ref{table:results}.
}
\label{fig:main-results}
\end{figure*}

When information from both directions of a process, $\Lambda$ and $\tilde{\Lambda}$, is available, the commonly used procedure to estimate $\fengdiff$ is via the Crooks relation \eqref{eq:crooks-rel}, as follows:\cite{crooks1999,seifert2012}
Constructing the forward and backward work histograms, $p(W|\Lambda)$ and $p(-W|\tilde{\Lambda})$, one identifies the point $W^*$ where they cross, i.e. $p(W^*|\Lambda)=p(-W^*|\tilde{\Lambda})$.  By virtue of the Crooks relation~\eqref{eq:crooks-rel} this gives an estimate~\cite{camunas2017} for the free energy as $\fengdiff = W^*$.
For the m4D2 protein this is illustrated in Fig.~\ref{fig:main-results}c), and in Fig.~\ref{fig:main-results}g) and Fig.~\figSthirteen~for the mutants. 
It should be noted that the work distributions in Figs.~\ref{fig:main-results}c), ~\ref{fig:main-results}g) and Fig.~\figSthirteen, corresponding to the nonequilibrium work values associated with instantaneous oxidation/reduction processes, can also be viewed as equilibrium distributions obtained from sampling the oxidized and reduced states.

However, this histogram-based method has the caveat that it requires a sufficiently large number of iterations to produce good estimates.\cite{jaynes2003}
Not only is finite statistics known to significantly impact the quality of standard free energy estimates,\cite{gore2003,pohorille2010,yunger-halpern2016} but using estimators whose validity depends on the size of the sample poses the risk of amplifying potential errors due to limited sampling in MD simulations.
To address these caveats, \citet{maragakis2008} put forward a \emph{Bayesian} framework for the estimation of free energy differences that combines the Crooks relation with Bayes theorem.\cite{jaynes2003, toussaint2011}
The key advantage of using Bayes theorem is that it can extract all the information available in a given sample regardless of its size.\cite{jaynes2003,maragakis2008,jesus2018} 
Therefore, it can help to prevent amplification of limited sampling. 

To use this approach, one first calculates the conditional probability density $p(\fengdiffhypo|\boldsymbol{W})$, 
where $\fengdiffhypo$ denotes a hypothesis about the true free energy difference $\fengdiff$ given the work values $\boldsymbol{W}$ provided by the MD simulations.
Following \citet{maragakis2008}, we use 
\begin{align}
    p(\fengdiffhypo|\boldsymbol{W}) \propto \prod_{i=1}^{\mu}\,
    & f(\beta \Delta \epsilon_i-\beta \fengdiffhypo) 
    \nonumber \\
    & \times\prod_{j=\mu+1}^{2\mu}\,f(\beta \Delta \epsilon_j+\beta \fengdiffhypo),
    \label{eq:posterior-main}
\end{align}
where $f(x)=1/[1+\mathrm{exp}(-x)]$ is the logistic function, $\beta = 1/(k_B T)$ is the inverse temperature used in the simulations (i.e., $T = 298\,\rm{K}$).
Eq.~\eqref{eq:posterior-main} is derived using minimal prior information, see details in the \SIn~subsection~\subsecsubcrooks.

The probability  distribution $p(\fengdiffhypo|\boldsymbol{W})$, shown in Fig.~\figSfourteen~for m4D2 and its mutants, contains all the information available to estimate $\fengdiff$.
To map this information into a concrete value for the redox potential $\redpot$, one uses the estimator (indicated by a tilde)
\begin{equation}
    \tilde{\redpot}(\boldsymbol{W}) =  - \frac{1}{F} \int d \fengdiffhypo \, \, p(\fengdiffhypo|\boldsymbol{W})  \,  \fengdiffhypo,
    \label{eq:crooks-bayes}
\end{equation}
where $F$ is the Faraday constant. 
This estimator is optimal under the square error criterion, \cite{jaynes2003, maragakis2008}
with error
\begin{equation}
    \sigma^2(\boldsymbol{W}) = \frac{1}{F^2}\int d \fengdiffhypo \,\, p(\fengdiffhypo|\boldsymbol{W}) \, (\fengdiffhypo)^2 - \tilde{\redpot}(\boldsymbol{W})^2.
    \label{eq:est-err}
\end{equation}
The use of energy differences $\Delta \epsilon$ obtained from MD simulations of proteins (see section \ref{subsec:DNEMD}), together with the Crooks-Bayes estimator given in Eq.~\eqref{eq:crooks-bayes} for probability distribution \eqref{eq:posterior-main}, constitutes a new computational method for the prediction of redox potentials. We refer to this method as the \MDCB~method, see illustration in Fig.~\ref{fig:main-results}. The redox potential shifts of the  mutants relative to m4D2 obtained with the \MDCB~method are reported in Tab.~\ref{table:results}.

\subsection{Note on other methods} \label{subsec:CEmethod}

Given the MD simulated work values, a variety of estimation methods exist to infer the free energy difference $\fengdiff$.
One of them is using Bennett's acceptance ratio (BAR).\cite{bennett1976, crooks2000path} The BAR estimation method has been shown to be akin to combining Crooks' relation with maximum likelihood estimation.\cite{shirts2003equilibrium, maragakis2008}
This makes its reliability generally justified only in the limit of an asymptotically large number of work measurements.\cite{kay1993}
In contrast, the Crooks-Bayes approach we use here leads to reliable estimates even for small data sets. 
This is a general property of Bayesian estimation techniques, enabled by the inclusion of prior information (or the absence of it), such as symmetries \cite{rubio2021}, in the calculations of estimators and errors. \cite{jaynes2003, toussaint2011} 

Another common method is using  Jarzynski's equality,\cite{jarzynski1997} $\langle \mathrm{exp}(-\beta W) \rangle = \mathrm{exp}(-\beta \fengdiff)$, an integral fluctuation relation that can be deduced from the Crooks relation. Jarzynski's equality provides a useful estimator for $\fengdiff$ for experiments conducted out of equilibrium that can only implement one  direction of the protocol.\cite{liphardt2002,kiet2020}
It is necessarily less informative than using the Crooks relation, because the average over a work probability density is less informative than the probability density itself.
Note that analogous considerations apply to any other method based on work averages for one direction of the protocol. 
This includes, e.g., Zwanzig's free energy perturbation (FEP) formula,\cite{zwanzig2004the} which coincides with Jarzynski's estimator for instantaneous changes, and the linear response (LR) approximation, which is based on the assumption that work distributions are Gaussian.

Relationships between various free energy estimation methods based on work averages or distributions have been discussed in the literature. \cite{jarzynski1997,shirts2003equilibrium,zwanzig2004the,maragakis2008} 
Hence, in addition to comparing the predictions of the new \MDCB~method to experimental results, we here chose to compare them to the predictions of a method that is  completely different. 
This approach, which we abbreviate \PBMC, is based on the widely used continuum electrostatic method\cite{baptista1999,teixeira2002} and is known to perform reasonably well for these systems, providing a baseline for practical protein engineering applications. 

The change in redox potential of the heme group between m4D2 and mutants has previously been calculated with this approach.\cite{baptista1999,teixeira2002}
This method involves simulating the joint binding equilibrium of proton and electrons. It uses a combination of Poisson-Boltzmann (PB) calculations, e.g. with MEAD (version2.2.9),\cite{bashford1990,bashford1992,bashford1997} and Metropolis Monte Carlo (MC) calculations, using the software PETIT (version 1.6).\cite{baptista2001} 
The PB calculations compute the individual and pairwise terms needed to obtain the free energies of protonation/reduction changes. These energies are then used in the MC calculations.
The changes in the redox potential of the heme group relative to m4D2 is determined from the corresponding reduction curve by extracting the $E$-shift values corresponding to a reduced fraction of 0.5 in Fig.~\figSeleven. 

The structural model predicted by Rosetta\cite{Hutchins2023} was used for the calculations of m4D2, while models for the mutants were constructed using PyMOL.\cite{pymol2015}
One structure for each system, namely m4D2, T19D, M23N, R34Q, R92Q and the DM,  was used for the calculation. We note that a slightly different structural preparation protocol prior to the calculations gives slightly different results.\cite{Hutchins2023}    
The charges for all the atoms in the protein (except the heme group) and radii were taken from the GROMOS 54A7 force field\cite{schmid2011} using a previously described procedure.\cite{teixeira2005}
The partial charges for the heme group were taken from our previous work.\cite{Hutchins2023} 
These calculations use the temperature of $298$~K and a molecular surface defined with a solvent probe radius of $1.4\,\si{\angstrom}$.\cite{teixeira2005} 
The dielectric constants used for the solvent (\(\varepsilon\)$_{\text{sol}}$) and for the protein (\(\varepsilon\)$_{\text{protein}}$) were 80 and 20, respectively.\cite{teixeira2005} An ionic strength of $0.05\,\si{M}$ was used. 
The finite-difference linear PB calculations used a three-step focusing procedure\cite{gilson1988} employing consecutive grid spacing of $1.0$, $0.5$ and $0.25\,\si{\angstrom}$. 
Each MC calculation comprised $10^{\text{5}}$ MC steps and the acceptance/rejection of each step followed a Metropolis criterion\cite{metropolis1953} using the PB free energies. 

We refer to this alternative method as the \PBMC~method, and report the predicted redox shifts of the m4D2 mutants in column 5 in Tab.~\ref{table:results}.

\subsection{Experimental redox potentials} \label{subsec:experiment}

The heme reduction potentials of the M23N, R34Q, R92Q variants were determined here (Fig.~\figSsixteen) using optically-transparent thin-layer electrochemistry,\cite{ost20044}  using methods previously used for m4D2 and other mutants.\cite{Hutchins2023}
$120\,\si{\mu L}$ of \emph{de novo} protein samples were mixed with $12\,\si{\mu L}$ of glycerol and $0.5\,\si{\mu L}$ each of indigotrisulfonic acid, 2-hydroxy-1,4-naphthoquinone, phenazine, anthroquinone-2-sulfonate and benzyl viologen mediators at approximately $10\,\si{\mu M}$ concentration. 
The mediators are used to facilitate electron transfer between the working electrode and the heme cofactor and therefore promote rapid equilibration in the electrochemical cell.\cite{dutton197823} 

To obtain the reduction potentials of the proteins, a Biologic SP-150 was used to apply stepwise potentials between a thin platinum gauze working electrode and a platinum counter electrode, typically over a range of $-525$~\si{mV} to $-225$~\si{mV} vs a Ag/AgCl reference electrode also in the electrochemical cell. The protein sample and  thin Pt gauze working electrode were housed in a modified quartz EPR cuvette (Wilmad), with a pathlength of $0.3\,\si{mm}$, and the counter and reference electrodes were held above in a glycerol-free buffer layer within a fused glass side arm tube. UV-Visible absorbance spectra were recorded between 200-800 nm after 30 mins of equilibration at each potential to measure the evolution of the heme absorbance spectrum as cycled between ferric and ferrous states during reductive and oxidative sweeps of potential.
Redox potential measurements of horse heart cyctochrome \emph{c} and m4D2 were used to calibrate the Ag/AgCl reference electrode for each round of redox measurements, enabling reduction potentials to be quoted versus the Nernst Hydrogen Electrode (NHE).
The experiments were conducted at room temperature (ca. $T = 298$~\si{K}).

For these \emph{b}-type heme proteins with bis-histidine coordination, A$_{416 nm}$ represents the position of the oxidized, ferric Soret band, and A$_{429 nm}$ represents the position of the reduced, ferrous Soret band. 
The $\Delta$A$_{429 nm}$ was plotted against applied potential (\si{mV}), and the Nernst equation to calculate the redox midpoint potential ($E$):
\begin{equation}\label{eq:general_nernst}
    E_{\text{app}} = E - 
    \left( \frac{R\,T}{n\,F} \right) 
    \ln 
    \left( \frac{[\text{red}]}{[\text{ox}]} \right).
\end{equation}
Here, $E_{\text{app}}$ and $E$ are the applied potential and the redox potential, respectively; $R$, the universal gas constant; $T$, the temperature; $n$, the number of electrons being transferred; [red], the concentration of the reduced ferrous heme; [ox], the concentration of the oxidised ferric heme; and $F$, Faraday's constant.
For a normal reduction reaction such as
\begin{equation}\label{eq:e_transfer}
    {\text{oxidized}} + n \, {e^-} \ce{<=>} {\text{reduced}},
\end{equation}
the general Nernst equation \eqref{eq:general_nernst} can be used to describe the redox potential under non-standard conditions such as those in which the data were collected.

For these experiments, data was collected in triplicate and then processed using a Jupyter Notebook. 
Normalised, mean data was fit to a 1-electron Nernst equation using the SciPy optimize \texttt{curve\_fit} function.\cite{scikit-learn}

%%%
\section{Results and discussion}

Multiple long MD simulations were performed for the oxidized and reduced states of m4D2, four single  (T19D, M23N, R34Q, R92Q) and the T19D-T77D double mutant (DM). The trajectories provided ensembles of conformations (4000 for m4D2 and single mutants and 8000 for the double mutant) to calculate the instantaneous oxidation and reduction process of the proteins as described in section \ref{subsec:DNEMD} using the approach illustrated in Fig.~\ref{fig:main-results}. 

As outlined in section \ref{subsec:FR}, fluctuation relations were then used to calculate redox potential shifts for the mutants relative to m4D2 (see results in Tab.~\ref{table:results}), using the energy differences described above. The energy changes between the reduced and oxidized states were used to determine the statistical work values associated with reduction and oxidation, for each of the six proteins.
These work values (Fig.~\figStwelve) were used to calculate the probability density \eqref{eq:posterior-main}. 
This encodes the information that the MD simulations provide about the redox potential according to the Crooks relation. All six densities are reported in full in Fig.~\figSfourteen. 
These probabilities were used to calculate the Crooks-Bayes estimates Eq.~\eqref{eq:crooks-bayes}. 
The redox potential changes calculated in this way are given (relative to m4D2) in Tab.~\ref{table:results} (column 4).
The uncertainties for these values were propagated from the mean square errors \eqref{eq:est-err} for each redox potential, and are indicated in brackets.
The results for all six proteins showed statistical convergence from $\mu \simeq 2000$ data points (Fig.~\figSfifteen).

While Eq.~\eqref{eq:crooks-bayes}
is in principle capable of predicting absolute redox potentials (see Tab.~\tabSone~in the \SIn), it should be noted that the energy differences from the simulations used here do not account for quantum effects (such as, polarization and ionization energy) as those are not captured by molecular mechanics approaches. Therefore, we report the redox potentials as changes relative to m4D2, denoted as $\delta E$;  see Tab.~\ref{table:results}.

The calculated redox potential shifts for the m4D2 mutants determined using the Crooks relation here generally correlate well with the experimental values Tab.~\ref{table:results} (columns 3, 4). 
For T19D, the change in redox potential predicted by \MDCB~result is $-4$~\si{mV} (experimental shift is $-28$~\si{mV}) whereas for M23N is $14$~\si{mV} (experimental shift is $+1$~\si{mV}). The \MDCB~ predictions for the two arginine-to-glutamine mutants, namely R34Q and R92Q, are very similar ($-12$~\si{mV} for R34Q and $-14$~\si{mV} for R92Q), as also observed experimentally (experimental shift is $-31$ and $-32$~\si{mV} for R34Q and R92Q, respectively). 
The \MDCB~ predictions show a relatively small shift ($-12$~\si{mV}) for the double mutant, which in experiments displays a large shift of $-56$~\si{mV}. The \MDCB~calculated value for the double mutant is notably less negative than the experimental result. Possible reasons for the differences observed between the \MDCB~ predicted and experimental redox shifts are discussed below.
 
\begin{table}[bt]
\centering
\begin{tabular}{|l|r|r|r|r|} 
    \hline
    ~ & \multicolumn{2}{c|}{\textbf{experiment}  (\si{mV})  } & \multicolumn{2}{c|}{ \textbf{predicted}  $\delta E $  (\si{mV}) } \\
    protein &  $ E $  &  $\delta E$  & \MDCB & \PBMC \\
    \hline
    \hline
    m4D2\cite{Hutchins2023} & -118 (1) & 0\, & 0\, & 0 \\
    T19D\cite{Hutchins2023} & -146 (1) &\ -28 (1) & -4 (2) & -35 \\
    M23N & -119 (1) & 1 (1) & 14 (2) & 0    \\
    R34Q & -149 (1) &  -31 (1) & -12 (2) & -11   \\
    R92Q & -150 (1) &  -32 (1) & -14 (2) & -14  \\
    DM\cite{Hutchins2023} & -174 (1) & -56 (1) & -12 (2) & -67  \\
    \hline
\end{tabular}
\caption{Experimental redox potentials $\redpot$ (column 2) and corresponding changes $\redpotshift$ relative to m4D2 (column 3), where the double mutant (DM) is T19D-T77D. Previously measured redox potentials \cite{Hutchins2023} are indicated in the table.
Calculated redox potential changes relative to m4D2 using the proposed \MDCB~method, which post-processes the data generated by the MD simulations via the Crooks-Bayes estimator \eqref{eq:crooks-bayes},  are shown in column 4.
Calculated redox potential changes relative to m4D2 using a well established CE approach (column 5), combining PB calculations and MC simulations (\PBMC).\cite{baptista1999,teixeira2002}  
The errors associated with $\redpotshift$ were propagated from those for $\redpot$. 
}
\label{table:results}

\end{table}

The \MDCB~calculations correctly predict the sign of the redox potential change for all of the mutants. They also give the correct order of the redox potentials shifts for all the single mutants simulated: M23N \textgreater\, T19D \textgreater\, R34Q $\approx$ R92Q. Indeed, for single mutants, the Pearson correlation coefficient is $\rho_{\mathrm{corr}} = 0.97$. Interestingly, for the single mutants, the \MDCB~ method predicts redox potential shifts that are consistently offset in comparison to experiment, by around~-18 mV.

These findings indicate that this approach may be usefully predictive of redox potential changes for single mutants. The performance of the \MDCB~method is overall comparable to that of the \PBMC~approach for these proteins.

The good agreement for the single mutants indicate that the \MDCB~method can give good results. The energy differences for the double mutant probably arise from issues of modelling the structure, protonation states and sampling the conformational landscapes of this mutant, in one or both redox states. 
For the \MDCB~method to give good results, it is essential that the MD simulations sample the conformations of each state adequately (giving a representative ensemble of structures for each), and that they overlap sufficiently. Although several microseconds of simulation were performed for each system, the sampling gathered may not be enough to explore the conformations of these mutants in one or both redox states.
This is suggested by simulations of, for example, the oxidized state of R92Q, in which we find an unusually persistent direct hydrogen bond between glutamine in position 92 and the heme propionates, present in more than 35\% of the total simulation time (Fig.~\figSnine~B). The uncommonly high frequency of this interaction in the oxidized R92Q system suggests that these simulations may be trapped in an energy minimum. This persistent hydrogen bond is indeed observed in five of the ten R92Q trajectories for the oxidized state.
Predictions for double mutants are generally more difficult because the structural changes induced by two mutations are generally significantly larger than for single mutations. The dynamics of the specific double mutant here (T19D-T77D) are significantly altered from m4D2, due to the two extra negative charges. The introduced aspartate residues form strong electrostatic interactions with nearby positively charged residues (K36 and K94) (Fig.~\figSten). These new salt bridges significantly affect the overall dynamics of the protein, rigidifying it (Fig.~\figSeight). This may mean that longer simulations or more repeats are required to sample the conformational space of this mutant properly. Overall (and despite the limitations discussed in section 2.3), our results clearly show that using fluctuation relations to post-process MD simulations data is a reasonably reliable approach to predict redox shifts in proteins, given sufficient sampling of both redox states in MD.

To benchmark the \MDCB~method, we also calculated the changes in redox potential relative to m4D2 using a well established CE approach, which combines Poisson-Boltzmann (PB) electrostatic calculations with Metropolis Monte Carlo (MC) simulations. \cite{baptista1999,teixeira2002} This method, which uses simplified models for both the solvent (dielectric continuum) and the protein (atomic point charges immersed in a low dielectric medium), allows fast calculation of the free-energy terms associated with redox and protonation changes.\cite{baptista1999,teixeira2002} 
These energy terms are then used to sample protonation and redox states using MC.\cite{baptista1999,teixeira2002}.This PB+MC method has already proven to be a valuable tool for the redox engineering of m4D2: it correctly predicted the order of reduction of the designed m4D2 mutants using predicted structures from Rosetta.\cite{Hutchins2023}  

The PB+MC results are presented in (Tab.~\ref{table:results}, column 5) and (Fig.~\figSeleven). Overall, we observe that for some mutants, e.g. T19D and M23N, the \PBMC~predictions are closer to the experimentally measured redox shifts while for others, e.g. R34Q and R92Q, the new method combining MD simulations and fluctuations relations (\MDCB) performs as well as the \PBMC~ approach. 

It should be noted that in the \PBMC~ calculations, the dynamic behaviour of the proteins is implicitly modelled using a dielectric constant. In contrast, a large set of conformations (4000 for m4D2, T19D, M23N, R34Q and R92Q, and 8000 conformations for the DM) was used for the \MDCB~ predictions, thus meaning that in this approach, protein dynamics is being explicitly factored into the calculations. This may be an advantage for systems that undergo conformational changes during the reduction/oxidation process. A potentially significant effect included in the \PBMC~and not in the \MDCB~is the inclusion of protonation state changes in the protein associated with redox changes. This may account for the better performance of \PBMC~in some cases. A potentially useful extension to the \MDCB~ method would be the inclusion of protonation state changes, e.g. through MC calculations or constant pH MD.

For all the mutants studied here, both methods give results with a similar correlation strength with the experimental values, namely $0.85$ and $0.84$ for the \MDCB~and \PBMC~methods, respectively. However, for the single mutants only, the \MDCB~method shows a correlation of $0.97$ with the experimental data compared with $0.61$ obtained for the \PBMC~method---thus reinforcing the earlier claim that fluctuation relations are, when combined with MD simulations, a valid predictive tool for calculating redox potential changes. 

\section{Conclusions}

Natural and designed redox proteins are increasingly widely used in technological applications (e.g., in biocatalysis and biomolecular electronics). For such applications, there is a need to be able to predict protein redox potentials, e.g. to aid in designing mutations to change the redox potential to optimize it for a particular application. Molecular simulation tools are potentially useful in this context, to suggest candidates for experimental characterization, and to understand the causes of observed redox potential changes. Here, we have proposed and tested a method for calculation of redox potentials from MD simulations with application of fluctuation relations and Bayesian inference. Comparison of the predictions of the \MDCB~approach against experimentally measured redox potentials for point variants of a \emph{de novo} heme protein indicate that the method is usefully predictive for relative potential shifts. Comparison with the completely different   \PBMC~method indicates that the approach proposed here performs similarly well for single point mutations. The \MDCB~method can be readily applied as a complement to standard equilibrium MD simulations of different redox states, and so may be useful in protein design and engineering applications.     
\paragraph{}

\textbf{Data availability statement} The files containing the energy differences extracted from the MD simulations for m4D2 and mutants, as well as the scripts needed for the prediction of the redox potential shifts using the \MDCB~approach, are available on GitHub (\href{https://github.com/jesus-rubiojimenez/FlucRels-CrooksBayes}{github.com/jesus-rubiojimenez/FlucRels-CrooksBayes}).

%%%%%%%%%%%%%%%%%%%%%%%%%%%%%%%%%%%%%%%%%%%%%%%%%%%%%%%%%%%%%%%%%%%%%
%% Acknowledgement
%%%%%%%%%%%%%%%%%%%%%%%%%%%%%%%%%%%%%%%%%%%%%%%%%%%%%%%%%%%%%%%%%%%%%
\begin{acknowledgement}

JR thanks F Cerisola, M Rider and WP Wardley for helpful discussions. 
This work is part of a project that has received funding from the European Research Council (ERC) under the European Union’s Horizon 2020 research and innovation programme (Grant agreement No. 101021207): AJM and ASFO thank ERC for funding for the PREDACTED Advanced grant. AJM and ASFO also thank EPSRC (grant number EP/M022609/1) and BBSRC (grant number (BB/R016445/1 and BB/X009831/1) for support, as well as BrisSynBio, a BBSRC/EPSRC Synthetic Biology Research Centre (Grant Number:BB/L01386X/1). JLRA, ASFO and AJM acknowledge the UKRI sLoLA grant BB/W003449/1. ASFO thanks Oracle for Research for funding. 
JA and JR acknowledge support from EPSRC (Grants Nos. EP/T002875/1 and EP/R045577/1), and JA thanks the Royal Society for support. 
JR also acknowledges support from the Surrey Future Fellowship Programme. 
MD simulations were carried out using the computational facilities of the Advanced Computing Research Centre, University of Bristol (http://www.bris.ac.uk/acrc). 

\end{acknowledgement}

\bigskip 

\noindent \textbf{ASSOCIATED CONTENT} \\
Supporting Information available. The supporting information file contains six sections, namely S.1-S.6. In section S.1, several plots evidencing the conformational stability of the MD simulations are shown (Fig. S1-S10). Section S.2 displays the reduction curves for the mutants relative to m4D2 determined using the PB+MC method (Fig. S11). Section S.3 discusses the mathematical procedure to estimate free energy differences by combining the Crooks fluctuation relation with Bayes theorem. 
Section S.4 contains several figures displaying the non-equilibrium work values computed from the simulation data (Fig S12), the work histograms for the forward and backward protocols (Fig. S13), the final posterior probabilities (Fig. S14), and the convergence of our redox potential estimator (Fig. S15). Section S.5 shows a figure with the experimentally determined redox potentials for the M23N, R34Q and R92Q mutants (Fig. 16). Section S.6 provides a table with the experimental and predicted redox potentials for m4D2, T19D, M23N, R34Q, R92Q, and T19D-T77D (Tab. S1).

%%%%%%%%%%%%%%%%%%%%%%%%%%%%%%%%%%%%%%%%%%%%%%%%%%%%%%%%%%%%%%%%%%%%%
%% Bibliography
%%%%%%%%%%%%%%%%%%%%%%%%%%%%%%%%%%%%%%%%%%%%%%%%%%%%%%%%%%%%%%%%%%%%%
\bibliography{refs2023oct}

\providecommand{\latin}[1]{#1}
\makeatletter
\providecommand{\doi}
  {\begingroup\let\do\@makeother\dospecials
  \catcode`\{=1 \catcode`\}=2 \doi@aux}
\providecommand{\doi@aux}[1]{\endgroup\texttt{#1}}
\makeatother
\providecommand*\mcitethebibliography{\thebibliography}
\csname @ifundefined\endcsname{endmcitethebibliography}
  {\let\endmcitethebibliography\endthebibliography}{}
\begin{mcitethebibliography}{86}
\providecommand*\natexlab[1]{#1}
\providecommand*\mciteSetBstSublistMode[1]{}
\providecommand*\mciteSetBstMaxWidthForm[2]{}
\providecommand*\mciteBstWouldAddEndPuncttrue
  {\def\EndOfBibitem{\unskip.}}
\providecommand*\mciteBstWouldAddEndPunctfalse
  {\let\EndOfBibitem\relax}
\providecommand*\mciteSetBstMidEndSepPunct[3]{}
\providecommand*\mciteSetBstSublistLabelBeginEnd[3]{}
\providecommand*\EndOfBibitem{}
\mciteSetBstSublistMode{f}
\mciteSetBstMaxWidthForm{subitem}{(\alph{mcitesubitemcount})}
\mciteSetBstSublistLabelBeginEnd
  {\mcitemaxwidthsubitemform\space}
  {\relax}
  {\relax}

\bibitem[Liu \latin{et~al.}(2014)Liu, Chakraborty, Hosseinzadeh, Yu, Tian,
  Petrik, Bhagi, and Lu]{Liu2014}
Liu,~J.; Chakraborty,~S.; Hosseinzadeh,~P.; Yu,~Y.; Tian,~S.; Petrik,~I.;
  Bhagi,~A.; Lu,~Y. Metalloproteins Containing Cytochrome, Iron–Sulfur, or
  Copper Redox Centers. \emph{Chemical Reviews} \textbf{2014}, \emph{114},
  4366--4469, PMID: 24758379\relax
\mciteBstWouldAddEndPuncttrue
\mciteSetBstMidEndSepPunct{\mcitedefaultmidpunct}
{\mcitedefaultendpunct}{\mcitedefaultseppunct}\relax
\EndOfBibitem
\bibitem[Ayuso-Fern\'{a}ndez \latin{et~al.}(2019)Ayuso-Fern\'{a}ndez, De~Lacey,
  Ca{\~n}ada, Ruiz-Due{\~n}as, and Mart\'{i}nez]{AyusoFernandez2019}
Ayuso-Fern\'{a}ndez,~I.; De~Lacey,~A.~L.; Ca{\~n}ada,~F.~J.;
  Ruiz-Due{\~n}as,~F.~J.; Mart\'{i}nez,~A.~T. {Increase of Redox Potential
  during the Evolution of Enzymes Degrading Recalcitrant Lignin}.
  \emph{Chemistry--A European Journal} \textbf{2019}, \emph{25},
  2708--2712\relax
\mciteBstWouldAddEndPuncttrue
\mciteSetBstMidEndSepPunct{\mcitedefaultmidpunct}
{\mcitedefaultendpunct}{\mcitedefaultseppunct}\relax
\EndOfBibitem
\bibitem[Hosseinzadeh and Lu(2016)Hosseinzadeh, and Lu]{hosseinzadeh2016}
Hosseinzadeh,~P.; Lu,~Y. Design and fine-tuning redox potentials of
  metalloproteins involved in electron transfer in bioenergetics.
  \emph{Biochimica et Biophysica Acta (BBA)-Bioenergetics} \textbf{2016},
  \emph{1857}, 557--581\relax
\mciteBstWouldAddEndPuncttrue
\mciteSetBstMidEndSepPunct{\mcitedefaultmidpunct}
{\mcitedefaultendpunct}{\mcitedefaultseppunct}\relax
\EndOfBibitem
\bibitem[Page \latin{et~al.}(1999)Page, Moser, Chen, and Dutton]{Page1999}
Page,~C.~C.; Moser,~C.~C.; Chen,~X.; Dutton,~P.~L. Natural engineering
  principles of electron tunnelling in biological oxidation–reduction.
  \emph{Nature} \textbf{1999}, \emph{402}, 47–52\relax
\mciteBstWouldAddEndPuncttrue
\mciteSetBstMidEndSepPunct{\mcitedefaultmidpunct}
{\mcitedefaultendpunct}{\mcitedefaultseppunct}\relax
\EndOfBibitem
\bibitem[Poulos(2014)]{poulos2014}
Poulos,~T.~L. Heme enzyme structure and function. \emph{Chemical reviews}
  \textbf{2014}, \emph{114}, 3919--3962\relax
\mciteBstWouldAddEndPuncttrue
\mciteSetBstMidEndSepPunct{\mcitedefaultmidpunct}
{\mcitedefaultendpunct}{\mcitedefaultseppunct}\relax
\EndOfBibitem
\bibitem[Yu \latin{et~al.}(2019)Yu, Liu, and Wang]{yu2019}
Yu,~Y.; Liu,~X.; Wang,~J. {Expansion of Redox Chemistry in Designer
  Metalloenzymes}. \emph{Accounts of Chemical Research} \textbf{2019},
  \emph{52}, 557--565\relax
\mciteBstWouldAddEndPuncttrue
\mciteSetBstMidEndSepPunct{\mcitedefaultmidpunct}
{\mcitedefaultendpunct}{\mcitedefaultseppunct}\relax
\EndOfBibitem
\bibitem[Koebke \latin{et~al.}(2022)Koebke, Pinter, Pitts, and
  Pecoraro]{Koebke2022}
Koebke,~K.~J.; Pinter,~T. B.~J.; Pitts,~W.~C.; Pecoraro,~V.~L. Catalysis and
  Electron Transfer in De Novo Designed Metalloproteins. \emph{Chemical
  Reviews} \textbf{2022}, \emph{122}, 12046--12109, PMID: 35763791\relax
\mciteBstWouldAddEndPuncttrue
\mciteSetBstMidEndSepPunct{\mcitedefaultmidpunct}
{\mcitedefaultendpunct}{\mcitedefaultseppunct}\relax
\EndOfBibitem
\bibitem[Prabhulkar \latin{et~al.}(2012)Prabhulkar, Tian, Wang, Zhu, and
  Li]{prabhulkar2012}
Prabhulkar,~S.; Tian,~H.; Wang,~X.; Zhu,~J.-J.; Li,~C.-Z. Engineered proteins:
  redox properties and their applications. \emph{Antioxidants \& redox
  signaling} \textbf{2012}, \emph{17}, 1796--1822\relax
\mciteBstWouldAddEndPuncttrue
\mciteSetBstMidEndSepPunct{\mcitedefaultmidpunct}
{\mcitedefaultendpunct}{\mcitedefaultseppunct}\relax
\EndOfBibitem
\bibitem[Bhagi-Damodaran \latin{et~al.}(2014)Bhagi-Damodaran, Petrik, Marshall,
  Robinson, and Lu]{bhagi2014}
Bhagi-Damodaran,~A.; Petrik,~I.~D.; Marshall,~N.~M.; Robinson,~H.; Lu,~Y.
  {Systematic Tuning of Heme Redox Potentials and Its Effects on O2 Reduction
  Rates in a Designed Oxidase in Myoglobin}. \emph{Journal of the American
  Chemical Society} \textbf{2014}, \emph{136}, 11882--11885\relax
\mciteBstWouldAddEndPuncttrue
\mciteSetBstMidEndSepPunct{\mcitedefaultmidpunct}
{\mcitedefaultendpunct}{\mcitedefaultseppunct}\relax
\EndOfBibitem
\bibitem[Bhagi-Damodaran \latin{et~al.}(2018)Bhagi-Damodaran, Reed, Zhu, Shi,
  Hosseinzadeh, Sandoval, Harnden, Wang, Sponholtz, Mirts, Dwaraknath, Zhang,
  Mo{\"e}nne-Loccoz, and Lu]{bhagi2018}
Bhagi-Damodaran,~A.; Reed,~J.~H.; Zhu,~Q.; Shi,~Y.; Hosseinzadeh,~P.;
  Sandoval,~B.~A.; Harnden,~K.~A.; Wang,~S.; Sponholtz,~M.~R.; Mirts,~E.~N.;
  Dwaraknath,~S.; Zhang,~Y.; Mo{\"e}nne-Loccoz,~P.; Lu,~Y. Heme redox
  potentials hold the key to reactivity differences between nitric oxide
  reductase and heme-copper oxidase. \emph{Proceedings of the National Academy
  of Sciences} \textbf{2018}, \emph{115}, 6195--6200\relax
\mciteBstWouldAddEndPuncttrue
\mciteSetBstMidEndSepPunct{\mcitedefaultmidpunct}
{\mcitedefaultendpunct}{\mcitedefaultseppunct}\relax
\EndOfBibitem
\bibitem[Bunzel \latin{et~al.}(2021)Bunzel, Anderson, and
  Mulholland]{bunzel2021}
Bunzel,~H.~A.; Anderson,~J.~R.; Mulholland,~A.~J. {Designing better enzymes:
  Insights from directed evolution}. \emph{Current Opinion in Structural
  Biology} \textbf{2021}, \emph{67}, 212--218\relax
\mciteBstWouldAddEndPuncttrue
\mciteSetBstMidEndSepPunct{\mcitedefaultmidpunct}
{\mcitedefaultendpunct}{\mcitedefaultseppunct}\relax
\EndOfBibitem
\bibitem[Breuer \latin{et~al.}(2017)Breuer, Rosso, and Blumberger]{breuer2017}
Breuer,~M.; Rosso,~K.~M.; Blumberger,~J. {Redox potentials in the decaheme
  cytochrome MtrF: Poisson{\textendash}Boltzmann vs. molecular dynamics
  simulations}. \emph{Proceedings of the National Academy of Sciences}
  \textbf{2017}, \emph{114}, E10028--E10028\relax
\mciteBstWouldAddEndPuncttrue
\mciteSetBstMidEndSepPunct{\mcitedefaultmidpunct}
{\mcitedefaultendpunct}{\mcitedefaultseppunct}\relax
\EndOfBibitem
\bibitem[Watanabe \latin{et~al.}(2017)Watanabe, Yamashita, and
  Ishikita]{watanabe2017}
Watanabe,~H.~C.; Yamashita,~Y.; Ishikita,~H. {Reply to Breuer et al.: Molecular
  dynamics simulations do not provide functionally relevant values of redox
  potential in MtrF}. \emph{Proceedings of the National Academy of Sciences}
  \textbf{2017}, \emph{114}, E10029--E10030\relax
\mciteBstWouldAddEndPuncttrue
\mciteSetBstMidEndSepPunct{\mcitedefaultmidpunct}
{\mcitedefaultendpunct}{\mcitedefaultseppunct}\relax
\EndOfBibitem
\bibitem[Cruzeiro \latin{et~al.}(2018)Cruzeiro, Amaral, and
  Roitberg]{cruzeiro2018}
Cruzeiro,~V. W.~D.; Amaral,~M.~S.; Roitberg,~A.~E. {Redox potential replica
  exchange molecular dynamics at constant pH in AMBER: Implementation and
  validation}. \emph{The Journal of Chemical Physics} \textbf{2018},
  \emph{149}, 072338\relax
\mciteBstWouldAddEndPuncttrue
\mciteSetBstMidEndSepPunct{\mcitedefaultmidpunct}
{\mcitedefaultendpunct}{\mcitedefaultseppunct}\relax
\EndOfBibitem
\bibitem[Warshel and Dryga(2011)Warshel, and Dryga]{warshel2011}
Warshel,~A.; Dryga,~A. {Simulating electrostatic energies in proteins:
  Perspectives and some recent studies of pKas, redox, and other crucial
  functional properties}. \emph{Proteins: Structure, Function, and
  Bioinformatics} \textbf{2011}, \emph{79}, 3469--3484\relax
\mciteBstWouldAddEndPuncttrue
\mciteSetBstMidEndSepPunct{\mcitedefaultmidpunct}
{\mcitedefaultendpunct}{\mcitedefaultseppunct}\relax
\EndOfBibitem
\bibitem[Jin \latin{et~al.}(2016)Jin, Goyal, Das, Gaus, Meuwly, and
  Cui]{haiyun2016}
Jin,~H.; Goyal,~P.; Das,~A.~K.; Gaus,~M.; Meuwly,~M.; Cui,~Q. {Copper
  Oxidation/Reduction in Water and Protein: Studies with DFTB3/MM and VALBOND
  Molecular Dynamics Simulations}. \emph{The journal of physical chemistry. B}
  \textbf{2016}, \emph{120}, 1894—1910\relax
\mciteBstWouldAddEndPuncttrue
\mciteSetBstMidEndSepPunct{\mcitedefaultmidpunct}
{\mcitedefaultendpunct}{\mcitedefaultseppunct}\relax
\EndOfBibitem
\bibitem[Formaneck \latin{et~al.}(2002)Formaneck, Li, Zhang, and
  Cui]{formaneck2002}
Formaneck,~M.~S.; Li,~G.; Zhang,~X.; Cui,~Q. {Calculating accurate redox
  potentials in enzymes with a combined QM/MM free energy perturbation
  approach}. \emph{Journal of Theoretical and Computational Chemistry}
  \textbf{2002}, \emph{01}, 53--67\relax
\mciteBstWouldAddEndPuncttrue
\mciteSetBstMidEndSepPunct{\mcitedefaultmidpunct}
{\mcitedefaultendpunct}{\mcitedefaultseppunct}\relax
\EndOfBibitem
\bibitem[Blumberger(2008)]{Blumberger2008}
Blumberger,~J. Free energies for biological electron transfer from QM/MM
  calculation: method{,} application and critical assessment. \emph{Physical
  Chemistry Chemical Physics} \textbf{2008}, \emph{10}, 5651--5667\relax
\mciteBstWouldAddEndPuncttrue
\mciteSetBstMidEndSepPunct{\mcitedefaultmidpunct}
{\mcitedefaultendpunct}{\mcitedefaultseppunct}\relax
\EndOfBibitem
\bibitem[Blumberger(2015)]{Blumberger2015}
Blumberger,~J. Recent Advances in the Theory and Molecular Simulation of
  Biological Electron Transfer Reactions. \emph{Chemical Reviews}
  \textbf{2015}, \emph{115}, 11191--11238, PMID: 26485093\relax
\mciteBstWouldAddEndPuncttrue
\mciteSetBstMidEndSepPunct{\mcitedefaultmidpunct}
{\mcitedefaultendpunct}{\mcitedefaultseppunct}\relax
\EndOfBibitem
\bibitem[Blumberger and Klein(2006)Blumberger, and Klein]{Blumberger2006}
Blumberger,~J.; Klein,~M.~L. Reorganization Free Energies for Long-Range
  Electron Transfer in a Porphyrin-Binding Four-Helix Bundle Protein.
  \emph{Journal of the American Chemical Society} \textbf{2006}, \emph{128},
  13854--13867, PMID: 17044714\relax
\mciteBstWouldAddEndPuncttrue
\mciteSetBstMidEndSepPunct{\mcitedefaultmidpunct}
{\mcitedefaultendpunct}{\mcitedefaultseppunct}\relax
\EndOfBibitem
\bibitem[Gunner and Baker(2016)Gunner, and Baker]{gunner2016}
Gunner,~M.; Baker,~N. In \emph{Computational Approaches for Studying Enzyme
  Mechanism Part B}; Voth,~G.~A., Ed.; Methods in Enzymology; Academic Press,
  2016; Vol. 578; pp 1--20\relax
\mciteBstWouldAddEndPuncttrue
\mciteSetBstMidEndSepPunct{\mcitedefaultmidpunct}
{\mcitedefaultendpunct}{\mcitedefaultseppunct}\relax
\EndOfBibitem
\bibitem[Warshel \latin{et~al.}(2006)Warshel, Sharma, Kato, and
  Parson]{warshel2006}
Warshel,~A.; Sharma,~P.~K.; Kato,~M.; Parson,~W.~W. Modeling electrostatic
  effects in proteins. \emph{Biochimica et Biophysica Acta (BBA) - Proteins and
  Proteomics} \textbf{2006}, \emph{1764}, 1647--1676\relax
\mciteBstWouldAddEndPuncttrue
\mciteSetBstMidEndSepPunct{\mcitedefaultmidpunct}
{\mcitedefaultendpunct}{\mcitedefaultseppunct}\relax
\EndOfBibitem
\bibitem[Teixeira \latin{et~al.}(2002)Teixeira, Soares, and
  Baptista]{teixeira2002}
Teixeira,~V.~H.; Soares,~C.~M.; Baptista,~A.~M. {Studies of the reduction and
  protonation behavior of tetraheme cytochromes using atomic detail}.
  \emph{Journal of Biological Inorganic Chemistry} \textbf{2002}, \emph{7},
  200–216\relax
\mciteBstWouldAddEndPuncttrue
\mciteSetBstMidEndSepPunct{\mcitedefaultmidpunct}
{\mcitedefaultendpunct}{\mcitedefaultseppunct}\relax
\EndOfBibitem
\bibitem[Cruzeiro \latin{et~al.}(2020)Cruzeiro, Feliciano, and
  Roitberg]{cruzeiro2020}
Cruzeiro,~V. W.~D.; Feliciano,~G.~T.; Roitberg,~A.~E. Exploring Coupled Redox
  and pH Processes with a Force-Field-Based Approach: Applications to Five
  Different Systems. \emph{Journal of the American Chemical Society}
  \textbf{2020}, \emph{142}, 3823--3835, PMID: 32011132\relax
\mciteBstWouldAddEndPuncttrue
\mciteSetBstMidEndSepPunct{\mcitedefaultmidpunct}
{\mcitedefaultendpunct}{\mcitedefaultseppunct}\relax
\EndOfBibitem
\bibitem[Machuqueiro and Baptista(2009)Machuqueiro, and
  Baptista]{machuqueiro2009}
Machuqueiro,~M.; Baptista,~A.~M. Molecular Dynamics at Constant pH and
  Reduction Potential: Application to Cytochrome c3. \emph{Journal of the
  American Chemical Society} \textbf{2009}, \emph{131}, 12586--12594, PMID:
  19685871\relax
\mciteBstWouldAddEndPuncttrue
\mciteSetBstMidEndSepPunct{\mcitedefaultmidpunct}
{\mcitedefaultendpunct}{\mcitedefaultseppunct}\relax
\EndOfBibitem
\bibitem[Evans and Searles(1994)Evans, and Searles]{Evans1994}
Evans,~D.~J.; Searles,~D.~J. Equilibrium microstates which generate second law
  violating steady states. \emph{Physical Review E} \textbf{1994}, \emph{50},
  1645--1648\relax
\mciteBstWouldAddEndPuncttrue
\mciteSetBstMidEndSepPunct{\mcitedefaultmidpunct}
{\mcitedefaultendpunct}{\mcitedefaultseppunct}\relax
\EndOfBibitem
\bibitem[Jarzynski(1997)]{jarzynski1997}
Jarzynski,~C. Nonequilibrium Equality for Free Energy Differences.
  \emph{Physical Review Letters} \textbf{1997}, \emph{78}, 2690--2693\relax
\mciteBstWouldAddEndPuncttrue
\mciteSetBstMidEndSepPunct{\mcitedefaultmidpunct}
{\mcitedefaultendpunct}{\mcitedefaultseppunct}\relax
\EndOfBibitem
\bibitem[Crooks(1999)]{crooks1999}
Crooks,~G.~E. Entropy production fluctuation theorem and the nonequilibrium
  work relation for free energy differences. \emph{Physical Review E}
  \textbf{1999}, \emph{60}, 2721--2726\relax
\mciteBstWouldAddEndPuncttrue
\mciteSetBstMidEndSepPunct{\mcitedefaultmidpunct}
{\mcitedefaultendpunct}{\mcitedefaultseppunct}\relax
\EndOfBibitem
\bibitem[Evans and Searles(2002)Evans, and Searles]{Evans2002}
Evans,~D.~J.; Searles,~D.~J. {The Fluctuation Theorem}. \emph{Advances in
  Physics} \textbf{2002}, \emph{51}, 1529--1585\relax
\mciteBstWouldAddEndPuncttrue
\mciteSetBstMidEndSepPunct{\mcitedefaultmidpunct}
{\mcitedefaultendpunct}{\mcitedefaultseppunct}\relax
\EndOfBibitem
\bibitem[Seifert(2012)]{seifert2012}
Seifert,~U. Stochastic thermodynamics, fluctuation theorems and molecular
  machines. \emph{Reports on progress in physics} \textbf{2012}, \emph{75},
  126001--126001\relax
\mciteBstWouldAddEndPuncttrue
\mciteSetBstMidEndSepPunct{\mcitedefaultmidpunct}
{\mcitedefaultendpunct}{\mcitedefaultseppunct}\relax
\EndOfBibitem
\bibitem[Liphardt \latin{et~al.}(2002)Liphardt, Dumont, Smith, Tinoco, and
  Bustamante]{liphardt2002}
Liphardt,~J.; Dumont,~S.; Smith,~S.~B.; Tinoco,~I.; Bustamante,~C. {Equilibrium
  Information from Nonequilibrium Measurements in an Experimental Test of
  Jarzynski's Equality}. \emph{Science} \textbf{2002}, \emph{296},
  1832--1835\relax
\mciteBstWouldAddEndPuncttrue
\mciteSetBstMidEndSepPunct{\mcitedefaultmidpunct}
{\mcitedefaultendpunct}{\mcitedefaultseppunct}\relax
\EndOfBibitem
\bibitem[Collin \latin{et~al.}(2005)Collin, Ritort, Jarzynski, Smith, Tinoco,
  and Bustamante]{Collin2005}
Collin,~D.; Ritort,~F.; Jarzynski,~C.; Smith,~S.~B.; Tinoco,~I.; Bustamante,~C.
  {Verification of the Crooks fluctuation theorem and recovery of RNA folding
  free energies}. \emph{Nature} \textbf{2005}, \emph{437}, 231--234\relax
\mciteBstWouldAddEndPuncttrue
\mciteSetBstMidEndSepPunct{\mcitedefaultmidpunct}
{\mcitedefaultendpunct}{\mcitedefaultseppunct}\relax
\EndOfBibitem
\bibitem[Camunas-Soler \latin{et~al.}(2017)Camunas-Soler, Alemany, and
  Ritort]{camunas2017}
Camunas-Soler,~J.; Alemany,~A.; Ritort,~F. Experimental measurement of binding
  energy, selectivity, and allostery using fluctuation theorems. \emph{Science}
  \textbf{2017}, \emph{355}, 412—415\relax
\mciteBstWouldAddEndPuncttrue
\mciteSetBstMidEndSepPunct{\mcitedefaultmidpunct}
{\mcitedefaultendpunct}{\mcitedefaultseppunct}\relax
\EndOfBibitem
\bibitem[Huggins \latin{et~al.}(2019)Huggins, Biggin, Dämgen, Essex, Harris,
  Henchman, Khalid, Kuzmanic, Laughton, Michel, Mulholland, Rosta, Sansom, and
  van~der Kamp]{huggins2019}
Huggins,~D.~J.; Biggin,~P.~C.; Dämgen,~M.~A.; Essex,~J.~W.; Harris,~S.~A.;
  Henchman,~R.~H.; Khalid,~S.; Kuzmanic,~A.; Laughton,~C.~A.; Michel,~J.;
  Mulholland,~A.~J.; Rosta,~E.; Sansom,~M. S.~P.; van~der Kamp,~M.~W.
  Biomolecular simulations: From dynamics and mechanisms to computational
  assays of biological activity. \emph{WIREs Computational Molecular Science}
  \textbf{2019}, \emph{9}, e1393\relax
\mciteBstWouldAddEndPuncttrue
\mciteSetBstMidEndSepPunct{\mcitedefaultmidpunct}
{\mcitedefaultendpunct}{\mcitedefaultseppunct}\relax
\EndOfBibitem
\bibitem[Huang \latin{et~al.}(2003)Huang, Gibney, Stayrook, {Leslie Dutton},
  and Lewis]{huang2003}
Huang,~S.~S.; Gibney,~B.~R.; Stayrook,~S.~E.; {Leslie Dutton},~P.; Lewis,~M.
  X-ray Structure of a Maquette Scaffold. \emph{Journal of Molecular Biology}
  \textbf{2003}, \emph{326}, 1219--1225\relax
\mciteBstWouldAddEndPuncttrue
\mciteSetBstMidEndSepPunct{\mcitedefaultmidpunct}
{\mcitedefaultendpunct}{\mcitedefaultseppunct}\relax
\EndOfBibitem
\bibitem[Koder \latin{et~al.}(2006)Koder, Valentine, Cerda, Noy, Smith, Wand,
  and Dutton]{koder2006}
Koder,~R.~L.; Valentine,~K.~G.; Cerda,~J.; Noy,~D.; Smith,~K.~M.; Wand,~A.~J.;
  Dutton,~P.~L. Nativelike Structure in Designed Four $\alpha$-Helix Bundles
  Driven by Buried Polar Interactions. \emph{Journal of the American Chemical
  Society} \textbf{2006}, \emph{128}, 14450--14451, PMID: 17090015\relax
\mciteBstWouldAddEndPuncttrue
\mciteSetBstMidEndSepPunct{\mcitedefaultmidpunct}
{\mcitedefaultendpunct}{\mcitedefaultseppunct}\relax
\EndOfBibitem
\bibitem[Fleishman \latin{et~al.}(2011)Fleishman, Leaver-Fay, Corn, Strauch,
  Khare, Koga, Ashworth, Murphy, Richter, Lemmon, Meiler, and
  Baker]{Fleishman2011}
Fleishman,~S.~J.; Leaver-Fay,~A.; Corn,~J.~E.; Strauch,~E.-M.; Khare,~S.~D.;
  Koga,~N.; Ashworth,~J.; Murphy,~P.; Richter,~F.; Lemmon,~G.; Meiler,~J.;
  Baker,~D. RosettaScripts: A Scripting Language Interface to the Rosetta
  Macromolecular Modeling Suite. \emph{PLOS ONE} \textbf{2011}, \emph{6},
  1--10\relax
\mciteBstWouldAddEndPuncttrue
\mciteSetBstMidEndSepPunct{\mcitedefaultmidpunct}
{\mcitedefaultendpunct}{\mcitedefaultseppunct}\relax
\EndOfBibitem
\bibitem[Hutchins \latin{et~al.}(2023)Hutchins, Noble, Bunzel, Williams,
  Dubiel, Yadav, Molinaro, Barringer, Blackburn, Hardy, Parnell, Landau, Race,
  Oliver, Koder, Crump, Schaffitzel, Oliveira, Mulholland, and
  Anderson]{Hutchins2023}
Hutchins,~G.~H.; Noble,~C. E.~M.; Bunzel,~H.~A.; Williams,~C.; Dubiel,~P.;
  Yadav,~S. K.~N.; Molinaro,~P.~M.; Barringer,~R.; Blackburn,~H.; Hardy,~B.~J.;
  Parnell,~A.~E.; Landau,~C.; Race,~P.~R.; Oliver,~T. A.~A.; Koder,~R.~L.;
  Crump,~M.~P.; Schaffitzel,~C.; Oliveira,~A. S.~F.; Mulholland,~A.~J.;
  Anderson,~J. L.~R. An expandable, modular de novo protein platform for
  precision redox engineering. \emph{Proceedings of the National Academy of
  Sciences} \textbf{2023}, \emph{120}, e2306046120\relax
\mciteBstWouldAddEndPuncttrue
\mciteSetBstMidEndSepPunct{\mcitedefaultmidpunct}
{\mcitedefaultendpunct}{\mcitedefaultseppunct}\relax
\EndOfBibitem
\bibitem[Goodin and McRee(1993)Goodin, and McRee]{goodin1993}
Goodin,~D.~B.; McRee,~D.~E. The Asp-His-iron triad of cytochrome c peroxidase
  controls the reduction potential electronic structure, and coupling of the
  tryptophan free radical to the heme. \emph{Biochemistry} \textbf{1993},
  \emph{32}, 3313--3324\relax
\mciteBstWouldAddEndPuncttrue
\mciteSetBstMidEndSepPunct{\mcitedefaultmidpunct}
{\mcitedefaultendpunct}{\mcitedefaultseppunct}\relax
\EndOfBibitem
\bibitem[Ortmayer \latin{et~al.}(2020)Ortmayer, Fisher, Basran, Wolde-Michael,
  Heyes, Levy, Lovelock, Anderson, Raven, Hay, \latin{et~al.}
  others]{ortmayer2020}
Ortmayer,~M.; Fisher,~K.; Basran,~J.; Wolde-Michael,~E.~M.; Heyes,~D.~J.;
  Levy,~C.; Lovelock,~S.~L.; Anderson,~J.~R.; Raven,~E.~L.; Hay,~S.,
  \latin{et~al.}  Rewiring the “Push-Pull” catalytic machinery of a heme
  enzyme using an expanded genetic code. \emph{ACS catalysis} \textbf{2020},
  \emph{10}, 2735--2746\relax
\mciteBstWouldAddEndPuncttrue
\mciteSetBstMidEndSepPunct{\mcitedefaultmidpunct}
{\mcitedefaultendpunct}{\mcitedefaultseppunct}\relax
\EndOfBibitem
\bibitem[Ghirlanda \latin{et~al.}(2004)Ghirlanda, Osyczka, Liu, Antolovich,
  Smith, Dutton, Wand, and DeGrado]{giovanna2004}
Ghirlanda,~G.; Osyczka,~A.; Liu,~W.; Antolovich,~M.; Smith,~K.~M.;
  Dutton,~P.~L.; Wand,~A.~J.; DeGrado,~W.~F. {De Novo Design of a
  D2-Symmetrical Protein that Reproduces the Diheme Four-Helix Bundle in
  Cytochrome bc1}. \emph{Journal of the American Chemical Society}
  \textbf{2004}, \emph{126}, 8141--8147, PMID: 15225055\relax
\mciteBstWouldAddEndPuncttrue
\mciteSetBstMidEndSepPunct{\mcitedefaultmidpunct}
{\mcitedefaultendpunct}{\mcitedefaultseppunct}\relax
\EndOfBibitem
\bibitem[Schrödinger and DeLano()Schrödinger, and DeLano]{pymol2015}
Schrödinger,~L.; DeLano,~W. PyMOL, \url{http://www.pymol.org/pymol}\relax
\mciteBstWouldAddEndPuncttrue
\mciteSetBstMidEndSepPunct{\mcitedefaultmidpunct}
{\mcitedefaultendpunct}{\mcitedefaultseppunct}\relax
\EndOfBibitem
\bibitem[Berendsen \latin{et~al.}(1995)Berendsen, {van der Spoel}, and {van
  Drunen}]{berendsen1995}
Berendsen,~H.; {van der Spoel},~D.; {van Drunen},~R. {GROMACS: A
  message-passing parallel molecular dynamics implementation}. \emph{Computer
  Physics Communications} \textbf{1995}, \emph{91}, 43--56\relax
\mciteBstWouldAddEndPuncttrue
\mciteSetBstMidEndSepPunct{\mcitedefaultmidpunct}
{\mcitedefaultendpunct}{\mcitedefaultseppunct}\relax
\EndOfBibitem
\bibitem[Van Der~Spoel \latin{et~al.}(2005)Van Der~Spoel, Lindahl, Hess,
  Groenhof, Mark, and Berendsen]{vanderspoel2005}
Van Der~Spoel,~D.; Lindahl,~E.; Hess,~B.; Groenhof,~G.; Mark,~A.~E.;
  Berendsen,~H. J.~C. GROMACS: Fast, flexible, and free. \emph{Journal of
  Computational Chemistry} \textbf{2005}, \emph{26}, 1701--1718\relax
\mciteBstWouldAddEndPuncttrue
\mciteSetBstMidEndSepPunct{\mcitedefaultmidpunct}
{\mcitedefaultendpunct}{\mcitedefaultseppunct}\relax
\EndOfBibitem
\bibitem[P{\'a}ll \latin{et~al.}(2015)P{\'a}ll, Abraham, Kutzner, Hess, and
  Lindahl]{pall2015}
P{\'a}ll,~S.; Abraham,~M.~J.; Kutzner,~C.; Hess,~B.; Lindahl,~E. {Tackling
  Exascale Software Challenges in Molecular Dynamics Simulations with GROMACS}.
  Solving Software Challenges for Exascale. Cham, 2015; pp 3--27\relax
\mciteBstWouldAddEndPuncttrue
\mciteSetBstMidEndSepPunct{\mcitedefaultmidpunct}
{\mcitedefaultendpunct}{\mcitedefaultseppunct}\relax
\EndOfBibitem
\bibitem[Abraham \latin{et~al.}(2015)Abraham, Murtola, Schulz, Páll, Smith,
  Hess, and Lindahl]{abraham2015}
Abraham,~M.~J.; Murtola,~T.; Schulz,~R.; Páll,~S.; Smith,~J.~C.; Hess,~B.;
  Lindahl,~E. {GROMACS: High performance molecular simulations through
  multi-level parallelism from laptops to supercomputers}. \emph{SoftwareX}
  \textbf{2015}, \emph{1-2}, 19--25\relax
\mciteBstWouldAddEndPuncttrue
\mciteSetBstMidEndSepPunct{\mcitedefaultmidpunct}
{\mcitedefaultendpunct}{\mcitedefaultseppunct}\relax
\EndOfBibitem
\bibitem[Schmid \latin{et~al.}(2011)Schmid, Eichenberger, Choutko, Riniker,
  Winger, Mark, and van Gunsteren]{schmid2011}
Schmid,~N.; Eichenberger,~A.~P.; Choutko,~A.; Riniker,~S.; Winger,~M.;
  Mark,~A.~E.; van Gunsteren,~W.~F. {Definition and testing of the GROMOS
  force-field versions 54A7 and 54B7}. \emph{European Biophysics Journal}
  \textbf{2011}, \emph{40}, 843–856\relax
\mciteBstWouldAddEndPuncttrue
\mciteSetBstMidEndSepPunct{\mcitedefaultmidpunct}
{\mcitedefaultendpunct}{\mcitedefaultseppunct}\relax
\EndOfBibitem
\bibitem[Hermans \latin{et~al.}(1984)Hermans, Berendsen, Van~Gunsteren, and
  Postma]{hermans1984}
Hermans,~J.; Berendsen,~H. J.~C.; Van~Gunsteren,~W.~F.; Postma,~J. P.~M. A
  consistent empirical potential for water–protein interactions.
  \emph{Biopolymers} \textbf{1984}, \emph{23}, 1513--1518\relax
\mciteBstWouldAddEndPuncttrue
\mciteSetBstMidEndSepPunct{\mcitedefaultmidpunct}
{\mcitedefaultendpunct}{\mcitedefaultseppunct}\relax
\EndOfBibitem
\bibitem[Bussi \latin{et~al.}(2007)Bussi, Donadio, and Parrinello]{bussi2007}
Bussi,~G.; Donadio,~D.; Parrinello,~M. Canonical sampling through velocity
  rescaling. \emph{The Journal of Chemical Physics} \textbf{2007}, \emph{126},
  014101\relax
\mciteBstWouldAddEndPuncttrue
\mciteSetBstMidEndSepPunct{\mcitedefaultmidpunct}
{\mcitedefaultendpunct}{\mcitedefaultseppunct}\relax
\EndOfBibitem
\bibitem[Parrinello and Rahman(1981)Parrinello, and Rahman]{parrinell01981}
Parrinello,~M.; Rahman,~A. Polymorphic transitions in single crystals: A new
  molecular dynamics method. \emph{Journal of Applied Physics} \textbf{1981},
  \emph{52}, 7182--7190\relax
\mciteBstWouldAddEndPuncttrue
\mciteSetBstMidEndSepPunct{\mcitedefaultmidpunct}
{\mcitedefaultendpunct}{\mcitedefaultseppunct}\relax
\EndOfBibitem
\bibitem[Nosé and Klein(1983)Nosé, and Klein]{nose1983}
Nosé,~S.; Klein,~M. Constant pressure molecular dynamics for molecular
  systems. \emph{Molecular Physics} \textbf{1983}, \emph{50}, 1055--1076\relax
\mciteBstWouldAddEndPuncttrue
\mciteSetBstMidEndSepPunct{\mcitedefaultmidpunct}
{\mcitedefaultendpunct}{\mcitedefaultseppunct}\relax
\EndOfBibitem
\bibitem[Hess \latin{et~al.}(1997)Hess, Bekker, Berendsen, and
  Fraaije]{hess1997}
Hess,~B.; Bekker,~H.; Berendsen,~H. J.~C.; Fraaije,~J. G. E.~M. {LINCS: A
  linear constraint solver for molecular simulations}. \emph{Journal of
  Computational Chemistry} \textbf{1997}, \emph{18}, 1463--1472\relax
\mciteBstWouldAddEndPuncttrue
\mciteSetBstMidEndSepPunct{\mcitedefaultmidpunct}
{\mcitedefaultendpunct}{\mcitedefaultseppunct}\relax
\EndOfBibitem
\bibitem[Miyamoto and Kollman(1992)Miyamoto, and Kollman]{miyamoto1992}
Miyamoto,~S.; Kollman,~P.~A. {Settle: An analytical version of the SHAKE and
  RATTLE algorithm for rigid water models}. \emph{Journal of Computational
  Chemistry} \textbf{1992}, \emph{13}, 952--962\relax
\mciteBstWouldAddEndPuncttrue
\mciteSetBstMidEndSepPunct{\mcitedefaultmidpunct}
{\mcitedefaultendpunct}{\mcitedefaultseppunct}\relax
\EndOfBibitem
\bibitem[Essmann \latin{et~al.}(1995)Essmann, Perera, Berkowitz, Darden, Lee,
  and Pedersen]{essmann1995}
Essmann,~U.; Perera,~L.; Berkowitz,~M.~L.; Darden,~T.; Lee,~H.; Pedersen,~L.~G.
  {A smooth particle mesh Ewald method}. \emph{The Journal of Chemical Physics}
  \textbf{1995}, \emph{103}, 8577--8593\relax
\mciteBstWouldAddEndPuncttrue
\mciteSetBstMidEndSepPunct{\mcitedefaultmidpunct}
{\mcitedefaultendpunct}{\mcitedefaultseppunct}\relax
\EndOfBibitem
\bibitem[Ciccotti and Jacucci(1975)Ciccotti, and Jacucci]{ciccotti1975}
Ciccotti,~G.; Jacucci,~G. Direct Computation of Dynamical Response by Molecular
  Dynamics: The Mobility of a Charged Lennard-Jones Particle. \emph{Phys. Rev.
  Lett.} \textbf{1975}, \emph{35}, 789--792\relax
\mciteBstWouldAddEndPuncttrue
\mciteSetBstMidEndSepPunct{\mcitedefaultmidpunct}
{\mcitedefaultendpunct}{\mcitedefaultseppunct}\relax
\EndOfBibitem
\bibitem[Ciccotti \latin{et~al.}(1979)Ciccotti, Jacucci, and
  McDonald]{ciccotti1979}
Ciccotti,~G.; Jacucci,~G.; McDonald,~I.~R. “Thought-experiments” by
  molecular dynamics. \emph{Journal of Statistical Physics} \textbf{1979},
  \emph{21}, 1--22\relax
\mciteBstWouldAddEndPuncttrue
\mciteSetBstMidEndSepPunct{\mcitedefaultmidpunct}
{\mcitedefaultendpunct}{\mcitedefaultseppunct}\relax
\EndOfBibitem
\bibitem[Ciccotti and Ferrario(2016)Ciccotti, and Ferrario]{ciccotti2016}
Ciccotti,~G.; Ferrario,~M. Non-equilibrium by molecular dynamics: a dynamical
  approach. \emph{Molecular Simulation} \textbf{2016}, \emph{42},
  1385--1400\relax
\mciteBstWouldAddEndPuncttrue
\mciteSetBstMidEndSepPunct{\mcitedefaultmidpunct}
{\mcitedefaultendpunct}{\mcitedefaultseppunct}\relax
\EndOfBibitem
\bibitem[Oliveira \latin{et~al.}(2021)Oliveira, Ciccotti, Haider, and
  Mulholland]{oliveira2021c}
Oliveira,~A. S.~F.; Ciccotti,~G.; Haider,~S.; Mulholland,~A.~J. Dynamical
  nonequilibrium molecular dynamics reveals the structural basis for allostery
  and signal propagation in biomolecular systems. \emph{European Physical
  Journal B} \textbf{2021}, \emph{94}, 144\relax
\mciteBstWouldAddEndPuncttrue
\mciteSetBstMidEndSepPunct{\mcitedefaultmidpunct}
{\mcitedefaultendpunct}{\mcitedefaultseppunct}\relax
\EndOfBibitem
\bibitem[Maragakis \latin{et~al.}(2008)Maragakis, Ritort, Bustamante, Karplus,
  and Crooks]{maragakis2008}
Maragakis,~P.; Ritort,~F.; Bustamante,~C.; Karplus,~M.; Crooks,~G.~E. Bayesian
  estimates of free energies from nonequilibrium work data in the presence of
  instrument noise. \emph{The Journal of Chemical Physics} \textbf{2008},
  \emph{129}, 024102\relax
\mciteBstWouldAddEndPuncttrue
\mciteSetBstMidEndSepPunct{\mcitedefaultmidpunct}
{\mcitedefaultendpunct}{\mcitedefaultseppunct}\relax
\EndOfBibitem
\bibitem[Atkins \latin{et~al.}(2022)Atkins, de~Paula, and Keeler]{atkins2022}
Atkins,~P.; de~Paula,~J.; Keeler,~J. \emph{Atkins' physical chemistry}; Oxford
  University Press, 2022\relax
\mciteBstWouldAddEndPuncttrue
\mciteSetBstMidEndSepPunct{\mcitedefaultmidpunct}
{\mcitedefaultendpunct}{\mcitedefaultseppunct}\relax
\EndOfBibitem
\bibitem[Jaynes(2003)]{jaynes2003}
Jaynes,~E.~T. \emph{{Probability Theory: The Logic of Science}}; Cambridge
  University Press, 2003\relax
\mciteBstWouldAddEndPuncttrue
\mciteSetBstMidEndSepPunct{\mcitedefaultmidpunct}
{\mcitedefaultendpunct}{\mcitedefaultseppunct}\relax
\EndOfBibitem
\bibitem[Gore \latin{et~al.}(2003)Gore, Ritort, and Bustamante]{gore2003}
Gore,~J.; Ritort,~F.; Bustamante,~C. Bias and error in estimates of equilibrium
  free-energy differences from nonequilibrium measurements. \emph{Proceedings
  of the National Academy of Sciences} \textbf{2003}, \emph{100},
  12564--12569\relax
\mciteBstWouldAddEndPuncttrue
\mciteSetBstMidEndSepPunct{\mcitedefaultmidpunct}
{\mcitedefaultendpunct}{\mcitedefaultseppunct}\relax
\EndOfBibitem
\bibitem[Pohorille \latin{et~al.}(2010)Pohorille, Jarzynski, and
  Chipot]{pohorille2010}
Pohorille,~A.; Jarzynski,~C.; Chipot,~C. Good Practices in Free-Energy
  Calculations. \emph{The Journal of Physical Chemistry B} \textbf{2010},
  \emph{114}, 10235--10253, PMID: 20701361\relax
\mciteBstWouldAddEndPuncttrue
\mciteSetBstMidEndSepPunct{\mcitedefaultmidpunct}
{\mcitedefaultendpunct}{\mcitedefaultseppunct}\relax
\EndOfBibitem
\bibitem[Yunger~Halpern and Jarzynski(2016)Yunger~Halpern, and
  Jarzynski]{yunger-halpern2016}
Yunger~Halpern,~N.; Jarzynski,~C. Number of trials required to estimate a
  free-energy difference, using fluctuation relations. \emph{Physical Review E}
  \textbf{2016}, \emph{93}, 052144\relax
\mciteBstWouldAddEndPuncttrue
\mciteSetBstMidEndSepPunct{\mcitedefaultmidpunct}
{\mcitedefaultendpunct}{\mcitedefaultseppunct}\relax
\EndOfBibitem
\bibitem[von Toussaint(2011)]{toussaint2011}
von Toussaint,~U. Bayesian inference in physics. \emph{Reviews of Modern
  Physics} \textbf{2011}, \emph{83}, 943--999\relax
\mciteBstWouldAddEndPuncttrue
\mciteSetBstMidEndSepPunct{\mcitedefaultmidpunct}
{\mcitedefaultendpunct}{\mcitedefaultseppunct}\relax
\EndOfBibitem
\bibitem[Rubio and Dunningham(2019)Rubio, and Dunningham]{jesus2018}
Rubio,~J.; Dunningham,~J.~A. Quantum metrology in the presence of limited data.
  \emph{New Journal of Physics} \textbf{2019}, \emph{21}, 043037\relax
\mciteBstWouldAddEndPuncttrue
\mciteSetBstMidEndSepPunct{\mcitedefaultmidpunct}
{\mcitedefaultendpunct}{\mcitedefaultseppunct}\relax
\EndOfBibitem
\bibitem[Bennett(1976)]{bennett1976}
Bennett,~C.~H. {Efficient estimation of free energy differences from Monte
  Carlo data}. \emph{Journal of Computational Physics} \textbf{1976},
  \emph{22}, 245--268\relax
\mciteBstWouldAddEndPuncttrue
\mciteSetBstMidEndSepPunct{\mcitedefaultmidpunct}
{\mcitedefaultendpunct}{\mcitedefaultseppunct}\relax
\EndOfBibitem
\bibitem[Crooks(2000)]{crooks2000path}
Crooks,~G.~E. Path-ensemble averages in systems driven far from equilibrium.
  \emph{Phys. Rev. E} \textbf{2000}, \emph{61}, 2361--2366\relax
\mciteBstWouldAddEndPuncttrue
\mciteSetBstMidEndSepPunct{\mcitedefaultmidpunct}
{\mcitedefaultendpunct}{\mcitedefaultseppunct}\relax
\EndOfBibitem
\bibitem[Shirts \latin{et~al.}(2003)Shirts, Bair, Hooker, and
  Pande]{shirts2003equilibrium}
Shirts,~M.~R.; Bair,~E.; Hooker,~G.; Pande,~V.~S. Equilibrium Free Energies
  from Nonequilibrium Measurements Using Maximum-Likelihood Methods.
  \emph{Phys. Rev. Lett.} \textbf{2003}, \emph{91}, 140601\relax
\mciteBstWouldAddEndPuncttrue
\mciteSetBstMidEndSepPunct{\mcitedefaultmidpunct}
{\mcitedefaultendpunct}{\mcitedefaultseppunct}\relax
\EndOfBibitem
\bibitem[Kay(1993)]{kay1993}
Kay,~S.~M. \emph{{Fundamentals of Statistical Signal Processing: Estimation
  Theory}}; Prentice-Hall, Inc., 1993\relax
\mciteBstWouldAddEndPuncttrue
\mciteSetBstMidEndSepPunct{\mcitedefaultmidpunct}
{\mcitedefaultendpunct}{\mcitedefaultseppunct}\relax
\EndOfBibitem
\bibitem[Rubio \latin{et~al.}(2021)Rubio, Anders, and Correa]{rubio2021}
Rubio,~J.; Anders,~J.; Correa,~L.~A. Global Quantum Thermometry. \emph{Phys.
  Rev. Lett.} \textbf{2021}, \emph{127}, 190402\relax
\mciteBstWouldAddEndPuncttrue
\mciteSetBstMidEndSepPunct{\mcitedefaultmidpunct}
{\mcitedefaultendpunct}{\mcitedefaultseppunct}\relax
\EndOfBibitem
\bibitem[Ho \latin{et~al.}(2020)Ho, Truong, and Li]{kiet2020}
Ho,~K.; Truong,~D.~T.; Li,~M.~S. {How Good is Jarzynski’s Equality for
  Computer-Aided Drug Design?} \emph{The Journal of Physical Chemistry B}
  \textbf{2020}, \emph{124}, 5338--5349, PMID: 32484689\relax
\mciteBstWouldAddEndPuncttrue
\mciteSetBstMidEndSepPunct{\mcitedefaultmidpunct}
{\mcitedefaultendpunct}{\mcitedefaultseppunct}\relax
\EndOfBibitem
\bibitem[Zwanzig(2004)]{zwanzig2004the}
Zwanzig,~R.~W. {High‐Temperature Equation of State by a Perturbation Method.
  I. Nonpolar Gases}. \emph{The Journal of Chemical Physics} \textbf{2004},
  \emph{22}, 1420--1426\relax
\mciteBstWouldAddEndPuncttrue
\mciteSetBstMidEndSepPunct{\mcitedefaultmidpunct}
{\mcitedefaultendpunct}{\mcitedefaultseppunct}\relax
\EndOfBibitem
\bibitem[Baptista \latin{et~al.}(1999)Baptista, Martel, and
  Soares]{baptista1999}
Baptista,~A.~M.; Martel,~P.~J.; Soares,~C.~M. {Simulation of Electron-Proton
  Coupling with a Monte Carlo Method: Application to Cytochrome c3 Using
  Continuum Electrostatics}. \emph{Biophysical Journal} \textbf{1999},
  \emph{76}, 2978--2998\relax
\mciteBstWouldAddEndPuncttrue
\mciteSetBstMidEndSepPunct{\mcitedefaultmidpunct}
{\mcitedefaultendpunct}{\mcitedefaultseppunct}\relax
\EndOfBibitem
\bibitem[Bashford and Karplus(1990)Bashford, and Karplus]{bashford1990}
Bashford,~D.; Karplus,~M. pKa's of ionizable groups in proteins: atomic detail
  from a continuum electrostatic model. \emph{Biochemistry} \textbf{1990},
  \emph{29}, 10219--10225, PMID: 2271649\relax
\mciteBstWouldAddEndPuncttrue
\mciteSetBstMidEndSepPunct{\mcitedefaultmidpunct}
{\mcitedefaultendpunct}{\mcitedefaultseppunct}\relax
\EndOfBibitem
\bibitem[Bashford and Gerwert(1992)Bashford, and Gerwert]{bashford1992}
Bashford,~D.; Gerwert,~K. Electrostatic calculations of the pKa values of
  ionizable groups in bacteriorhodopsin. \emph{Journal of Molecular Biology}
  \textbf{1992}, \emph{224}, 473--486\relax
\mciteBstWouldAddEndPuncttrue
\mciteSetBstMidEndSepPunct{\mcitedefaultmidpunct}
{\mcitedefaultendpunct}{\mcitedefaultseppunct}\relax
\EndOfBibitem
\bibitem[Bashford(1997)]{bashford1997}
Bashford,~D. An object-oriented programming suite for electrostatic effects in
  biological molecules An experience report on the MEAD project. Scientific
  Computing in Object-Oriented Parallel Environments. Berlin, Heidelberg, 1997;
  pp 233--240\relax
\mciteBstWouldAddEndPuncttrue
\mciteSetBstMidEndSepPunct{\mcitedefaultmidpunct}
{\mcitedefaultendpunct}{\mcitedefaultseppunct}\relax
\EndOfBibitem
\bibitem[Baptista and Soares(2001)Baptista, and Soares]{baptista2001}
Baptista,~A.~M.; Soares,~C.~M. Some Theoretical and Computational Aspects of
  the Inclusion of Proton Isomerism in the Protonation Equilibrium of Proteins.
  \emph{The Journal of Physical Chemistry B} \textbf{2001}, \emph{105},
  293--309\relax
\mciteBstWouldAddEndPuncttrue
\mciteSetBstMidEndSepPunct{\mcitedefaultmidpunct}
{\mcitedefaultendpunct}{\mcitedefaultseppunct}\relax
\EndOfBibitem
\bibitem[Teixeira \latin{et~al.}(2005)Teixeira, Cunha, Machuqueiro, Oliveira,
  Victor, Soares, and Baptista]{teixeira2005}
Teixeira,~V.~H.; Cunha,~C.~A.; Machuqueiro,~M.; Oliveira,~A. S.~F.;
  Victor,~B.~L.; Soares,~C.~M.; Baptista,~A.~M. On the Use of Different
  Dielectric Constants for Computing Individual and Pairwise Terms in
  Poisson-Boltzmann Studies of Protein Ionization Equilibrium. \emph{The
  Journal of Physical Chemistry B} \textbf{2005}, \emph{109}, 14691--14706,
  PMID: 16852854\relax
\mciteBstWouldAddEndPuncttrue
\mciteSetBstMidEndSepPunct{\mcitedefaultmidpunct}
{\mcitedefaultendpunct}{\mcitedefaultseppunct}\relax
\EndOfBibitem
\bibitem[Gilson \latin{et~al.}(1988)Gilson, Sharp, and Honig]{gilson1988}
Gilson,~M.~K.; Sharp,~K.~A.; Honig,~B.~H. Calculating the electrostatic
  potential of molecules in solution: Method and error assessment.
  \emph{Journal of Computational Chemistry} \textbf{1988}, \emph{9},
  327--335\relax
\mciteBstWouldAddEndPuncttrue
\mciteSetBstMidEndSepPunct{\mcitedefaultmidpunct}
{\mcitedefaultendpunct}{\mcitedefaultseppunct}\relax
\EndOfBibitem
\bibitem[Metropolis \latin{et~al.}(1953)Metropolis, Rosenbluth, Rosenbluth,
  Teller, and Teller]{metropolis1953}
Metropolis,~N.; Rosenbluth,~A.~W.; Rosenbluth,~M.~N.; Teller,~A.~H.; Teller,~E.
  Equation of State Calculations by Fast Computing Machines. \emph{The Journal
  of Chemical Physics} \textbf{1953}, \emph{21}, 1087--1092\relax
\mciteBstWouldAddEndPuncttrue
\mciteSetBstMidEndSepPunct{\mcitedefaultmidpunct}
{\mcitedefaultendpunct}{\mcitedefaultseppunct}\relax
\EndOfBibitem
\bibitem[Ost \latin{et~al.}(2004)Ost, Clark, Anderson, Yellowlees, Daff, and
  Chapman]{ost20044}
Ost,~T.~W.; Clark,~J.~P.; Anderson,~J.~R.; Yellowlees,~L.~J.; Daff,~S.;
  Chapman,~S.~K. 4-Cyanopyridine, a versatile spectroscopic probe for
  cytochrome P450 BM3. \emph{Journal of Biological Chemistry} \textbf{2004},
  \emph{279}, 48876--48882\relax
\mciteBstWouldAddEndPuncttrue
\mciteSetBstMidEndSepPunct{\mcitedefaultmidpunct}
{\mcitedefaultendpunct}{\mcitedefaultseppunct}\relax
\EndOfBibitem
\bibitem[{Leslie Dutton}(1978)]{dutton197823}
{Leslie Dutton},~P. \emph{Biomembranes - Part E: Biological Oxidations};
  Methods in Enzymology; Academic Press, 1978; Vol.~54; pp 411--435\relax
\mciteBstWouldAddEndPuncttrue
\mciteSetBstMidEndSepPunct{\mcitedefaultmidpunct}
{\mcitedefaultendpunct}{\mcitedefaultseppunct}\relax
\EndOfBibitem
\bibitem[Pedregosa \latin{et~al.}(2011)Pedregosa, Varoquaux, Gramfort, Michel,
  Thirion, Grisel, Blondel, Prettenhofer, Weiss, Dubourg, Vanderplas, Passos,
  Cournapeau, Brucher, Perrot, and Duchesnay]{scikit-learn}
Pedregosa,~F.; Varoquaux,~G.; Gramfort,~A.; Michel,~V.; Thirion,~B.;
  Grisel,~O.; Blondel,~M.; Prettenhofer,~P.; Weiss,~R.; Dubourg,~V.;
  Vanderplas,~J.; Passos,~A.; Cournapeau,~D.; Brucher,~M.; Perrot,~M.;
  Duchesnay,~E. Scikit-learn: Machine Learning in {P}ython. \emph{Journal of
  Machine Learning Research} \textbf{2011}, \emph{12}, 2825--2830\relax
\mciteBstWouldAddEndPuncttrue
\mciteSetBstMidEndSepPunct{\mcitedefaultmidpunct}
{\mcitedefaultendpunct}{\mcitedefaultseppunct}\relax
\EndOfBibitem
\bibitem[Kabsch and Sander(1983)Kabsch, and Sander]{kabsch1983}
Kabsch,~W.; Sander,~C. {Dictionary of protein secondary structure: Pattern
  recognition of hydrogen-bonded and geometrical features}. \emph{Biopolymers}
  \textbf{1983}, \emph{22}, 2577--2637\relax
\mciteBstWouldAddEndPuncttrue
\mciteSetBstMidEndSepPunct{\mcitedefaultmidpunct}
{\mcitedefaultendpunct}{\mcitedefaultseppunct}\relax
\EndOfBibitem
\end{mcitethebibliography}

%%%%%%%%%%%%%%%%%%%%%%%%%%%%%%%%%%%%%%%%%%%%%%%%%%%%%%%%%%%%%%%%%%%%%
%% Supporting Information
%%%%%%%%%%%%%%%%%%%%%%%%%%%%%%%%%%%%%%%%%%%%%%%%%%%%%%%%%%%%%%%%%%%%%
\appendix
\onecolumn
\newpage

\setcounter{equation}{0}
\renewcommand\theequation{S\arabic{equation}}
\setcounter{figure}{0}
\renewcommand\thefigure{S\arabic{figure}}    
\renewcommand{\thesection}{S}
\renewcommand{\thetable}{S1}

\begin{center}
    \section*{-- Supporting information --}
\end{center} 

\begin{figure}[h!] % figure S5
\centering
\includegraphics[trim={0cm 0cm 0cm 0cm},clip,width=0.5\textwidth]{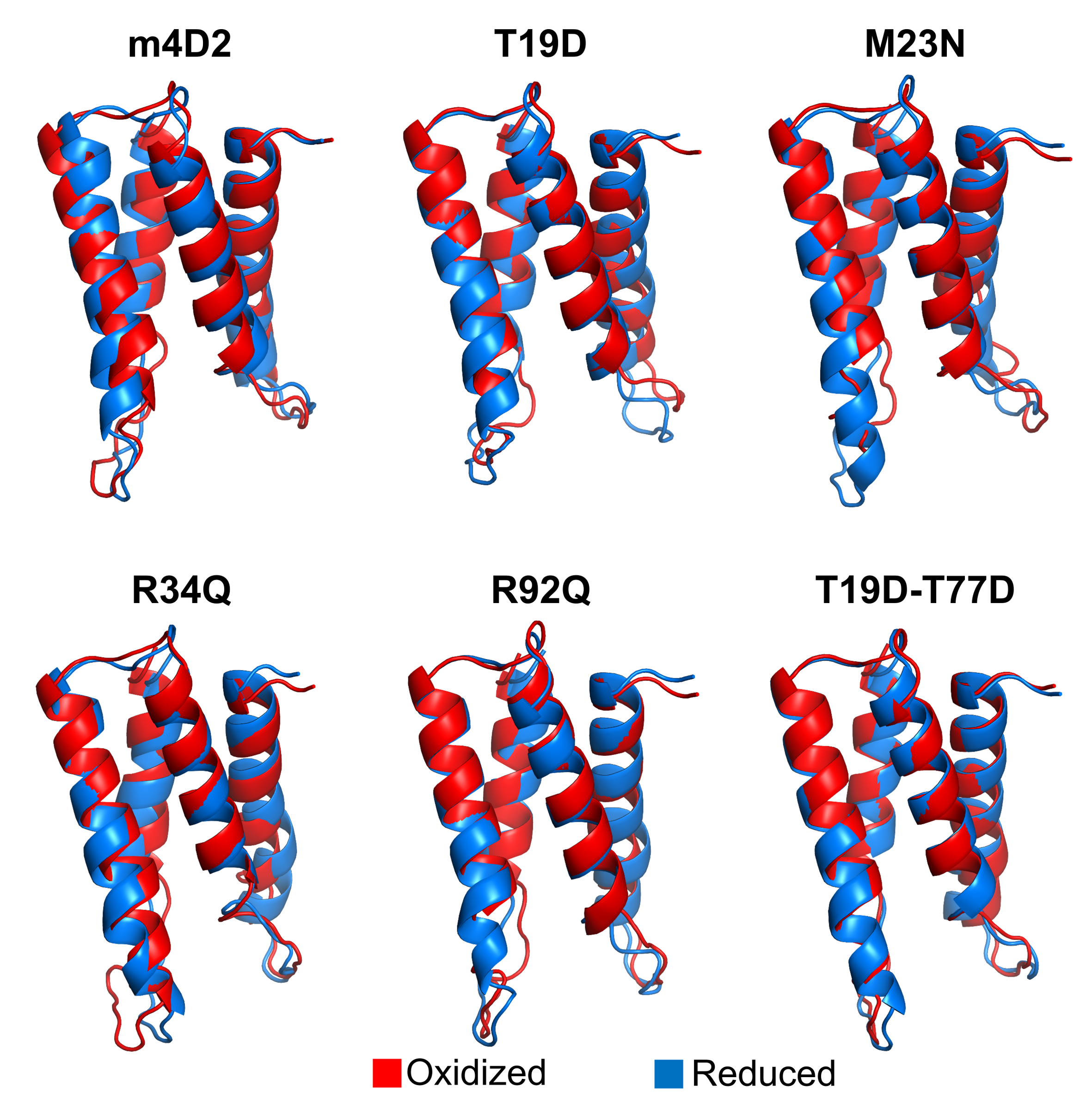}
	\caption{Average structures for the oxidized and reduced m4D2, T19D, M23N, R34Q, R92Q and T19D-T77D. The average structures were calculated using the last 400 ns of all replicates for each system (10 replicates for m4D2 and single mutants and 20 replicates for the double mutant).} 
\label{fig:avg-struc}
\end{figure}

\begin{figure*}[h!] % figure S1
\centering
\includegraphics[trim={0cm 0cm 0cm 0cm},clip,width=0.9\textwidth]{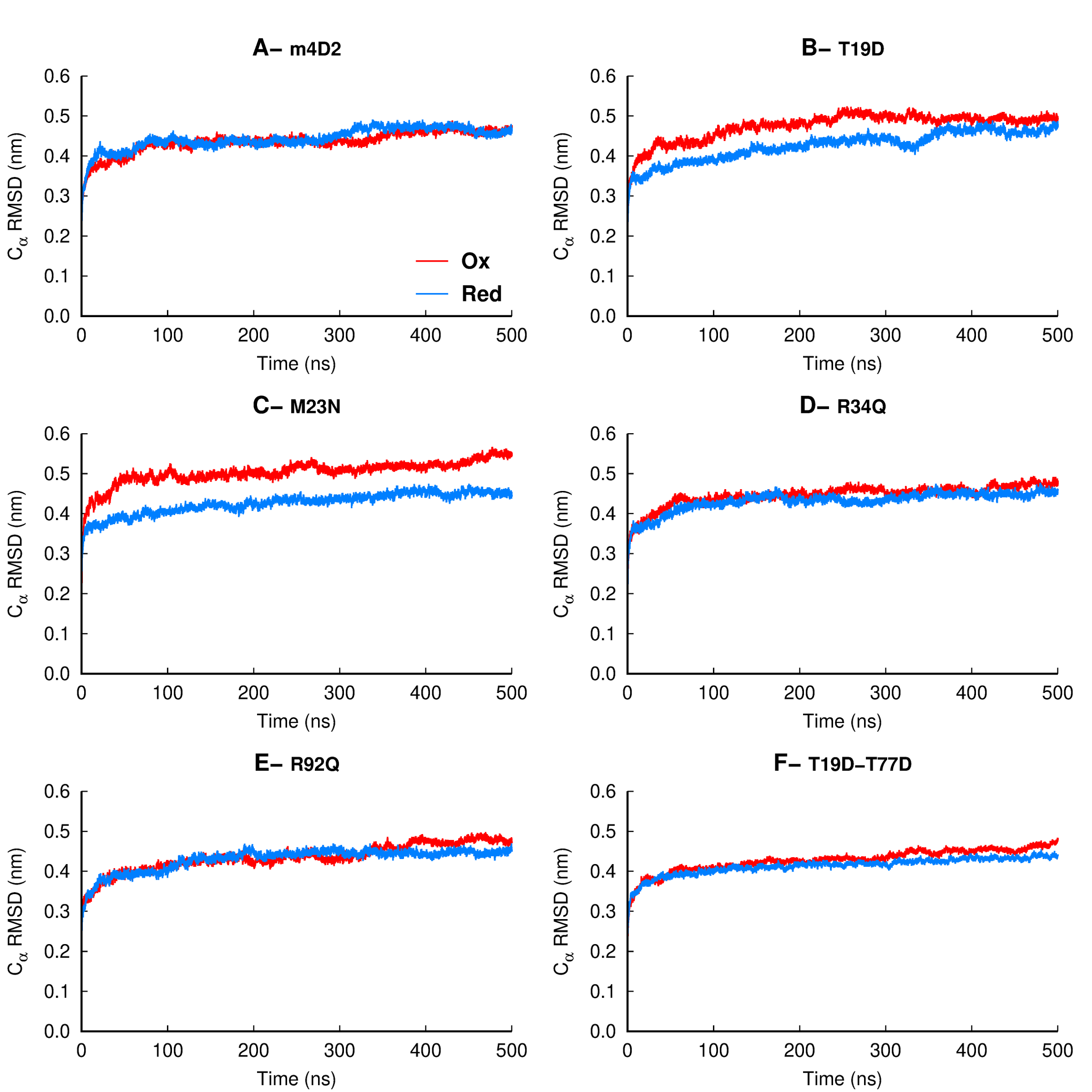}
	\caption{Time evolution of the average C\(\alpha\) RMSD for m4D2 (\textbf{A}), T19D (\textbf{B}), M23N (\textbf{C}), R34Q (\textbf{D}), R92Q (\textbf{E}) and T19D-T77D (\textbf{F}). The C$\alpha$ RMSD was calculated relative to the starting structures and averaged over all replicates ($10$ replicates for m4D2 and single mutants and 20 replicates for the double mutant). All the systems were considered equilibrated after $100\,\si{ns}$.} 
\label{fig:time-evolv}
\end{figure*}

\begin{figure*}[h!] % new_figure_S3
\centering
\includegraphics[trim={0cm 0cm 0cm 0cm},clip,width=0.9\textwidth]{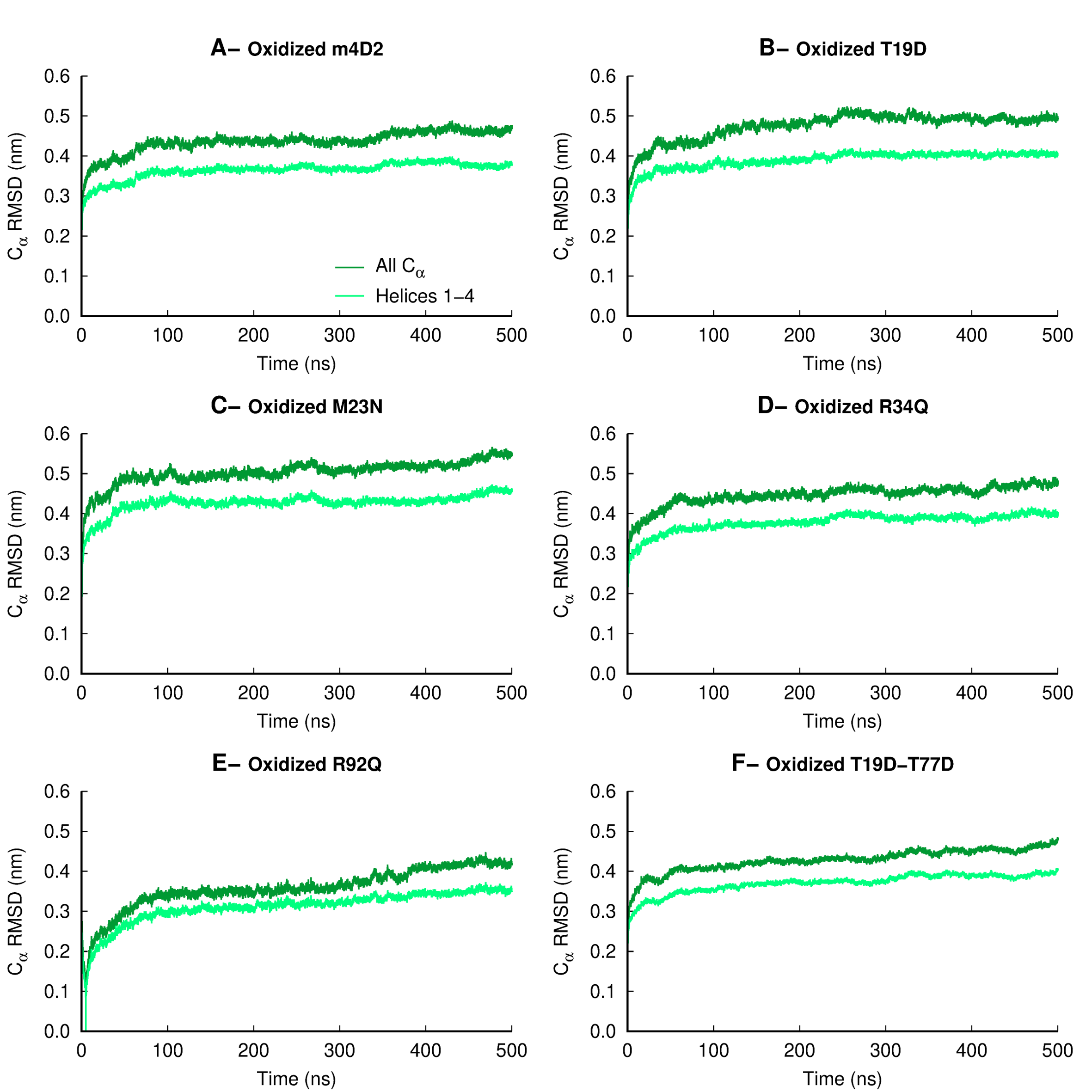}
	\caption{Time evolution of the average C\(\alpha\) RMSD for oxidized m4D2 (\textbf{A}), T19D (\textbf{B}), M23N (\textbf{C}), R34Q (\textbf{D}), R92Q (\textbf{E}) and T19D-T77D (\textbf{F}) relative to the starting structures for all C\(\alpha\) atoms (dark green line) and for the structured regions (light green line) of the proteins. The C$\alpha$ RMSD was calculated relative to the starting structures and averaged over all replicates ($10$ replicates for m4D2 and single mutants and 20 replicates for the double mutant). The structured region includes \(\alpha\)-helices 1-4. Please note that helices 2 and 4 contain the two histidine residues axially coordinating the heme Fe atom.} 
\label{fig:time-evolv-ox}
\end{figure*}

\begin{figure*}[h!] % new_figure_S4
\centering
\includegraphics[trim={0cm 0cm 0cm 0cm},clip,width=0.9\textwidth]{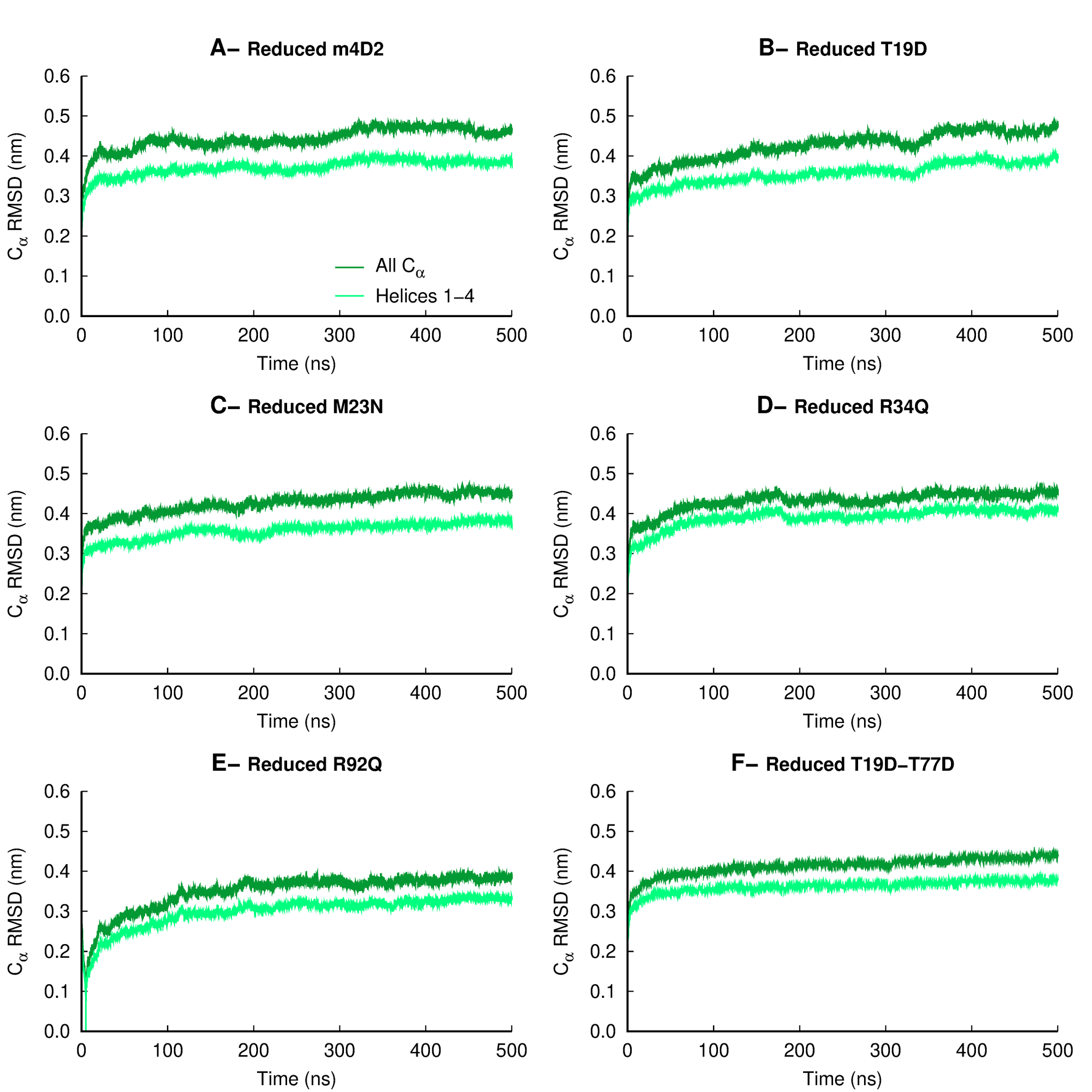}
	\caption{Time evolution of the average C\(\alpha\) RMSD for reduced m4D2 (\textbf{A}), T19D (\textbf{B}), M23N (\textbf{C}), R34Q (\textbf{D}), R92Q (\textbf{E}) and T19D-T77D (\textbf{F}) relative to the starting structures for all C\(\alpha\) atoms (dark green line) and for the structured regions (light green line) of the proteins. The C$\alpha$ RMSD was calculated relative to the starting structures and averaged over all replicates ($10$ replicates for m4D2 and single mutants and 20 replicates for the double mutant). The structured region includes \(\alpha\)-helices 1-4. Please note that helices 2 and 4 contain the two histidine residues axially coordinating the heme Fe atom.} 
\label{fig:time-evolv-red}
\end{figure*}

\begin{figure*}[h!] % figure S2
\centering
\includegraphics[trim={0cm 0cm 0cm 0cm},clip,width=0.99\textwidth]{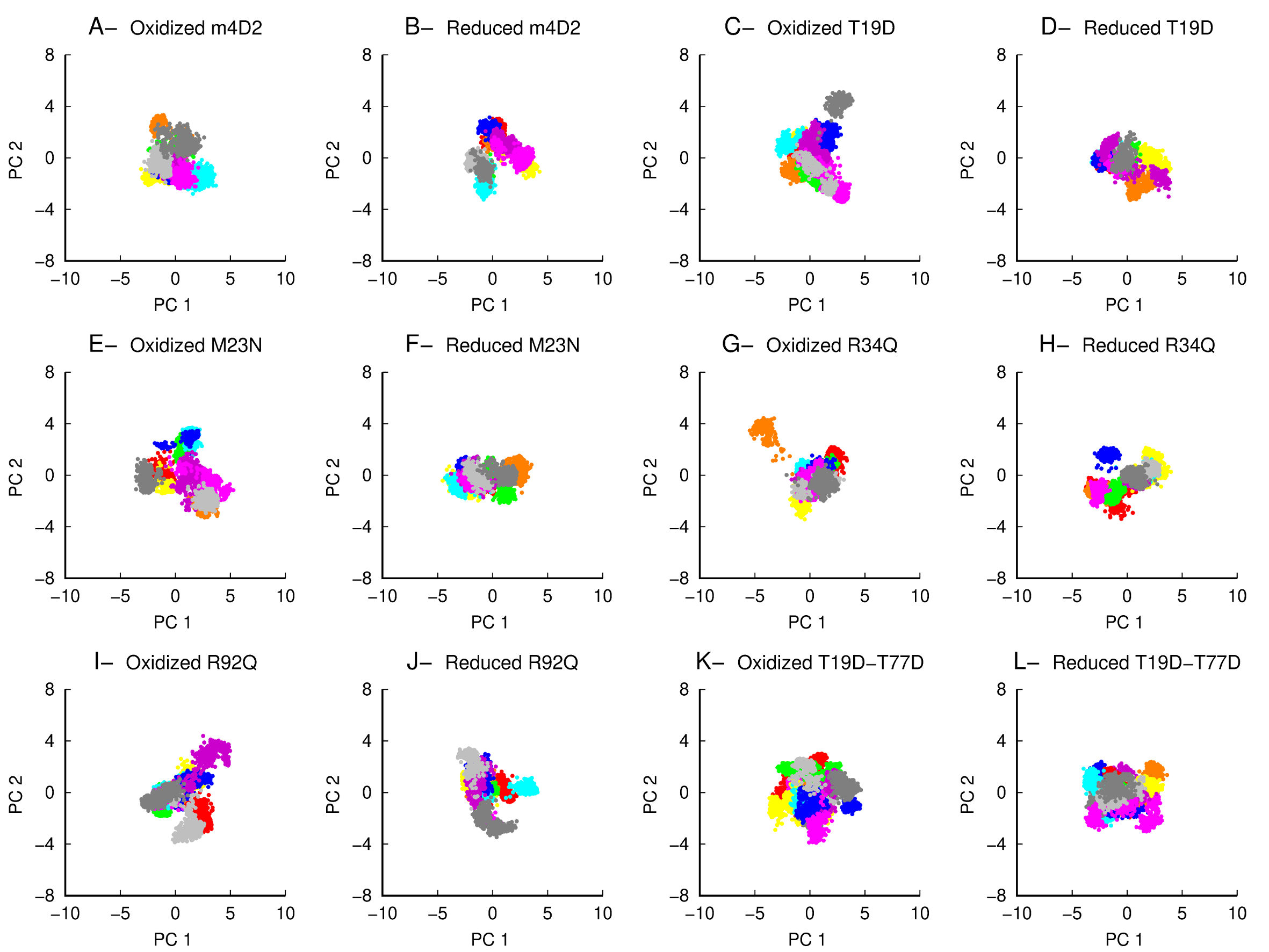}
	\caption{PCA of all replicates for the oxidized and reduced m4D2 (\textbf{A-B}), T19D (\textbf{C-D}), M23N (\textbf{E-F}), R34Q (\textbf{G-H}), R92Q (\textbf{I-J}) and T19D-T77D (\textbf{K-L}). All oxidized and reduced replicates for each system were combined before the analysis, and each trajectory contained one conformation per nanosecond per replicate (totaling 10001 frames for the m4D2 and single mutants and 20001 frames for the double mutant) with all the C\(\alpha\) atoms of the protein. Please zoom into the image for detailed visualization. Note that different replicates sample different regions of conformational space, thus improving the overall sampling for each system.} 
\label{fig:PCA-natural}
\end{figure*}

\begin{figure}[h!] % figure S3
\centering
\includegraphics[trim={0cm 0cm 0cm 0cm},clip,width=0.5\textwidth]{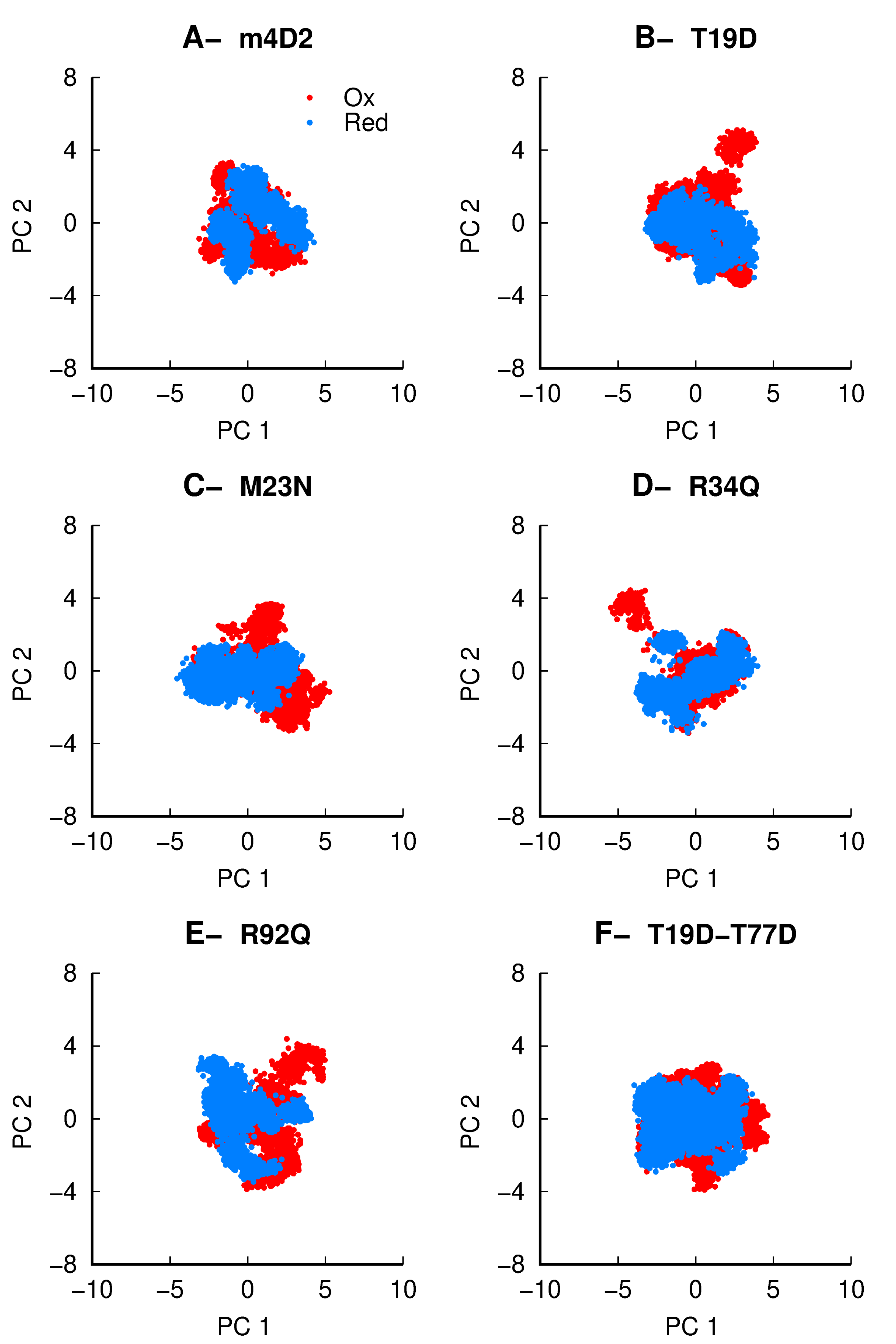}
\caption{PCA for the m4D2 (\textbf{A}), T19D (\textbf{B}), M23N (\textbf{C}), R34Q (\textbf{D}), R92Q (\textbf{E}) and T19D-T77D (\textbf{F}). All oxidized and reduced replicates for each system were combined before the analysis, and each trajectory contained one conformation per nanosecond per replicate (totaling 10001 frames for the m4D2 and single mutants and 20001 frames for the double mutant) with all the C\(\alpha\) atoms of the protein. Note that the oxidized and reduced systems sample different regions of conformational space.} 
\label{fig:PCA-ox+red}
\end{figure}

\begin{figure}[h!] % figure S4
\centering
\includegraphics[trim={0cm 0cm 0cm 0cm},clip,width=0.8\textwidth]{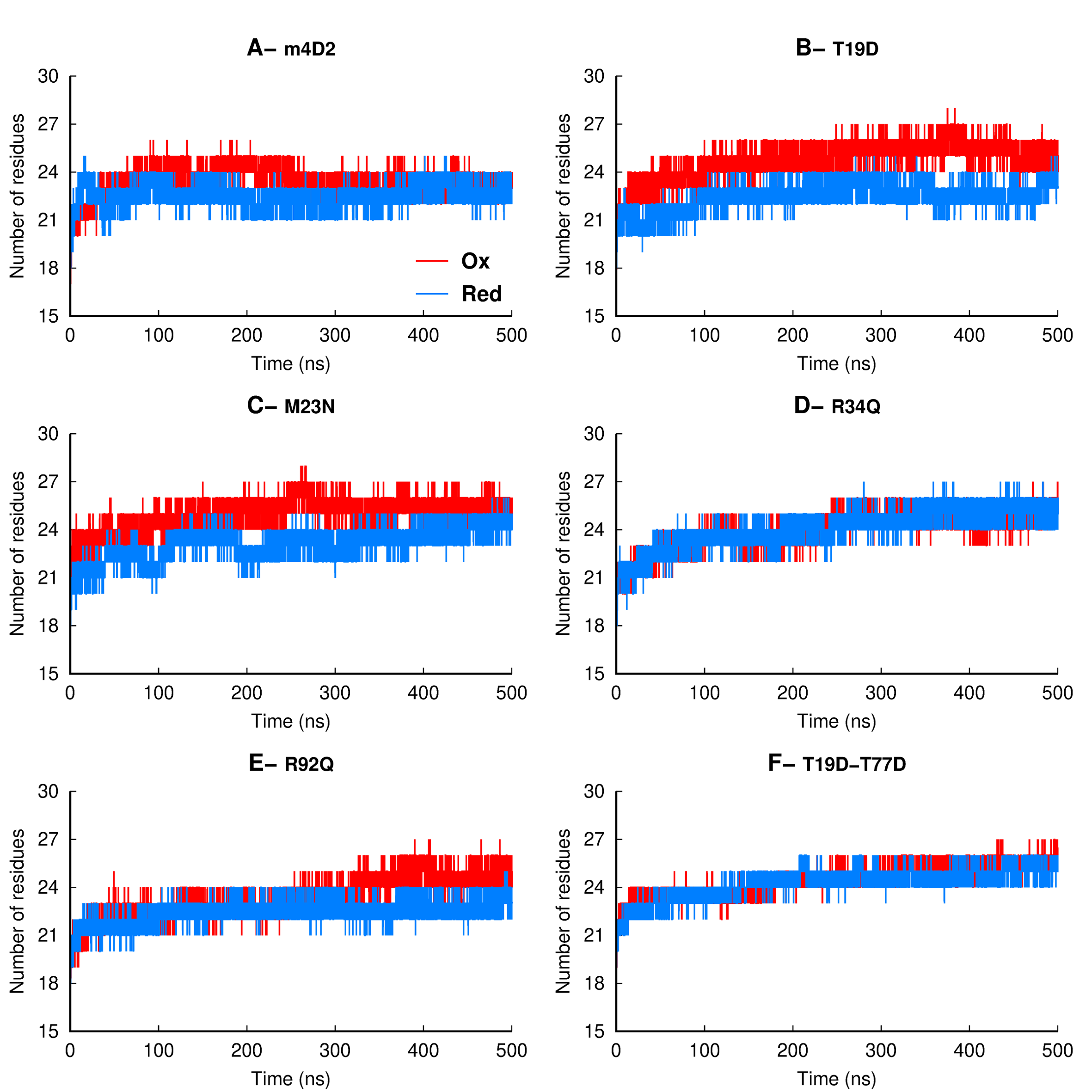}
\caption{Number of residues with coil secondary structure along the simulation time for the m4D2 (\textbf{A}), T19D (\textbf{B}), M23N (\textbf{C}), R34Q (\textbf{D}), R92Q (\textbf{E}) and T19D-T77D (\textbf{F}). The secondary structure assignment was performed with the DSSP\cite{kabsch1983} software.}
\label{fig:Numb-residu}
\end{figure}

\begin{figure}[h!] % figure S6
\centering
\includegraphics[trim={0cm 0cm 0cm 0cm},clip,width=0.9\textwidth]{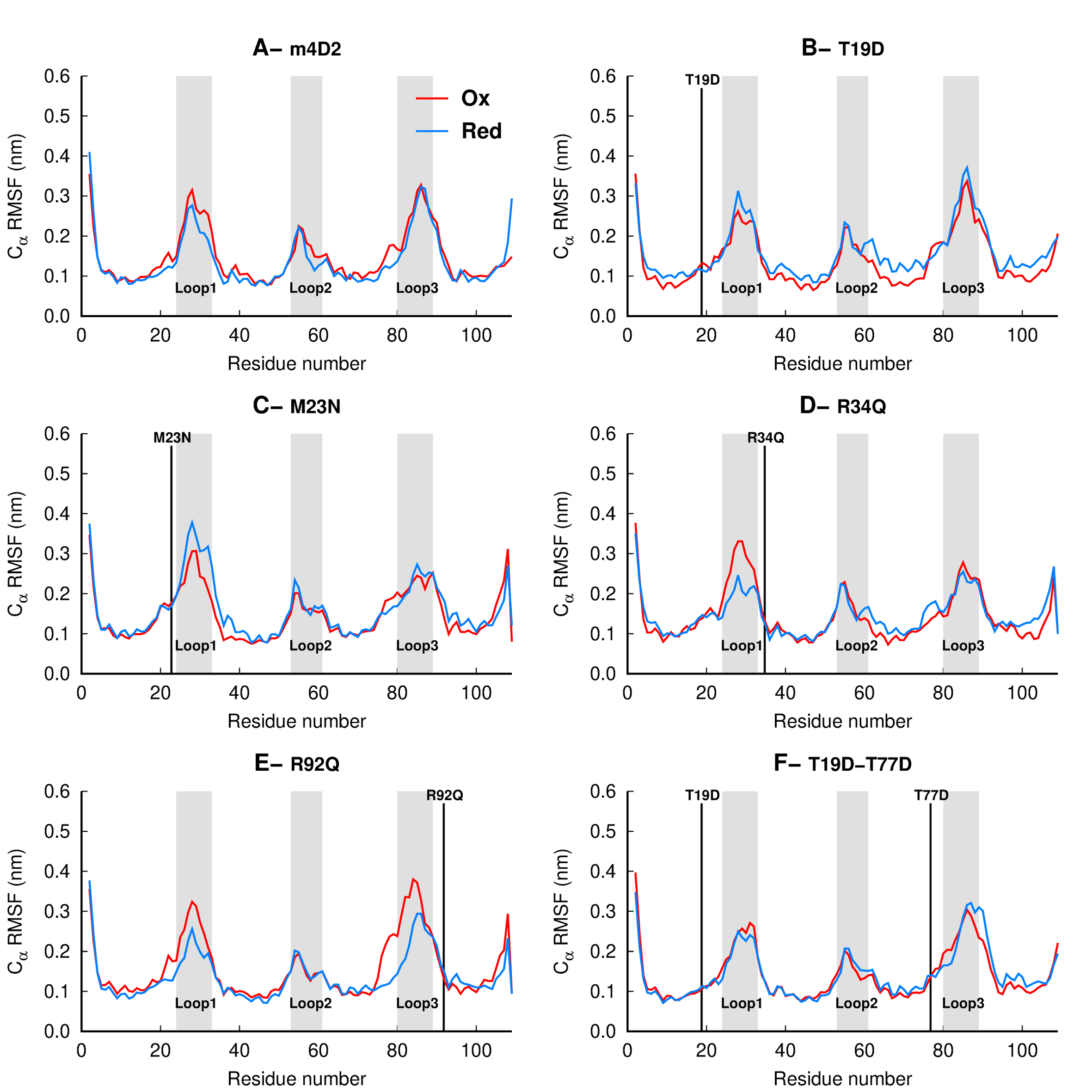}
	\caption{Average C\(\alpha\) RMSF for m4D2, T19D, M23N, R34Q, R92Q and T19D-T77D. The C\(\alpha\) RMSF was calculated for the last 400 ns of simulation and averaged over all replicates (10 replicates for m4D2 and single mutants and 20 replicates for the double mutant). The vertical black lines highlight the mutation site(s). The grey boxes identify the position of the loop regions, namely loop 1 (connecting \(\alpha\)-helix 1 to \(\alpha\)-helix 2), loop 2 (connecting \(\alpha\)-helix 2 to \(\alpha\)-helix 3) and loop 3 (connecting \(\alpha\)-helix 3 to \(\alpha\)-helix 4). Please zoom into the image for a detailed visualisation.} 
\label{fig:avg-Calpha}
\end{figure}

\begin{figure}[h!] % figure S7
\centering
\includegraphics[trim={0cm 0cm 0cm 0cm},clip,width=0.8\textwidth]{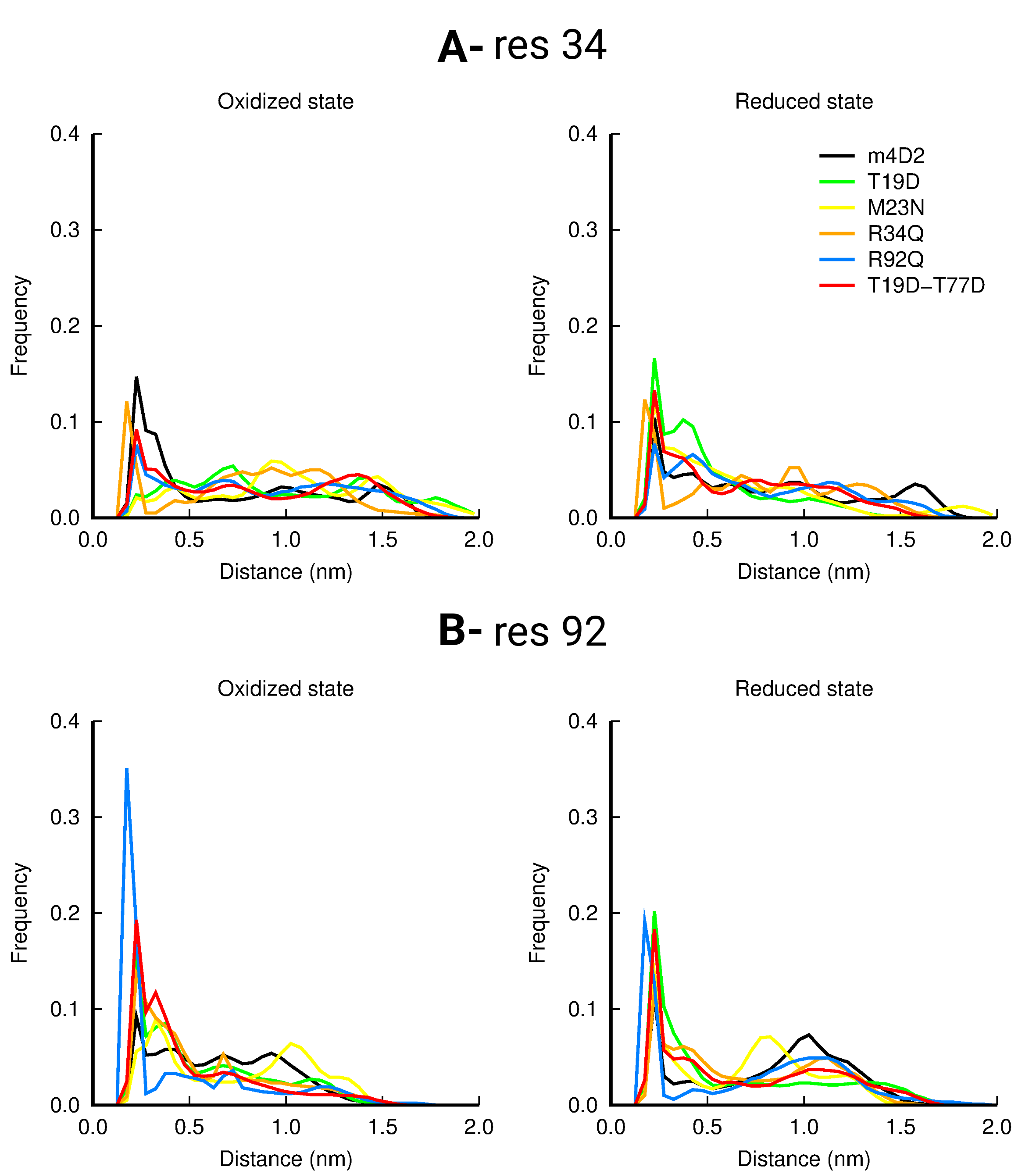}
	\caption{Minimum distance between residue 34 and the heme propionates (\textbf{A}) and residue 92 and the heme propionates (\textbf{B}) in the MD simulations of m4D2, T19D, M23N, R34Q, R92Q and T19D-T77D. Overall distribution of the minimum distance between the sidechain of residue 34 and 92 and the propionates. The histograms reflect the distances over the last 400 ns of each simulation.}
\label{fig:dist-34+92}
\end{figure}

\begin{figure}[h!] % figure S8
\centering
\includegraphics[trim={0cm 0cm 0cm 0cm},clip,width=0.7\textwidth]{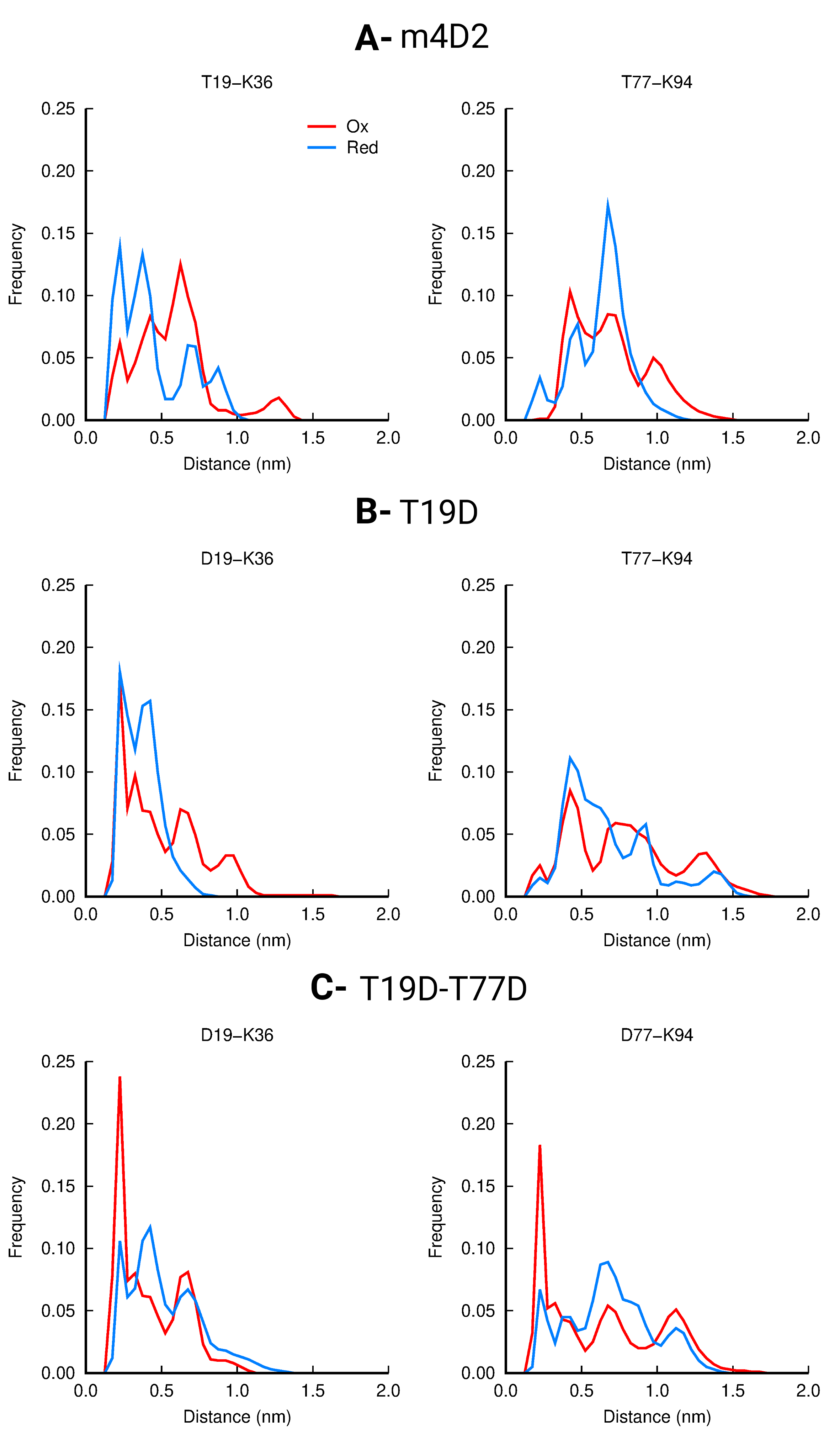}
	\caption{Minimum distance between residue in position $19$ and $36$, and $77$ and $94$ for (\textbf{A}) m4D2, (\textbf{B}) T19D and (\textbf{C}) T19D-T77D. Overall distribution of the minimum distance between the sidechain of residue $19$ and $36$, and $77$ and $94$. The histograms reflect the distances over the last $400\,\rm{ns}$ of each simulation.}
\label{fig:dist-m4D2+19+DM}
\end{figure}

\noindent In section \ref{subsec:stability} below, we display a range of figures to evidence the conformational stability of the MD simulations: 

\medskip 

\noindent Fig.~\ref{fig:avg-struc}: Average structures of oxidized and reduced m4D2 and its mutants. \\
Fig.~\ref{fig:time-evolv}: Time evolution of the average C\(\alpha\) RMSD for m4D2 and mutants showing that all systems are equilibrated after 100 ns. \\
Fig.~\ref{fig:time-evolv-ox}: Time evolution of the average C\(\alpha\) RMSD for all the C\(\alpha\) atoms and for helices 1-4 in the oxidized systems. \\
Fig.~\ref{fig:time-evolv-red}: Time evolution of the average C\(\alpha\) RMSD for all the C\(\alpha\) atoms and for helices 1-4 in the reduced systems. \\
Fig.~\ref{fig:PCA-natural}: PCA for m4D2 and mutants showing that the replicates sample different regions of the conformational landscape. \\ 
Fig.~\ref{fig:PCA-ox+red}: PCA for the oxidized and reduced m4D2 and mutants showing that the two states sample different regions of conformational space. \\
Fig.~\ref{fig:Numb-residu}: Time evolution of the number of residues without secondary structure showing that the protein structures were stable during the simulation time. \\ 
Fig.~\ref{fig:avg-Calpha}: Average fluctuations for the C\(\alpha\) atoms in m4D2 and mutants showing that in general the mutations affect the dynamics of their surrounding regions. \\ 
Fig.~\ref{fig:dist-34+92}: Histogram of the distance between residues in position 34 and 92 and the heme propionates for m4D2 and its mutants. \\ 
Fig.~\ref{fig:dist-m4D2+19+DM}: Histogram of the distance between residue in position 19 and 36, and 77 and 94 for m4D2, T19D and T19D-T77D.

\bigskip 

\noindent In section \ref{subsec:subCEmethod} below, we display the following figure:

\medskip 

\noindent Fig.~\ref{fig:PBMC}: Reduction curves for T19D, M23N, R34Q, R92Q and T19D-T77D relative to m4D2 highlighting the mutant's $ E $ shift relative to m4D2.

\bigskip 

\noindent In section \ref{subsec:subcrooks} below, we discuss the mathematical procedure to estimate free energy differences by combining the Crooks fluctuation relation with Bayes theorem.

\bigskip

\noindent In section \ref{subsec:workdata} below, we display the following figures:

\medskip 

\noindent Fig.~\ref{fig:raw-data}: Non-equilibrium work values computed from the MD simulation data for m4D2 and its mutants. \\
Fig.~\ref{fig:raw-histograms}: Work histograms for forward and backward protocols obtained from the data in Fig.~\ref{fig:raw-data}.  \\ 
Fig.~\ref{fig:posteriors}: Final posterior probabilities obtained from the data in Fig.~\ref{fig:raw-data}. \\
Fig.~\ref{fig:convergence}: Convergence of our redox potential estimator $\tilde{E}$ towards the values given in the main text as the number of input data $\mu$ increases. Such values are based on the final posterior probabilities shown in Fig.~\ref{fig:posteriors}.

\bigskip 

\noindent In section \ref{subsec:exp} below, we display the following figure:

\medskip 

\noindent Fig.~\ref{fig:exp-fig}: Experimentally determined redox potentials for the M23N, R34Q and R92Q mutants of m4D2.

\bigskip 

\noindent In section \ref{subsec:abs} below, we provide the following table:

\medskip 

\noindent Tab.~\ref{table:abs}: Experimental and predicted redox potentials $E$ for m4D2, T19D, M23N, R34Q, R92Q, and T19D-T77D.

\newpage

\subsection{Conformational stability of the MD simulations} \label{subsec:stability}

\newpage

\newpage

\newpage

\newpage

\newpage

\newpage

\newpage
 
\newpage

\newpage

\cleardoublepage

\subsection{Continuum electrostatics calculations} \label{subsec:subCEmethod}

\begin{figure}[h!] % figure S9
\centering
\includegraphics[clip,width=11.5cm]{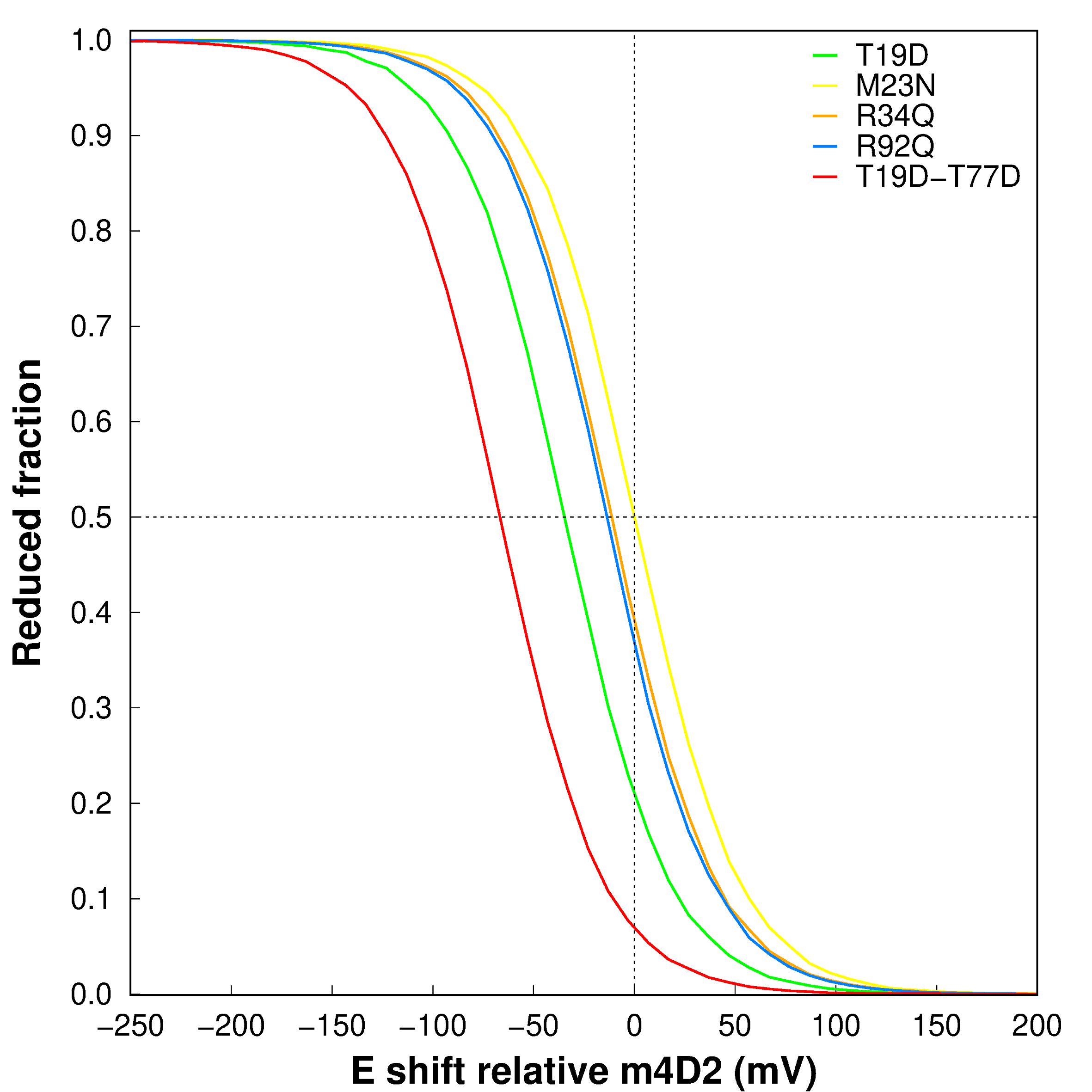}
	\caption{Individual reduction curve for T19D, M23N, R34Q, R92Q and T19D-T77D relative to m4D2. These curves were obtained using the \PBMC~method, a combination of Poisson-Boltzmann (PB) calculations and Monte Carlo (MC) simulations.\cite{teixeira2002}}
\label{fig:PBMC}
\end{figure}

\cleardoublepage

\subsection{Theory of estimation of free energy differences using the Crooks-Bayes method} \label{subsec:subcrooks}

Consider a protocol $\Lambda_i$ connecting two thermodynamic states, $A_i$ and $B_i$, whose difference in free energy is $\Delta G_{\Lambda_i} = G_{B_i} - G_{A_i}$.
Let $W_i$ be the work cost of implementing such a protocol.
\citet{maragakis2008} showed that, in the event of implementing $i=1,\dots,\nu$ of these protocols, the probability density $p(\boldsymbol{g}|\boldsymbol{W},\boldsymbol{\Lambda})$ for the vector of hypotheses $\boldsymbol{g} = (g_{A_1},g_{B_1}, \dots, g_{A_\nu},g_{B_\nu})$ about the true free energies $\boldsymbol{G} = (G_{A_1},G_{B_1}, \dots, G_{A_\nu},G_{B_\nu})$, given the work vector $\boldsymbol{W} = (W_1, \dots, W_{\nu})$ and the protocol vector $\boldsymbol{\Lambda} = (\Lambda_1, \dots, \Lambda_{\nu})$, can be taken as
\begin{equation}
    p(\boldsymbol{g}|\boldsymbol{W},\boldsymbol{\Lambda}) \propto \prod_{i=1}^{\nu} f(\beta W_i - \beta \Delta g_{\Lambda_i} + M_{\Lambda_i}),
    \label{eq:gen-BayesCrooks}
\end{equation}
where $\beta$ is an inverse temperature, $f(x)=1/[1+\mathrm{exp}(-x)]$ is the logistic function, and each $M_{\Lambda_i}$ is defined as
\begin{equation}
    M_{\Lambda_i} = \log\left[\frac{p(\Lambda_i|\Delta g_{\Lambda_i})}{p(\tilde{\Lambda}_i|\Delta g_{\tilde{\Lambda}_i})}\right].
\end{equation}
Here, $\Delta g_{\tilde{\Lambda}_i} = - \Delta g_{\Lambda_i}$ and $\tilde{\Lambda}_i$ labels a backward protocol connecting $B_i$ and $A_i$.

Eq.~\eqref{eq:gen-BayesCrooks} rests on four pillars:\cite{maragakis2008} (i) Bayes theorem, written as $p(\Delta g_{\Lambda_i}|W_i,\Lambda_i) \propto p(\Delta g_{\Lambda_i}) \,p (W_i,\Lambda_i|\Delta g_{\Lambda_i})$, where $p(\Delta g_{\Lambda_i})$ and $p(\Delta g_{\Lambda_i}|W_i,\Lambda_i)$ are the prior and posterior probabilities, respectively, and $p (W_i,\Lambda_i|\Delta g_{\Lambda_i})$ is the likelihood function; (ii) an uninformed state of information about both $\Delta G_{\Lambda_i}$ and the work distribution associated with each protocol, represented by $p(\Delta g_{\Lambda_i}) \propto 1$ and  $ p(W_i,\Lambda_i|\Delta g_{\Lambda_i}) + p(-W_i,\tilde{\Lambda}_i|\Delta g_{\tilde{\Lambda}_i}) \propto 1$, respectively; (iii) Crooks' relation\cite{crooks1999,maragakis2008,seifert2012}
\begin{equation}
    \frac{p(W_i|\Delta g_{\Lambda_i},\Lambda_i)}{p(- W_i |\Delta g_{\tilde{\Lambda}_i}, \tilde{\Lambda}_i)} = \mathrm{e}^{\beta\left(W_i - \Delta g_{\Lambda_i}\right)},
    \label{eq:crooks-relApp}
\end{equation}
which is a detailed fluctuating work symmetry; and (iv) uncorrelated work values.
We thus refer to the application of Eq.~\eqref{eq:gen-BayesCrooks} as the Crooks-Bayes method. 
Assumptions (ii - iv) encapsulate the information available prior to collecting the simulated work data.
We further note that, unlike in the main text, the work probabilities in Eq.~\eqref{eq:crooks-relApp} are written as conditional on the free energy difference.
This is a matter of notation; in the main text, Crooks' relation is first discussed in the context of histogram-based estimation, where the aforementioned conditioning can be ignored. 
This cannot however be done when employing Bayesian estimation.\cite{jaynes2003} 

To predict redox potentials via Eq.~\eqref{eq:gen-BayesCrooks}, we first need to adapt it to our particular scenario, as follows. 
Imagine $\nu = 2\mu$ experiments such that half of them correspond to the repetitions of a single protocol, connecting the thermodynamic states $A$ and $B$, while the the other half corresponds to repeating the backward protocol, i.e., connecting $B$ and $A$. 
In that case, the vector of free energies is given as 
\begin{equation*}
    \boldsymbol{G} = (\smash[t]{\underbrace{G_{A},G_{B}, \dots, G_{A},G_{B}}_\text{$\mu$ times}}, \smash[t]{\underbrace{G_{B}, G_{A},\dots,G_{B}, G_{A}}_\text{$\mu$ times}}).
\end{equation*}
By defining $\Delta G \equiv G_B - G_A$, we further see that $\Delta G_{\Lambda_i} = \Delta G$ for all $i$.
It is then appropriate to rename $p(\boldsymbol{g}|\boldsymbol{W},\boldsymbol{\Lambda})$ as $p(\Delta g|\boldsymbol{W},\boldsymbol{\Lambda})$, where, in this case, $\boldsymbol{W} = (W_1, \dots, W_{2\mu})$ and $\boldsymbol{\Lambda} = (\Lambda_1, \dots, \Lambda_\mu, \tilde{\Lambda}_{\mu+1},\dots,\tilde{\Lambda}_{2\mu})$.
The fact that we have the same number of forward and backward iterations further allows each $M_{\Lambda_i}$ to be approximated as $M_{\Lambda_i} \approx 0$. \cite{maragakis2008}
Putting all the pieces together, Eq.~\eqref{eq:gen-BayesCrooks} becomes
\begin{align}
    p(\Delta g|\boldsymbol{W},\boldsymbol{\Lambda}) \propto 
    \prod_{i=1}^{\mu}\,
    & f(\beta W_i-\beta\Delta g_{\Lambda_i}) 
    \prod_{j=\mu+1}^{2\mu}\,f(\beta W_j-\beta \Delta g_{\tilde{\Lambda}_i})
    \nonumber \\
    = \prod_{i=1}^{\mu}\,&f(\beta W_i-\beta\Delta g) 
    \prod_{j=\mu+1}^{2\mu}\,f(\beta W_j+\beta\Delta g).
    \label{eq:posterior-supp-final}
\end{align}
To simplify the notation, we may now drop the dependence on the protocol vector as $p(\Delta g|\boldsymbol{W},\boldsymbol{\Lambda}) \mapsto p(\Delta g|\boldsymbol{W})$.
Inserting the work values $\boldsymbol{W}$ associated with the protocols $\boldsymbol{\Lambda}$, where $W_i = \epsilon_{\rm{red}}^i - \epsilon_{\rm{ox}}^i =  \Delta \epsilon_i$ for $i=1, ..., \mu$, and $W_i = \epsilon_{\rm{ox}}^i - \epsilon_{\rm{red}}^i =  \Delta \epsilon_i$ for $i=\mu+1, ..., 2 \mu$, in Eq.~\eqref{eq:posterior-supp-final}, produces Eq.~(2) in the main text. 
%~\eqref{eq:posterior-main} 
This allows us to calculate the likelihood of different hypotheses $\Delta g$, as shown in Fig.~\ref{fig:posteriors} for the oxidation and reduction of the heme group. 
This then leads to the redox potential estimates reported in the main text, here shown in Fig.~\ref{fig:convergence}. 

\cleardoublepage 

\subsection{Estimation of redox potentials from non-acknowledgement work data} \label{subsec:workdata}

\begin{figure}[h!]
\centering
\includegraphics[trim={0.2cm -0.75cm 1.55cm 0cm},clip,width=8.5cm]{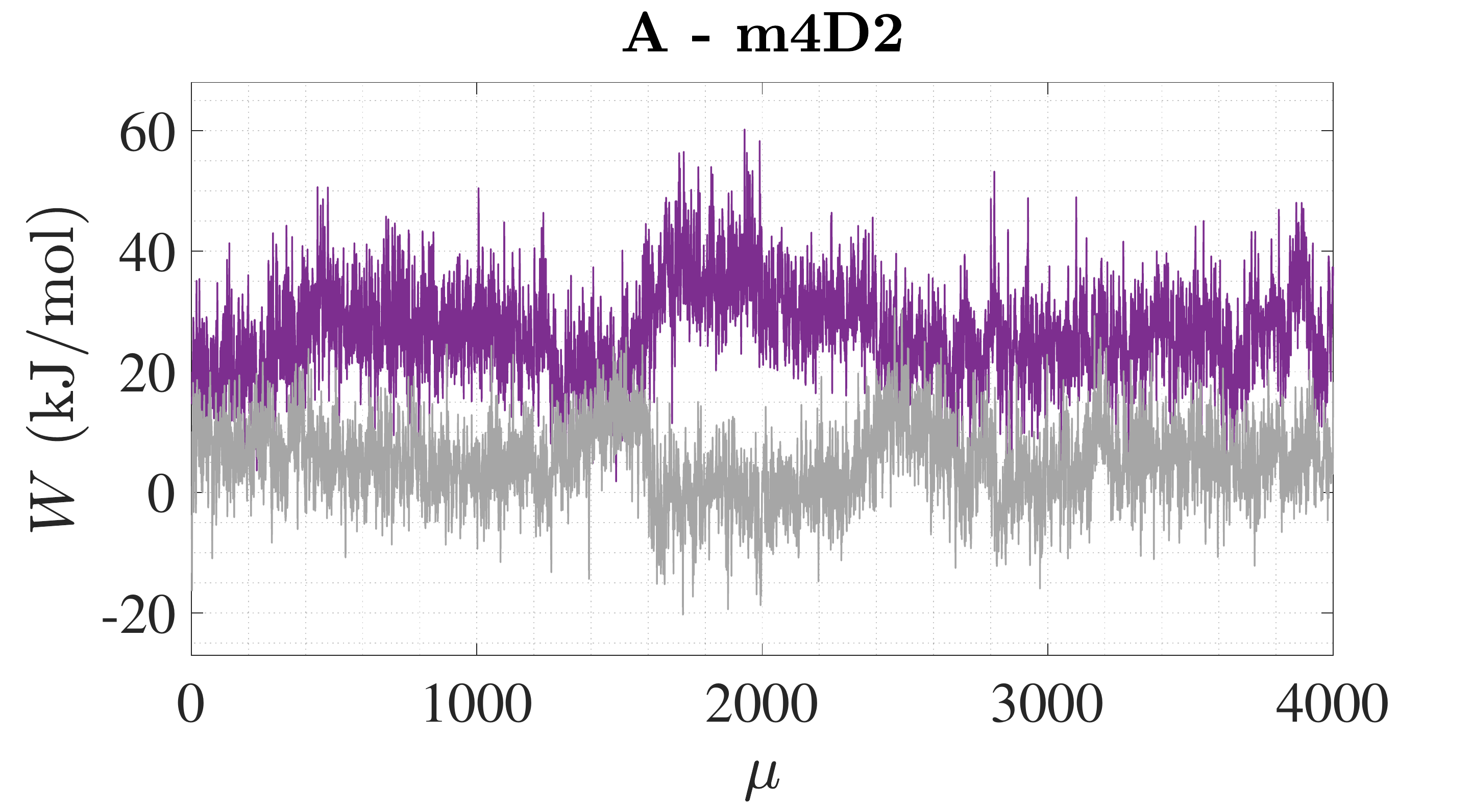}\includegraphics[trim={0.2cm -0.75cm 1.55cm 0cm},clip,width=8.5cm]{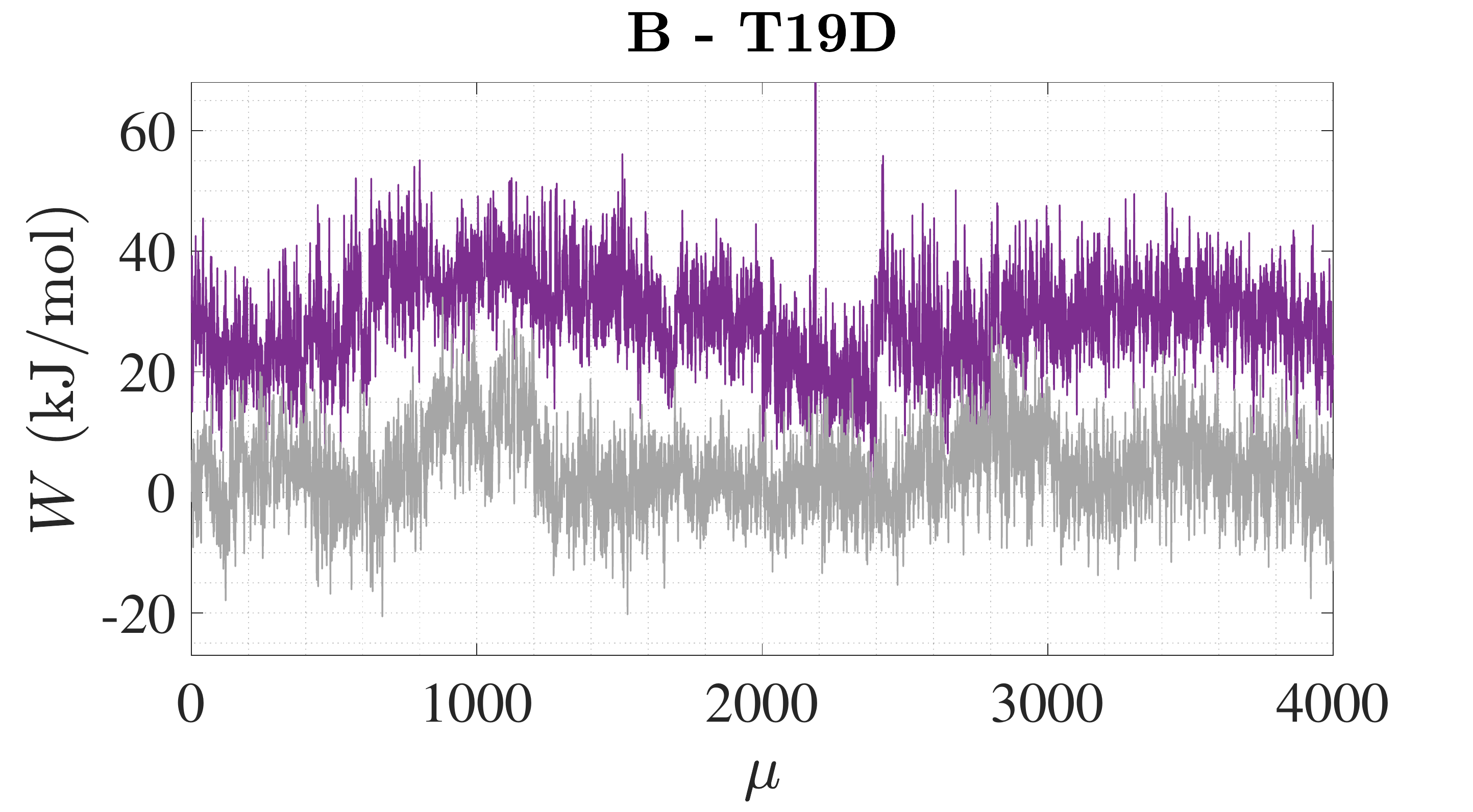}
\includegraphics[trim={0.2cm -0.75cm 1.55cm 0cm},clip,width=8.5cm]{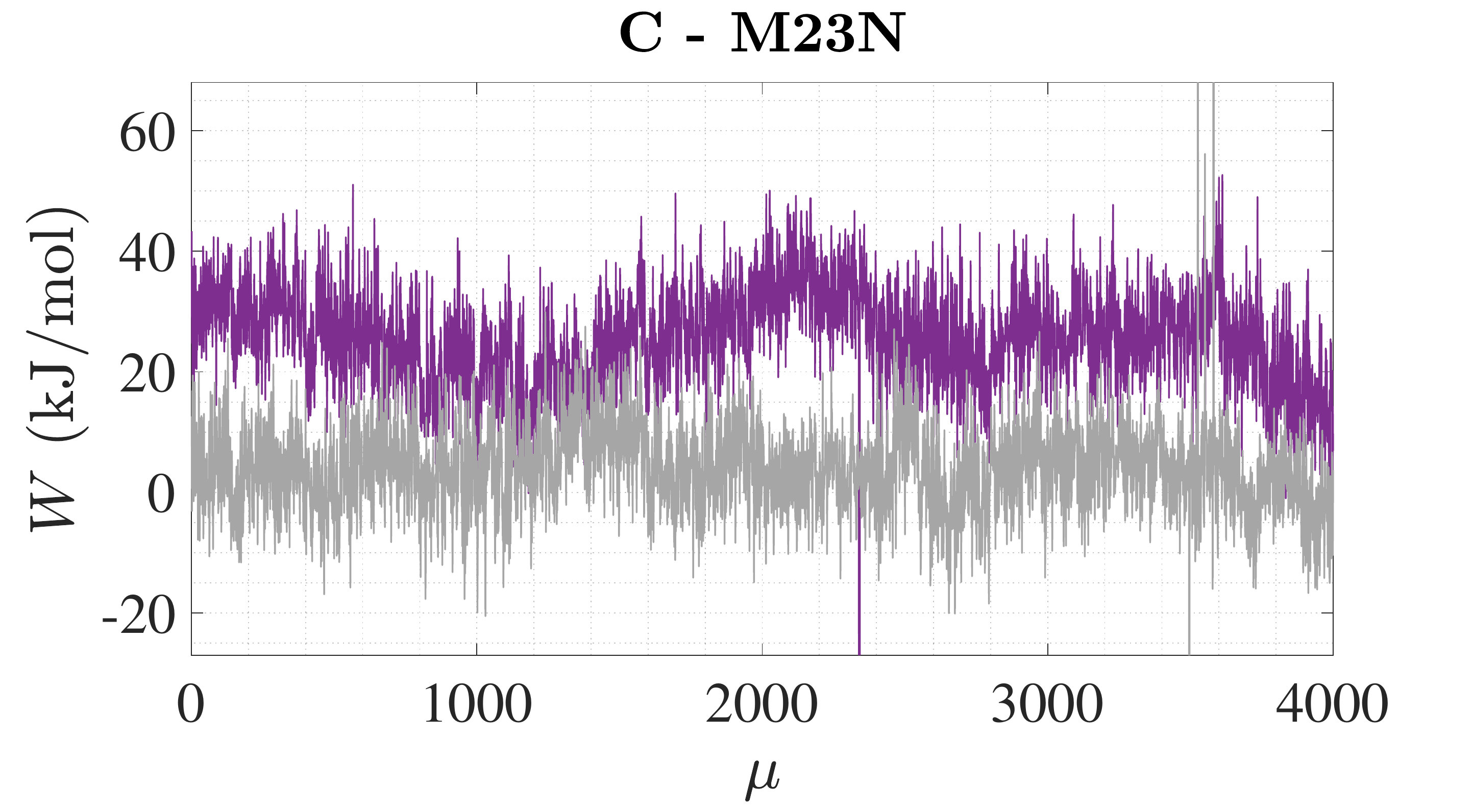}\includegraphics[trim={0.2cm -0.75cm 1.55cm 0cm},clip,width=8.5cm]{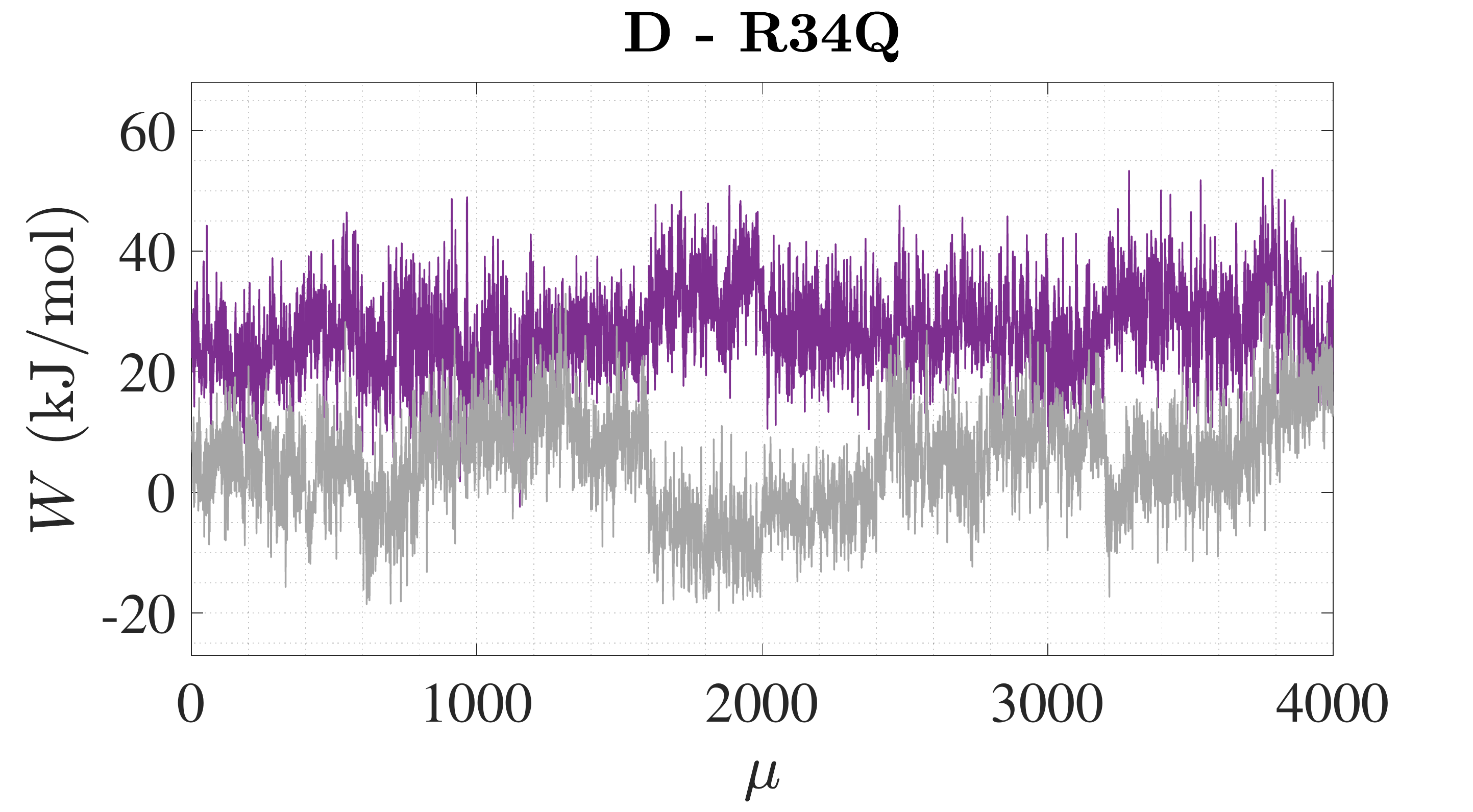}
\includegraphics[trim={0.2cm -0.75cm 1.55cm 0cm},clip,width=8.5cm]{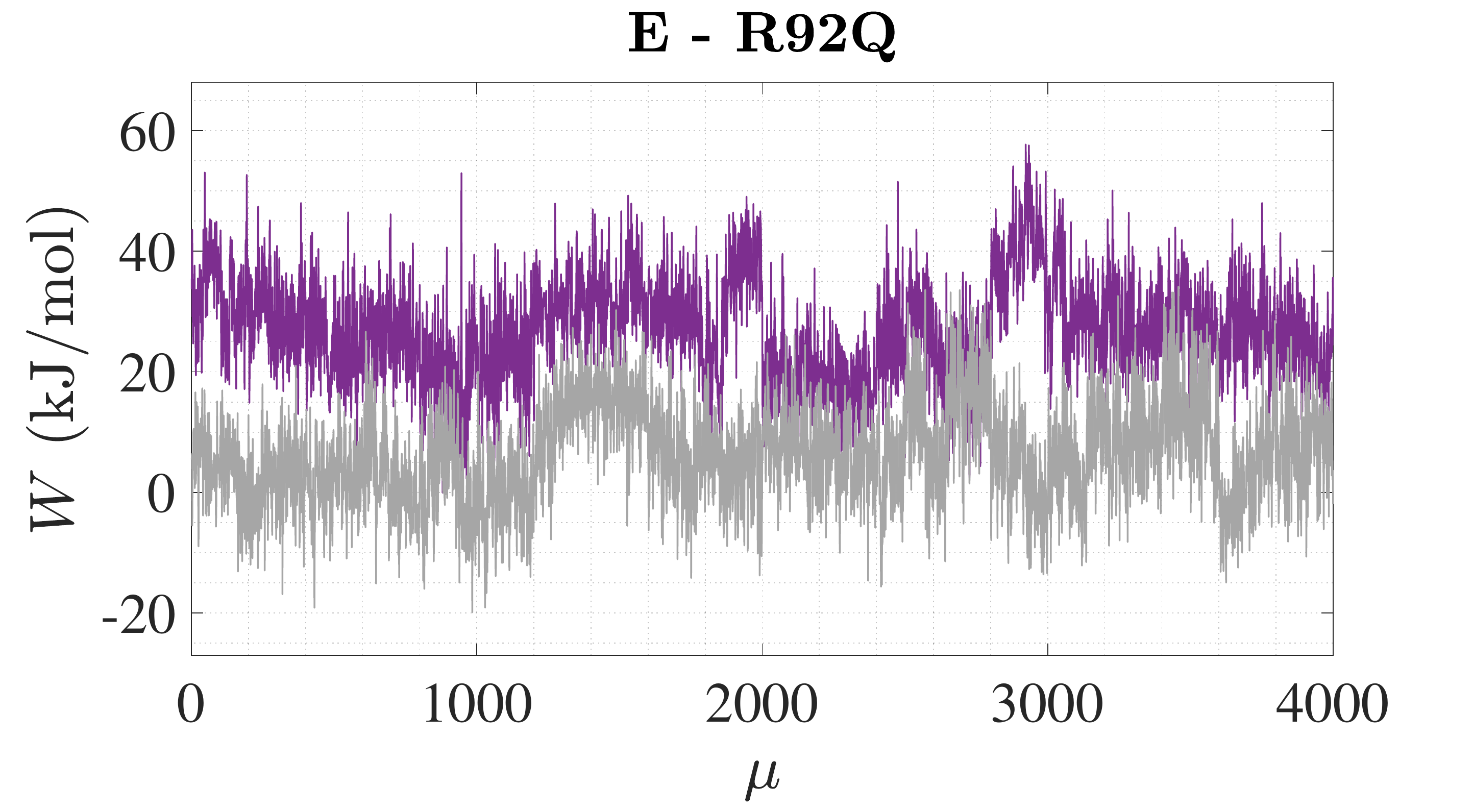}\includegraphics[trim={0.2cm -0.75cm 1.55cm 0cm},clip,width=8.5cm]{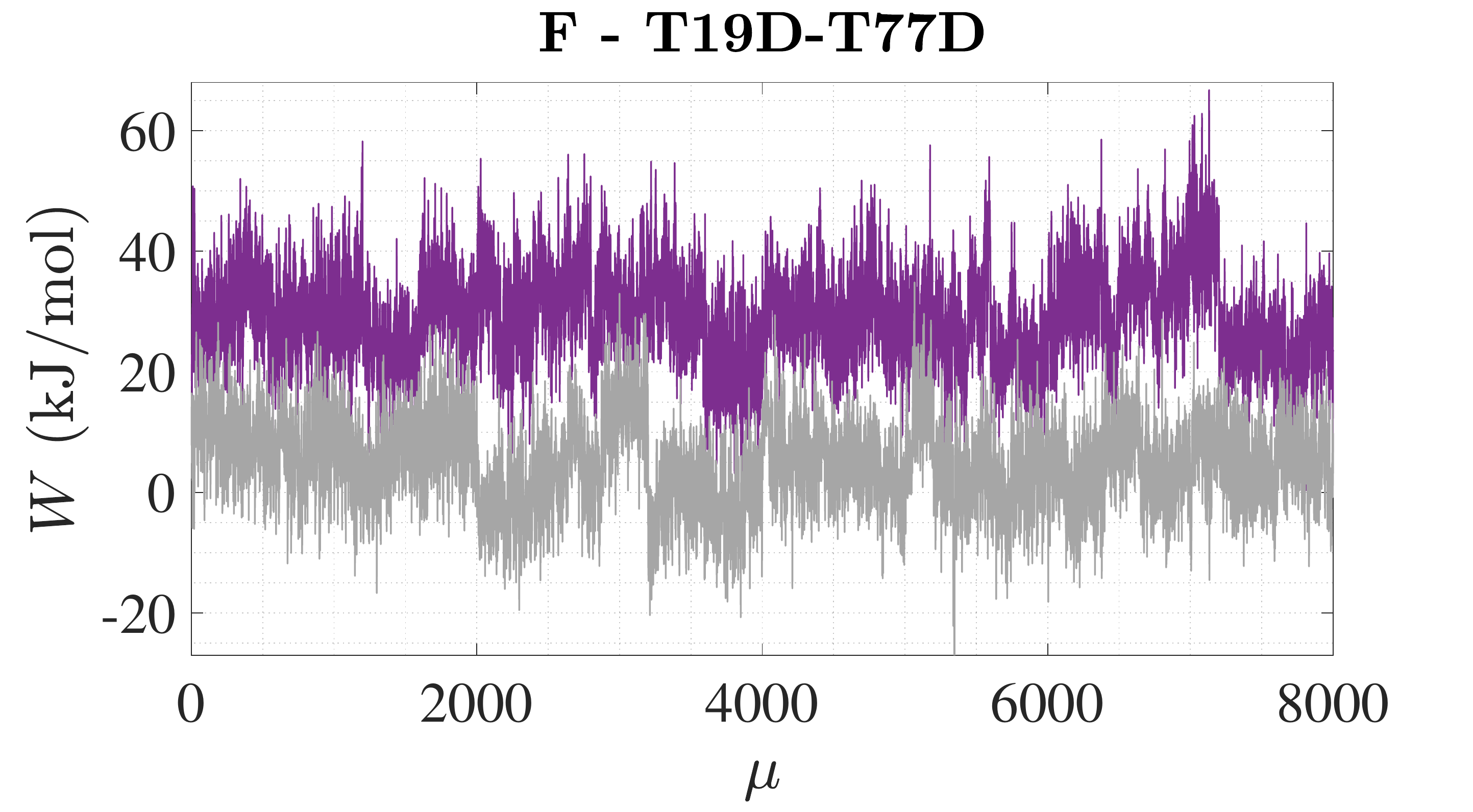}
	\caption{
Statistical work values $W$ and $-W$ for the forward/reduction (purple) and backward/oxidation (grey) processes, respectively. Each $W$ is given as $W =  \Delta \epsilon$ where  $\Delta \epsilon$ denotes the change of energy of the protein\cite{jarzynski1997,seifert2012}.
 }
\label{fig:raw-data}
\end{figure}

\newpage

\begin{figure}[th!]
\centering
\includegraphics[trim={0.2cm -0.75cm 2.1cm 0cm},clip,width=8.5cm]{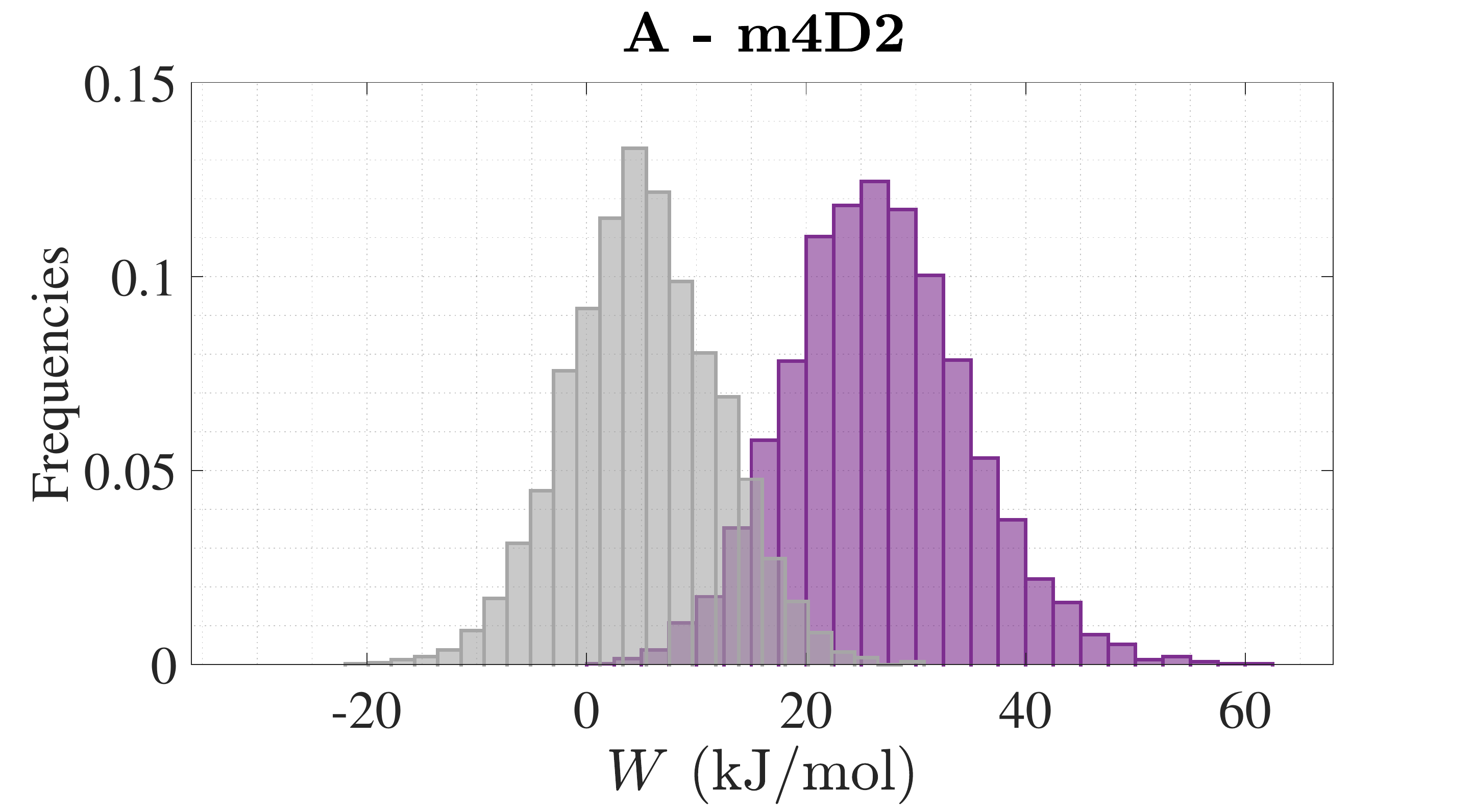}\includegraphics[trim={0.2cm -0.75cm 2.1cm 0cm},clip,width=8.5cm]{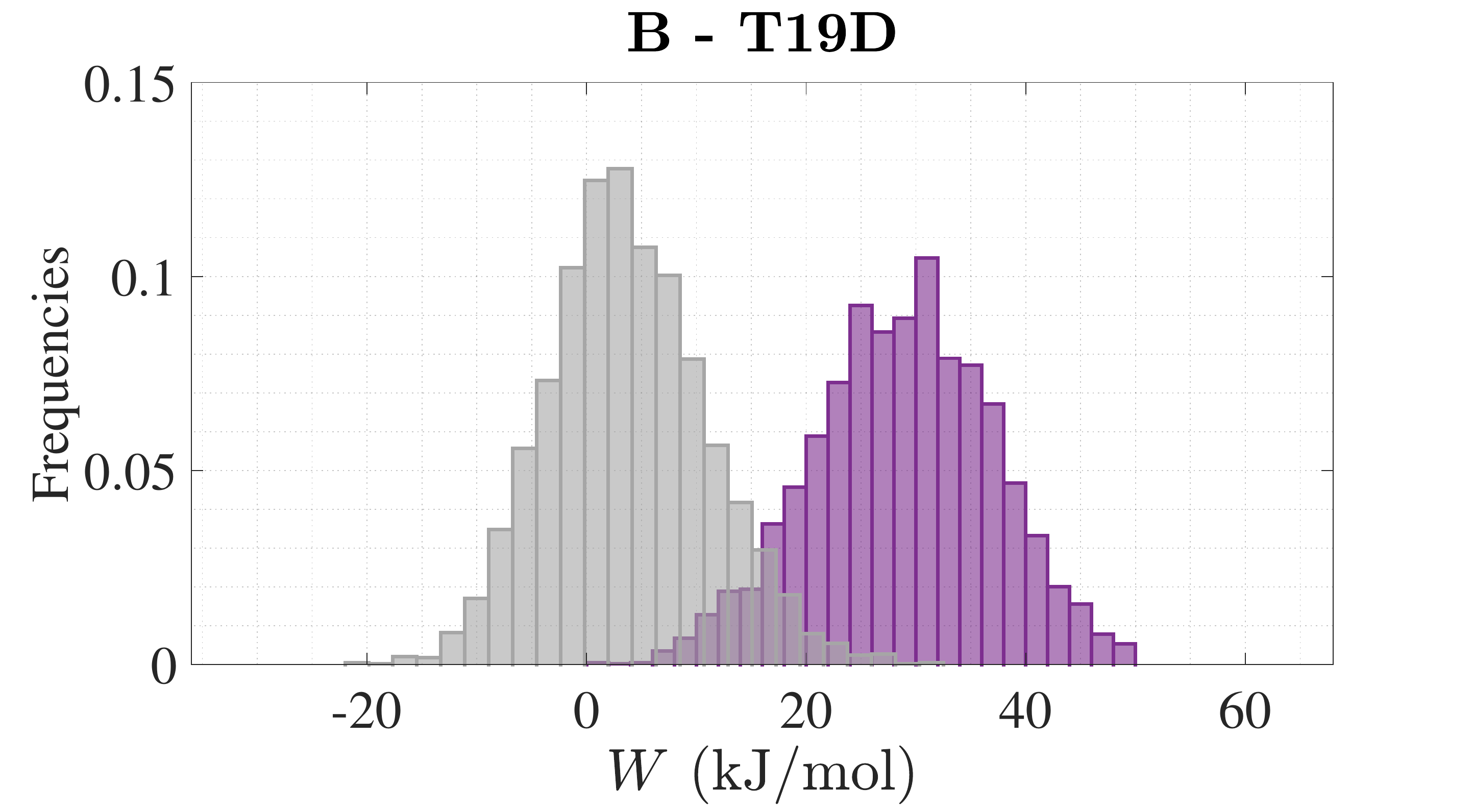}
\includegraphics[trim={0.2cm -0.75cm 2.1cm 0cm},clip,width=8.5cm]{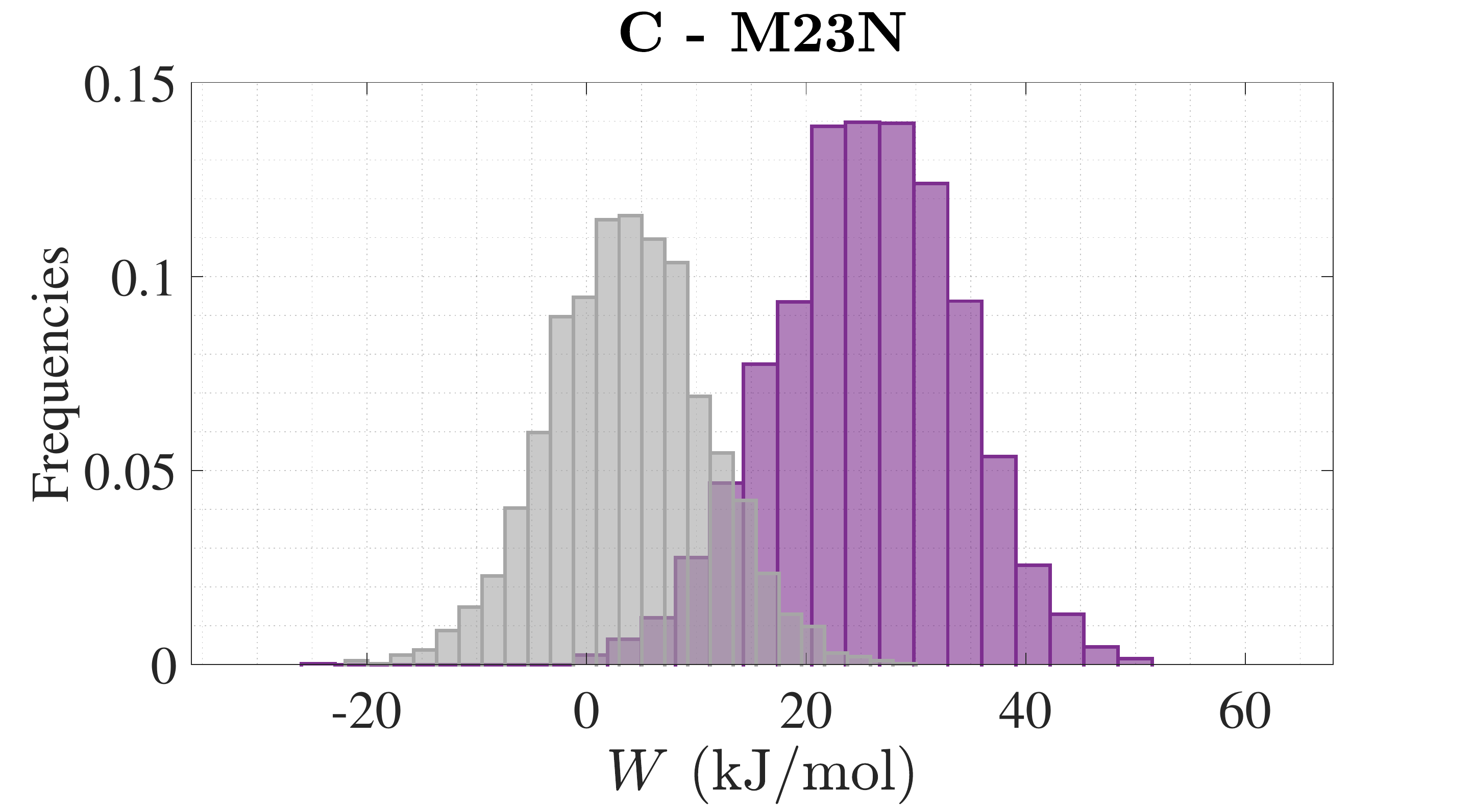}\includegraphics[trim={0.2cm -0.75cm 2.1cm 0cm},clip,width=8.5cm]{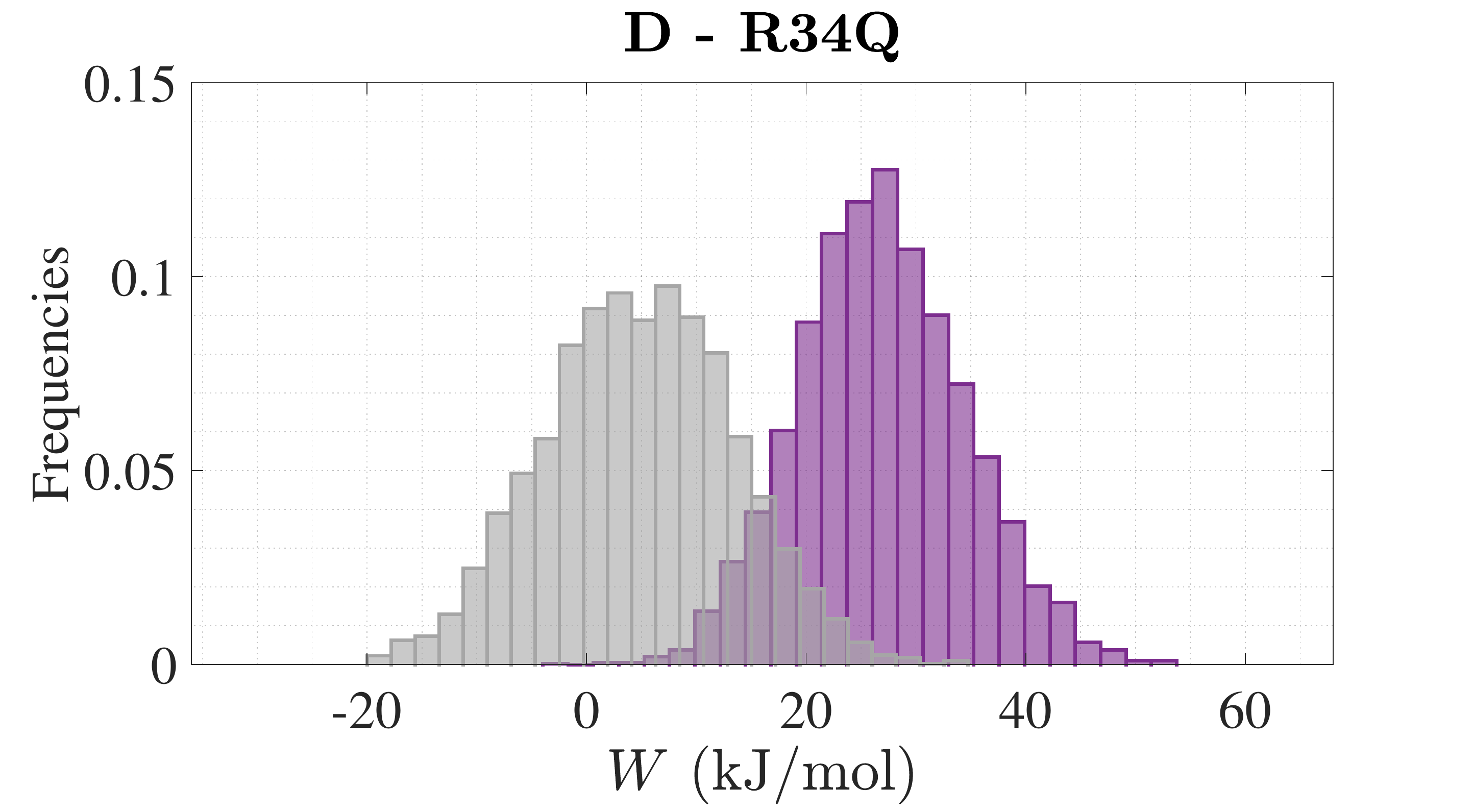}
\includegraphics[trim={0.2cm -0.75cm 2.1cm 0cm},clip,width=8.5cm]{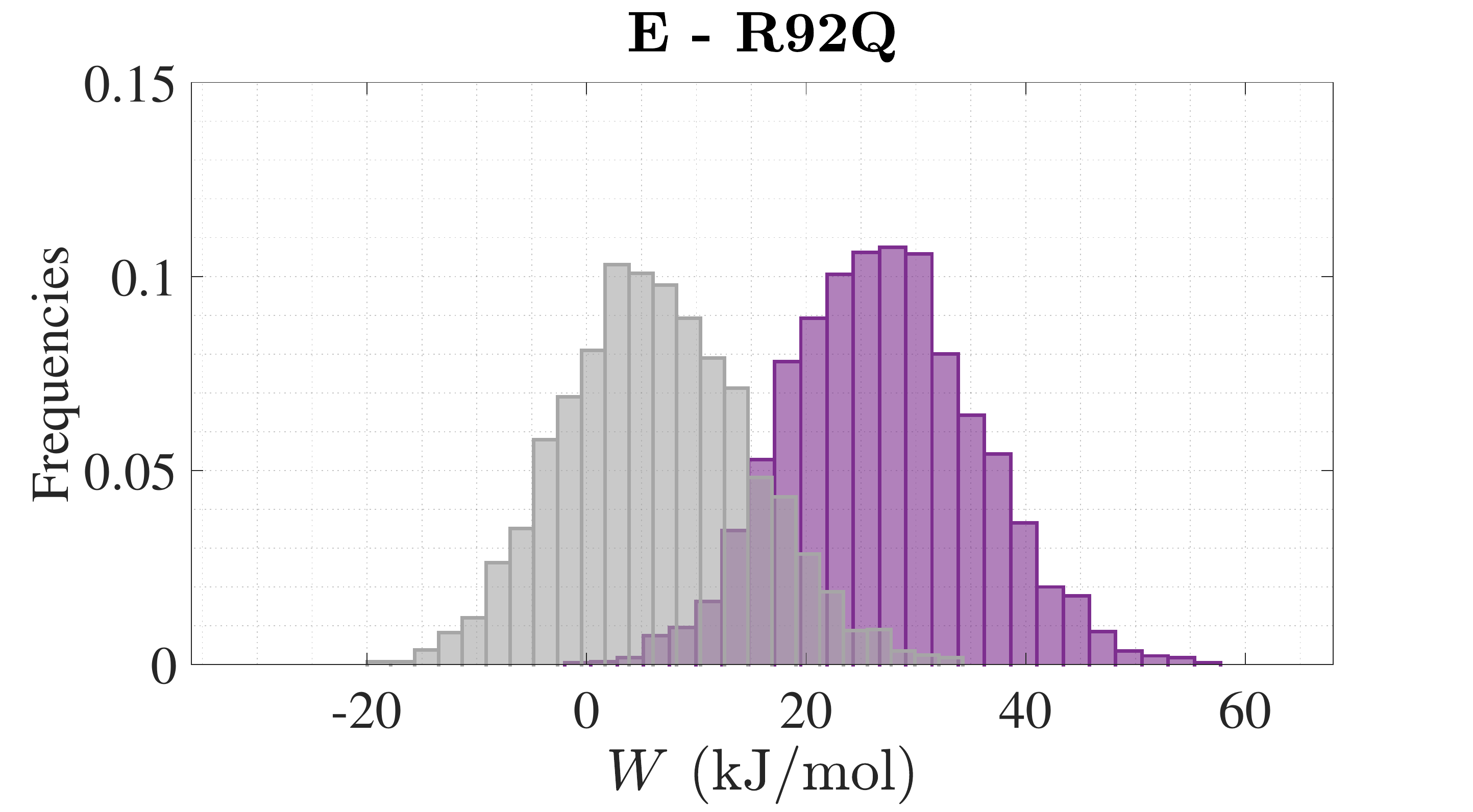}\includegraphics[trim={0.2cm -0.75cm 2.1cm 0cm},clip,width=8.5cm]{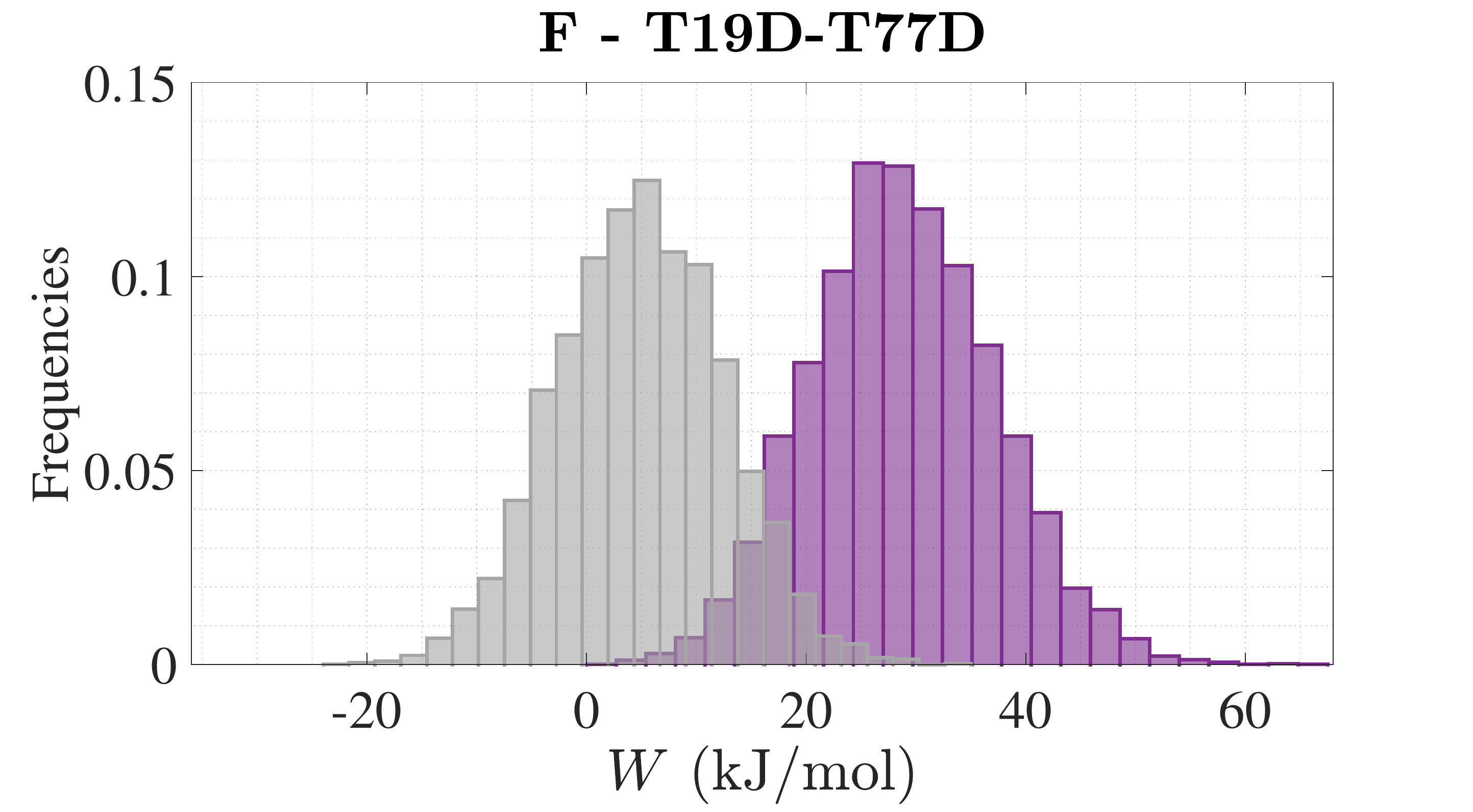}
\caption{
Work histograms $p(W|\Lambda)$ and $p(-W|\tilde{\Lambda})$ for the forward/reduction ($\Lambda$; purple) and backward/oxidation ($\tilde{\Lambda}$; grey) processes obtained from the data in Fig.~\ref{fig:raw-data}, respectively.
These have been generated using the \texttt{histogram} function in MATLAB.
Note that the Crooks-Bayes method discussed in Sec.~\ref{subsec:subcrooks} does not make use of these histograms, which are shown here for illustrative purposes only.
While histogram-based estimation coupled with statistical bootstrapping is the current standard in the literature, we found that such a method was less precise and computationally slower for a given data set, thus justifying our choice of using the Crooks-Bayes approach in this work.  
}
\label{fig:raw-histograms}
\end{figure}

\newpage

\begin{figure}[h!]
\centering
\includegraphics[trim={0.2cm -0.75cm 2.1cm 0cm},clip,width=8.5cm]{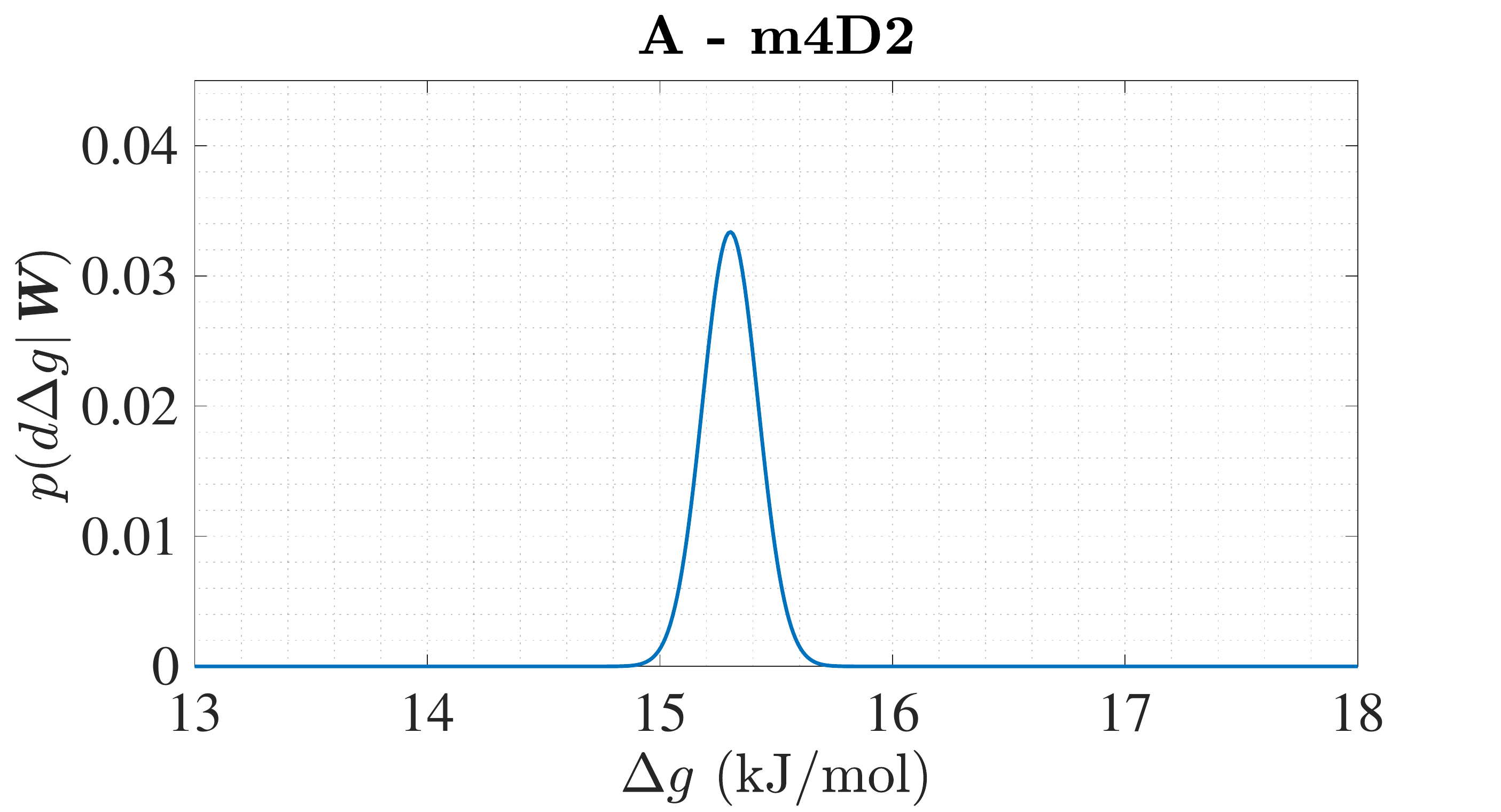}\includegraphics[trim={0.2cm -0.75cm 2.1cm 0cm},clip,width=8.5cm]{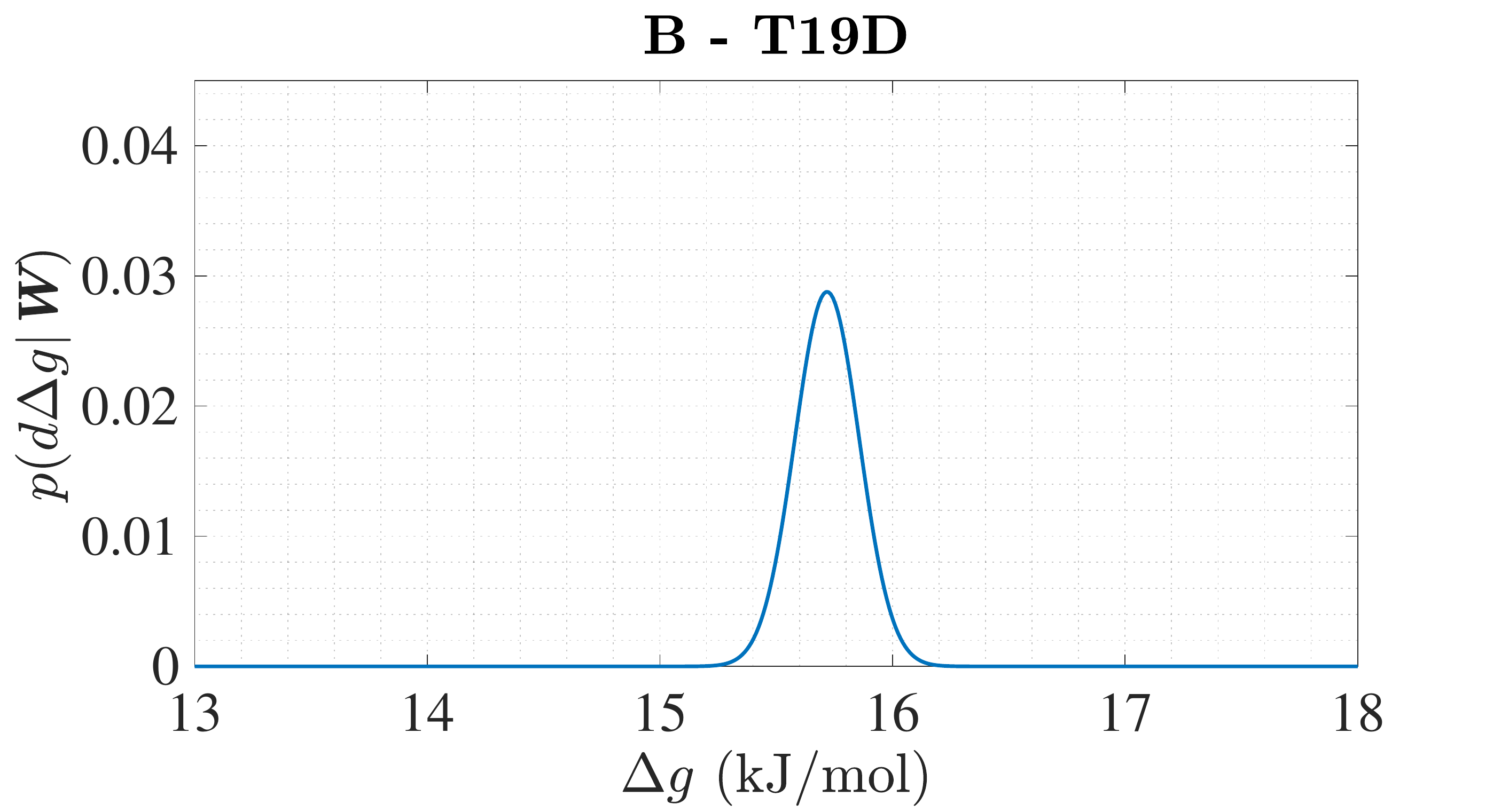}
\includegraphics[trim={0.2cm -0.75cm 2.1cm 0cm},clip,width=8.5cm]{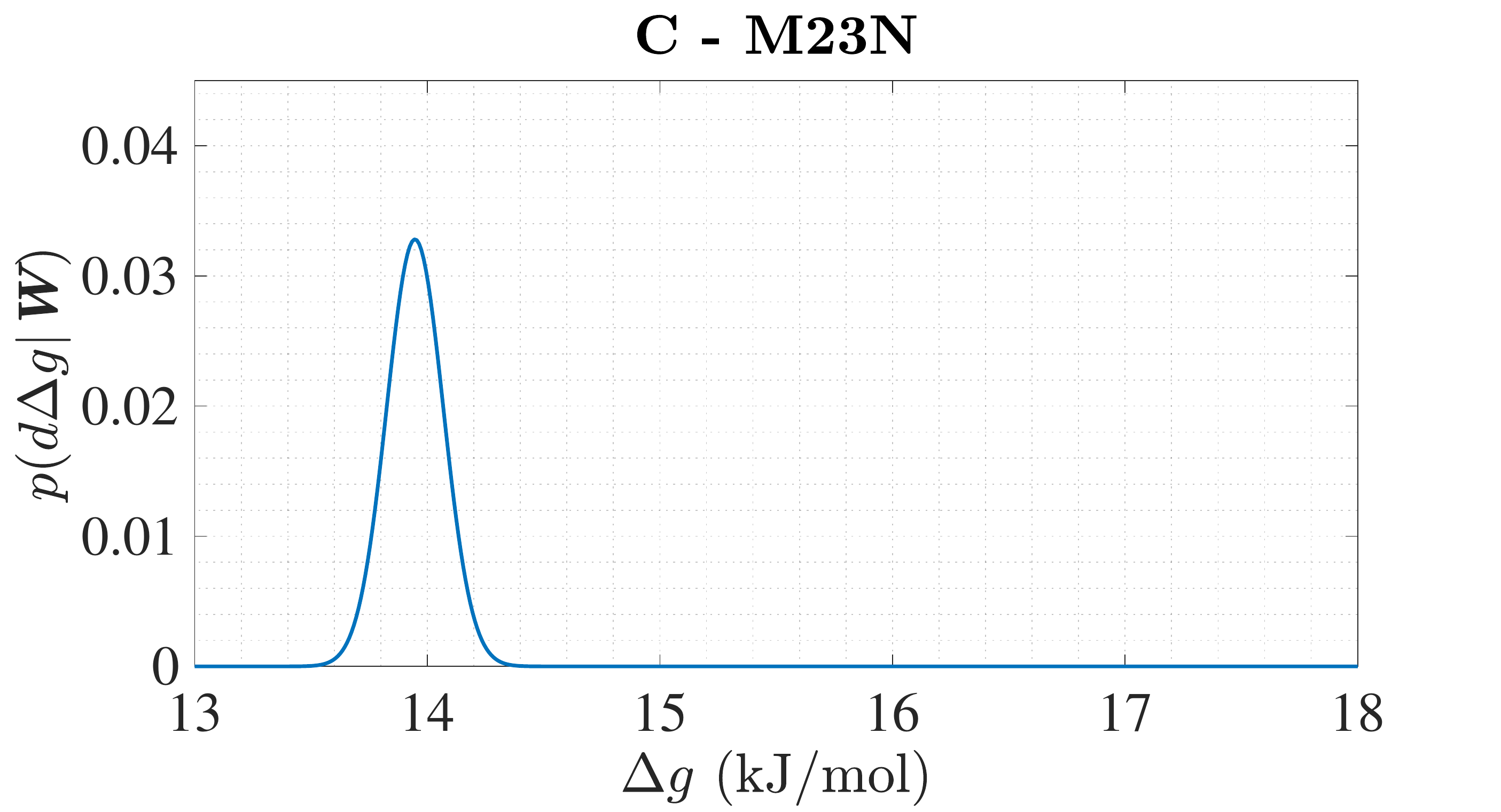}\includegraphics[trim={0.2cm -0.75cm 2.1cm 0cm},clip,width=8.5cm]{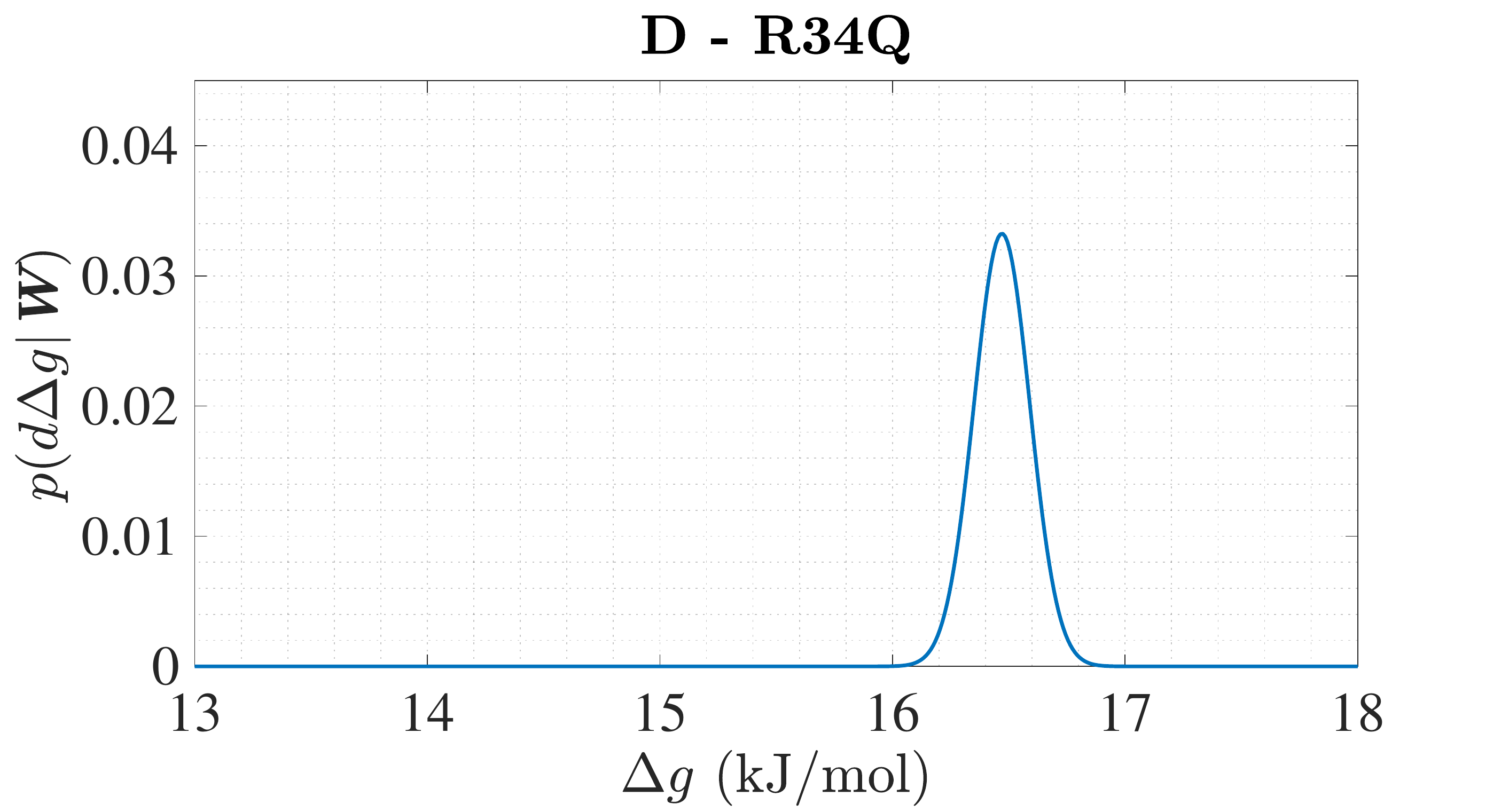}
\includegraphics[trim={0.2cm -0.75cm 2.1cm 0cm},clip,width=8.5cm]{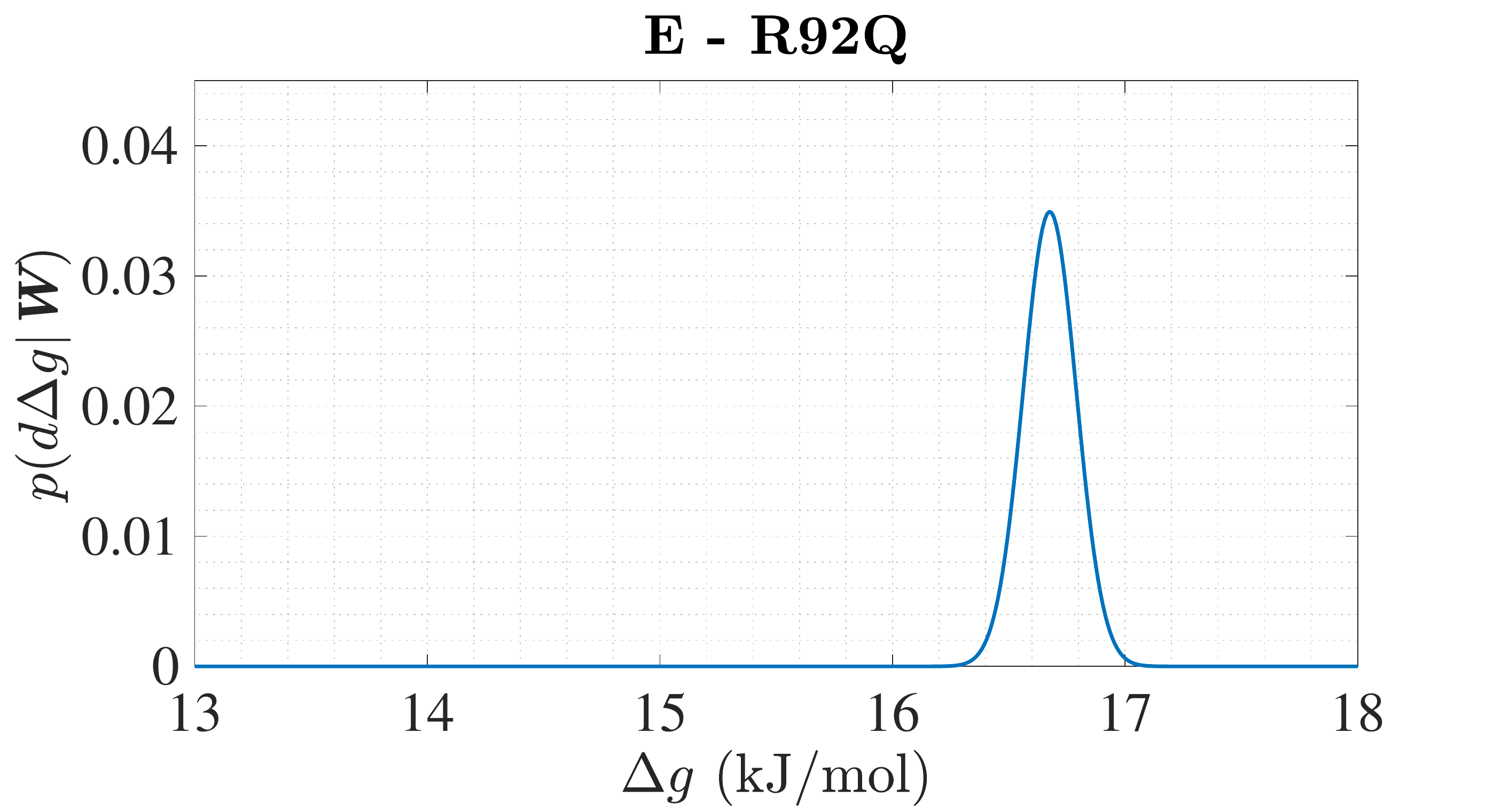}\includegraphics[trim={0.2cm -0.75cm 2.1cm 0cm},clip,width=8.5cm]{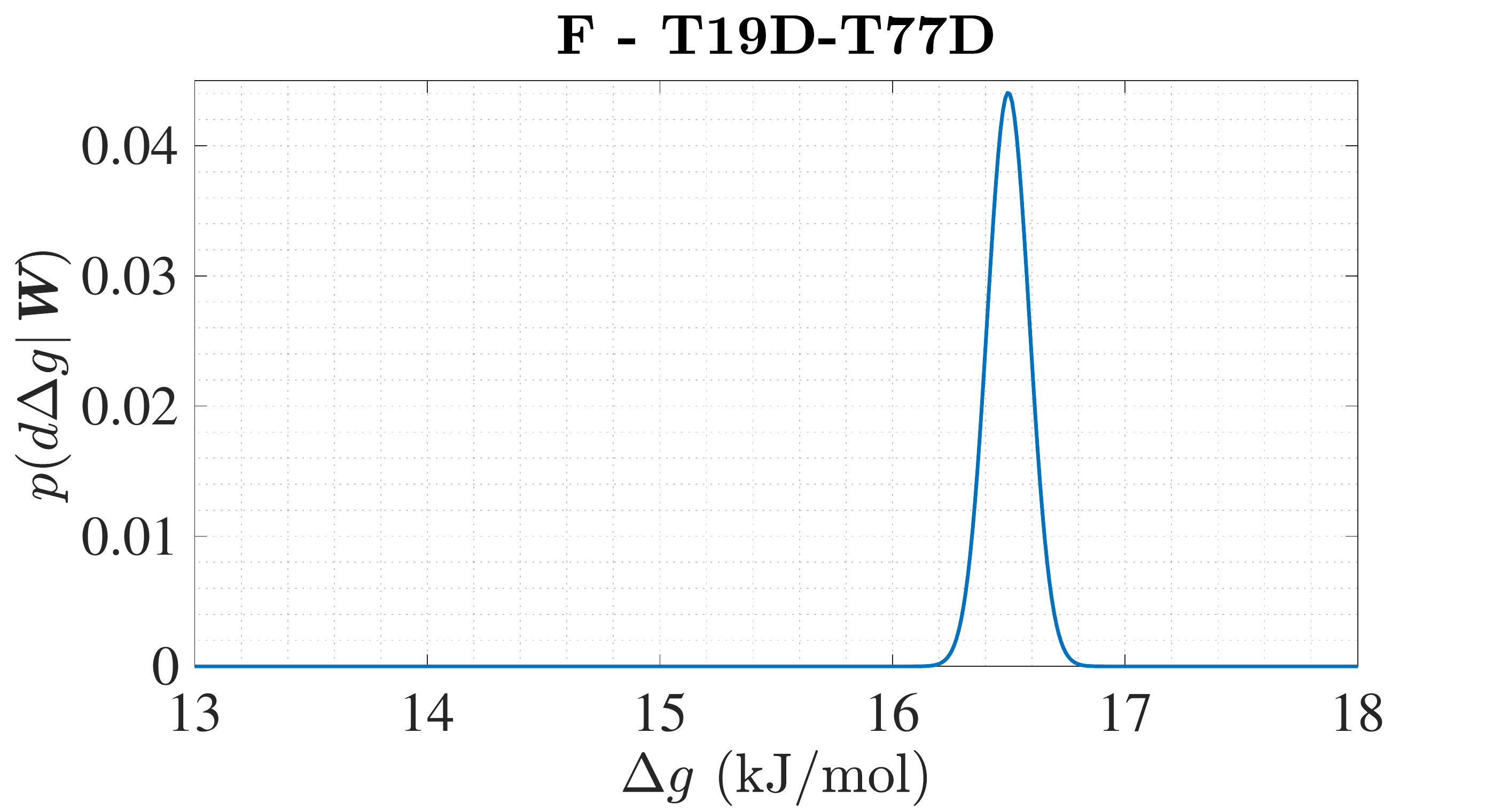}
	\caption{Numerical posterior probability $p(d\Delta g|\boldsymbol{W}) = p(\Delta g|\boldsymbol{W}) d\Delta g$ calculated via Eq.~\eqref{eq:posterior-supp-final} using the work data $\boldsymbol{W}$ in Fig.~\ref{fig:raw-data}.
    This amounts to multiplying a large number of logistic functions with their mirror image, thus leading to the bell-like shape that can be observed.\cite{maragakis2008}  
    More importantly, each profile encodes all the available information needed to calculate an estimate for the true free energy difference $\Delta G$.
    To normalise these profiles, we chose a conservative range of $\Delta g \in [-396, 531]\,\rm{kJ}/\rm{mol}$, which covered the most likely work values produced by our MD simulations.
    As shown in Fig.~\ref{fig:convergence}, these estimates enable the means to predict redox potentials $E$, given the relation $E = - \Delta G/F$. Here, $E$ and $\Delta G$ are measured in units of voltage and of energy, respectively, while $F$ stands for Faraday's constant.
} 
\label{fig:posteriors}
\end{figure}

\newpage

\begin{figure}[h!]
\centering
\includegraphics[trim={0.2cm -0.75cm 1.55cm 0cm},clip,width=8.5cm]{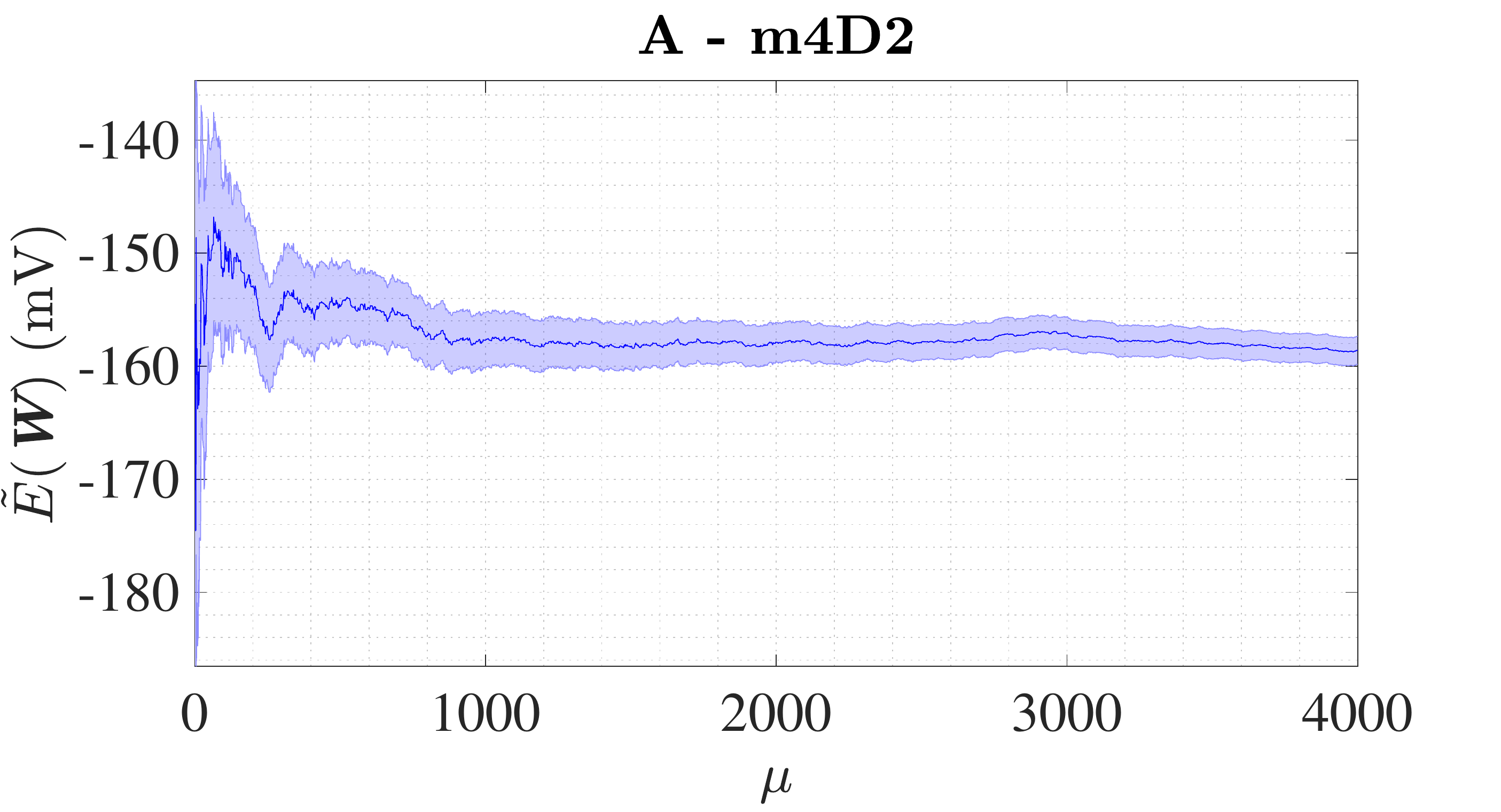}\includegraphics[trim={0.2cm -0.75cm 1.55cm 0cm},clip,width=8.75cm]{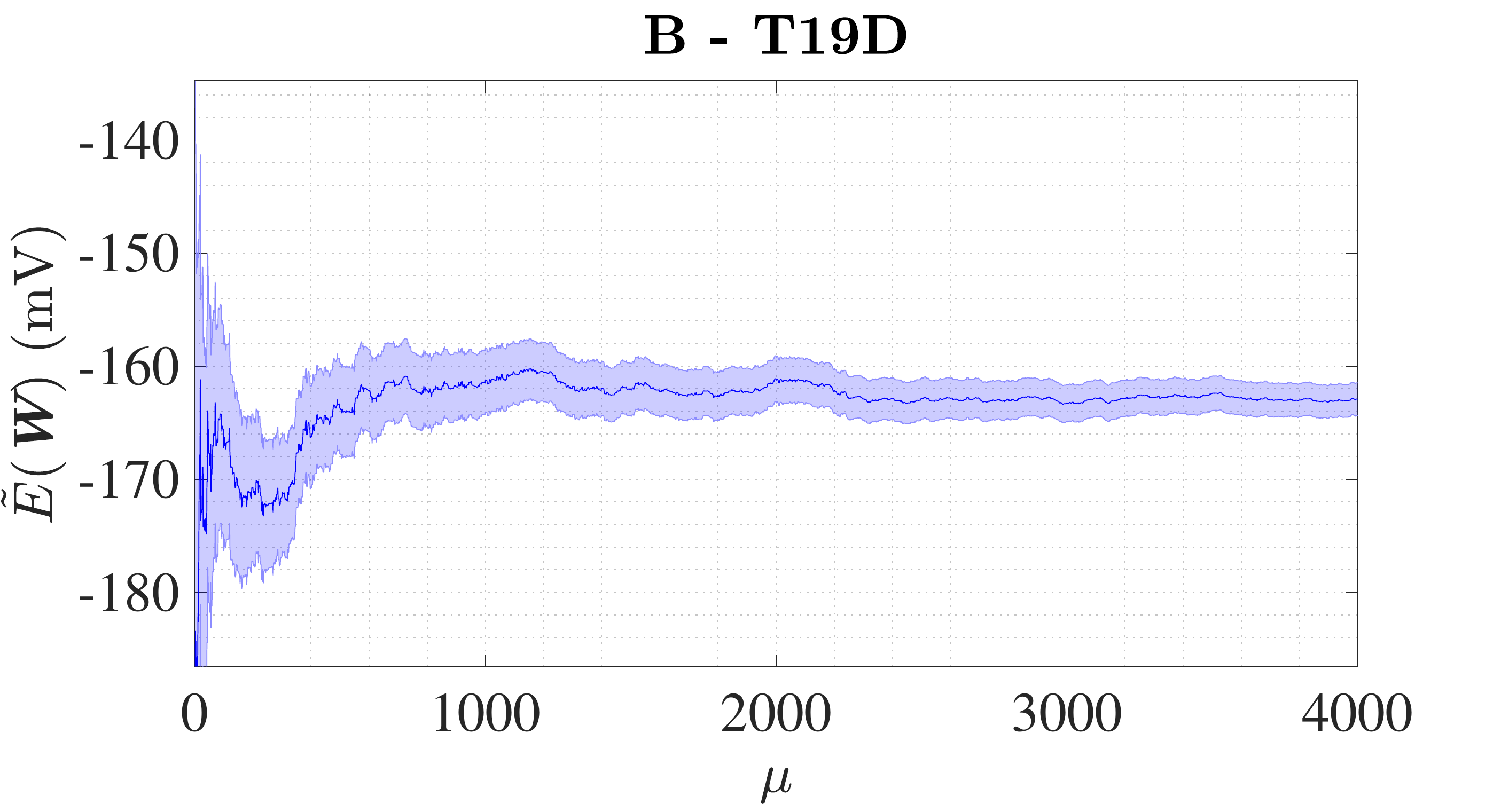}
\includegraphics[trim={0.2cm -0.75cm 1.55cm 0cm},clip,width=8.5cm]{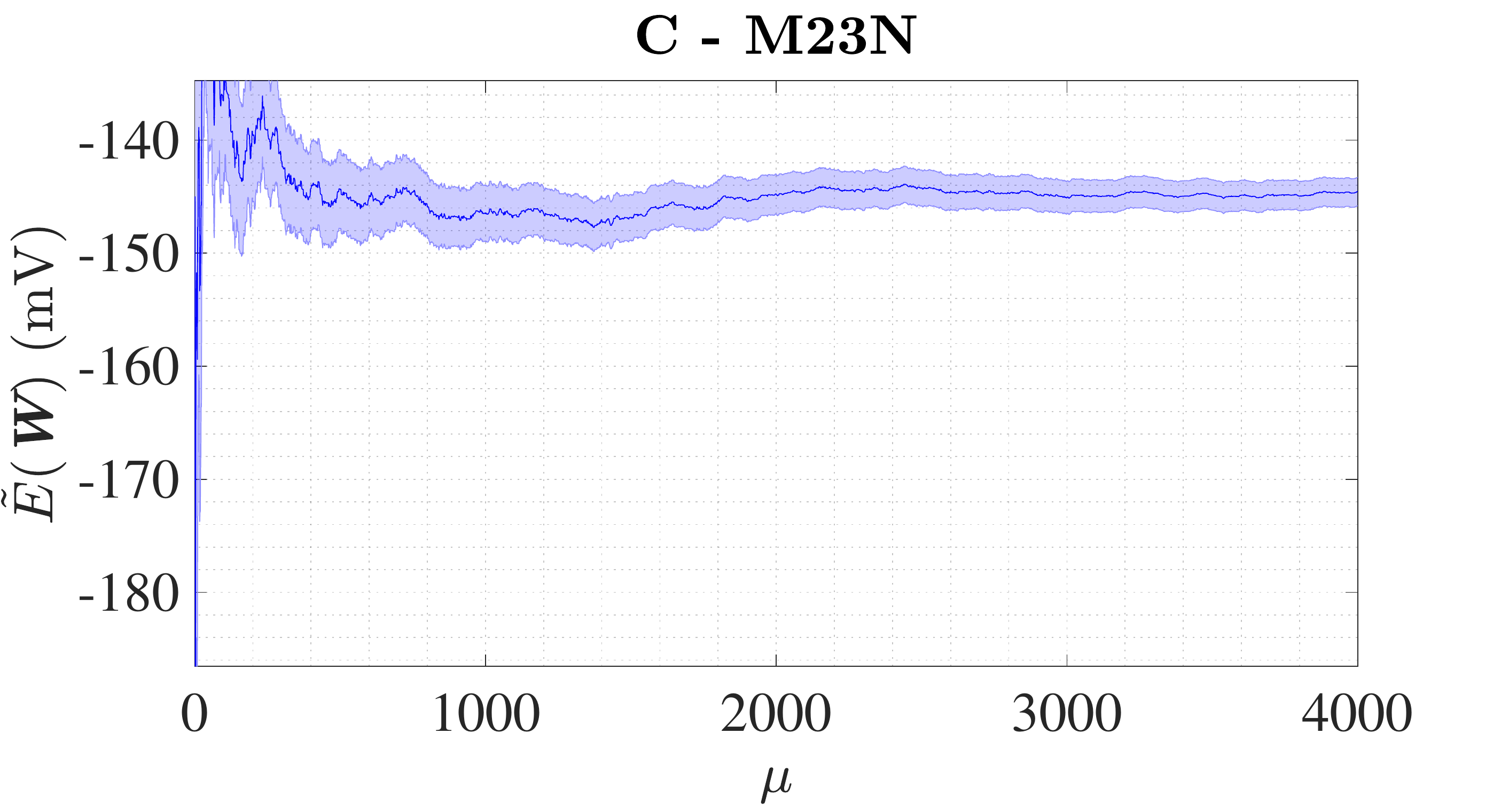}\includegraphics[trim={0.2cm -0.75cm 1.55cm 0cm},clip,width=8.5cm]{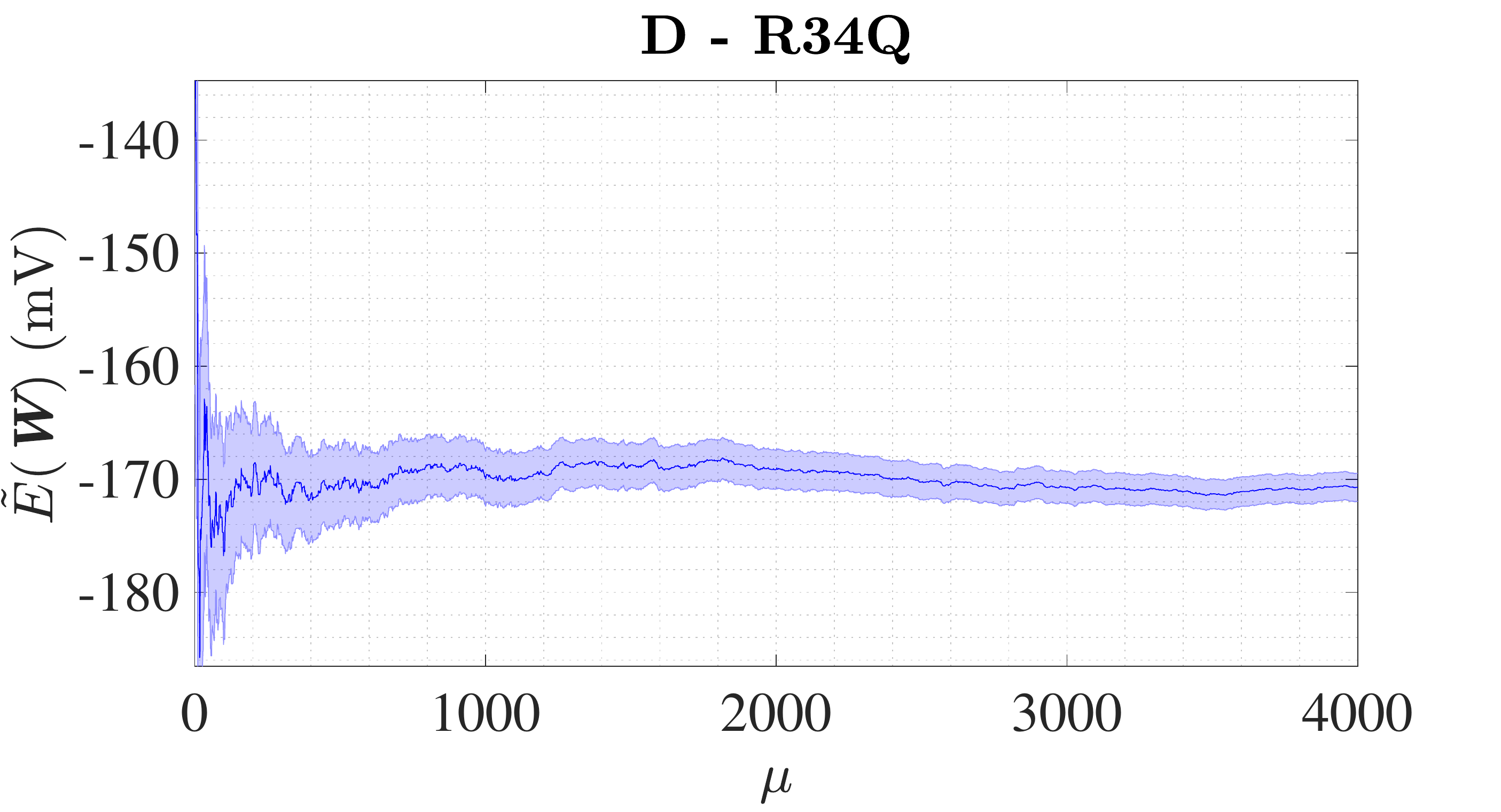}
\includegraphics[trim={0.2cm -0.75cm 1.55cm 0cm},clip,width=8.5cm]{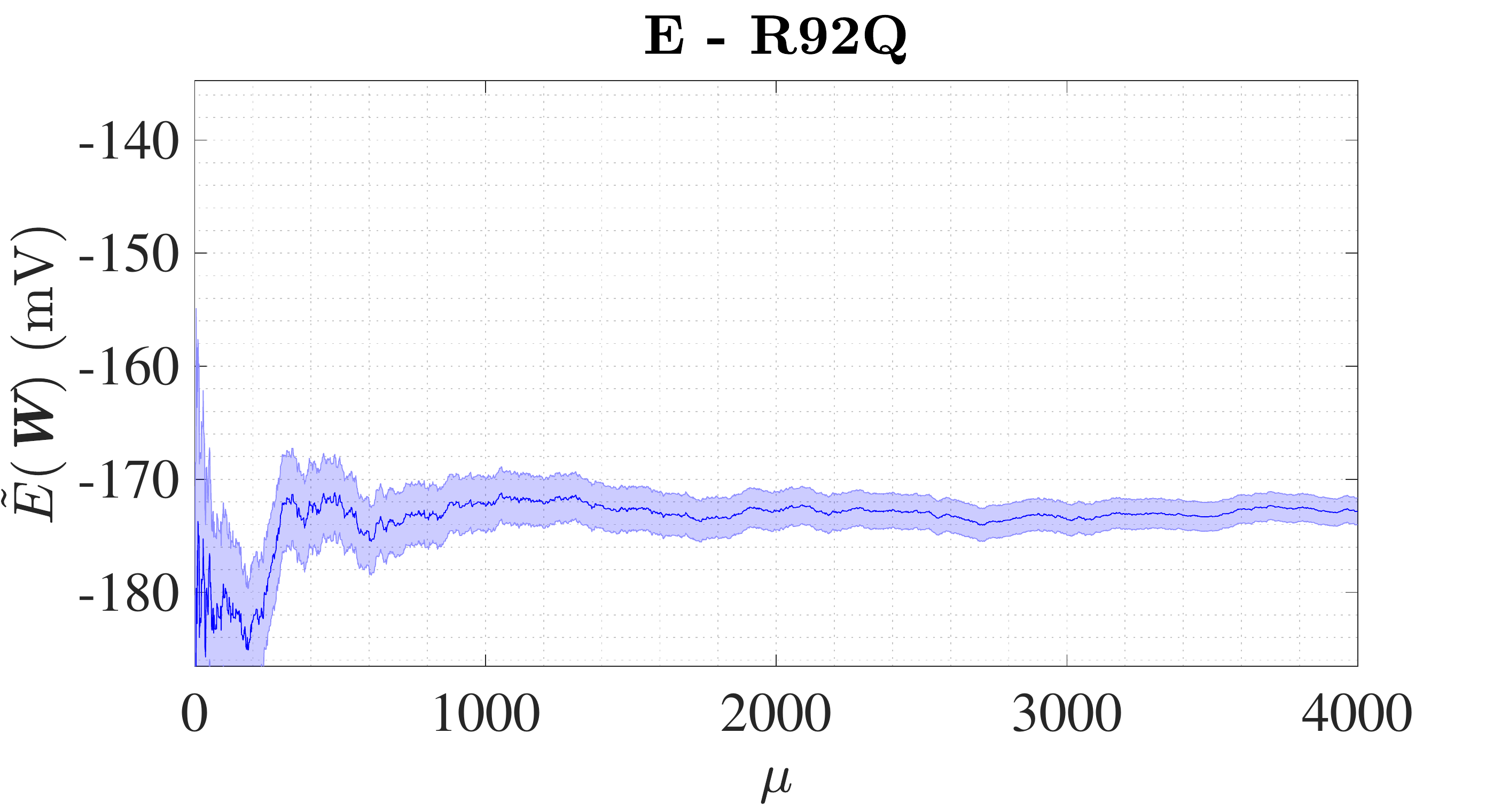}\includegraphics[trim={0.2cm -0.75cm 1.55cm 0cm},clip,width=8.5cm]{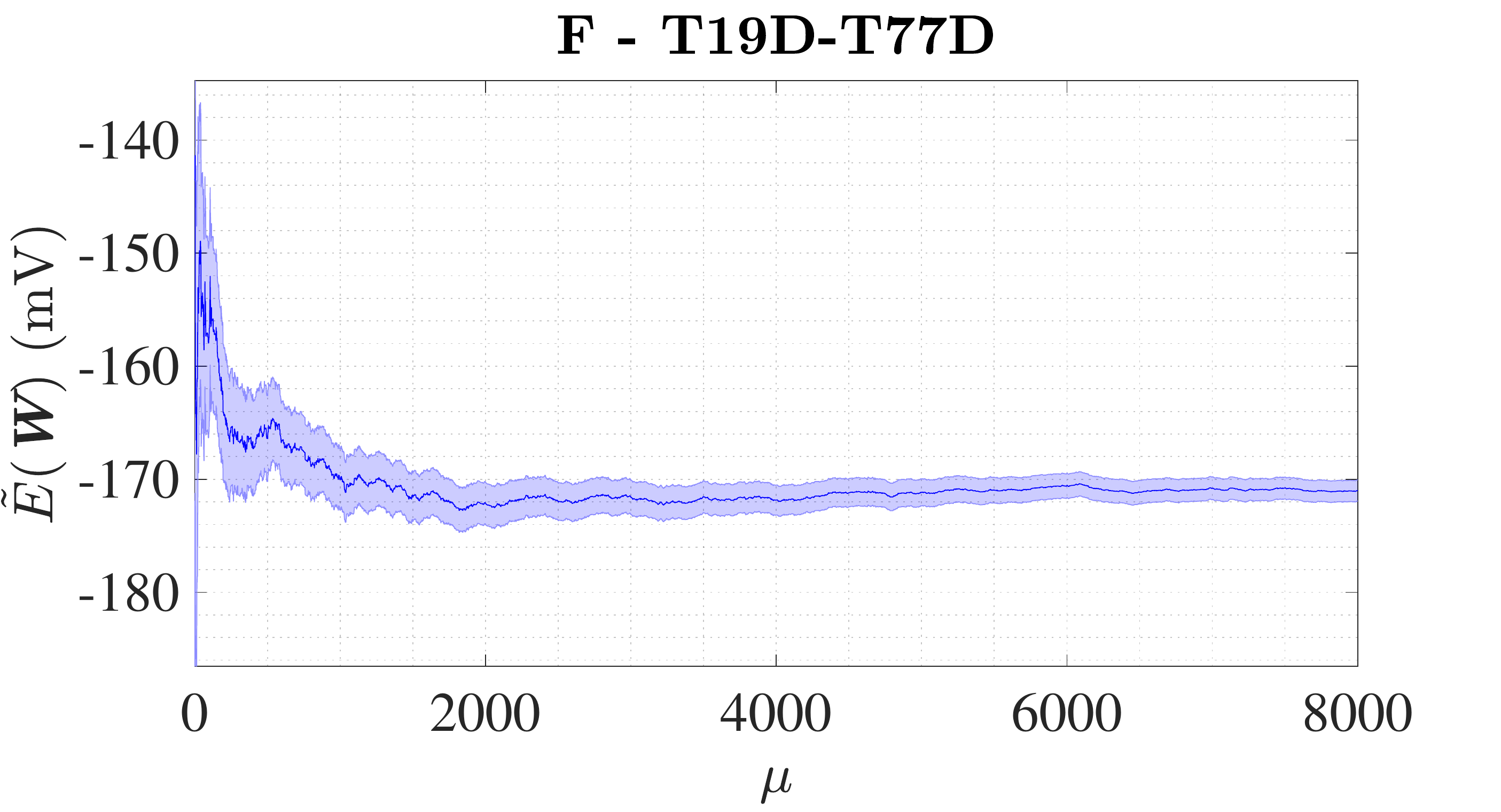}
    \caption{
Crooks-Bayes estimates $\tilde{E}(\boldsymbol{W})\,\pm\,\sigma (\boldsymbol{W})$ from $\mu$ iterations of the forward/reduction and backward/oxidation protocols. These are calculated using the estimator $E(\boldsymbol{W}) = - \int d\Delta g\,p(\Delta g|\boldsymbol{W}) \Delta g/F$, which is optimal under the square error criterion.\cite{jaynes2003, toussaint2011}
    Here, $F$ is Faraday's constant.
    The error in such an estimate is given by $\sigma^2 (\boldsymbol{W}) = \int d\Delta g\,p(\Delta g|\boldsymbol{W}) [-\Delta g/F-E(\boldsymbol{W})]^2$.
    The posterior probabilities $p(\Delta g|\boldsymbol{W})$ are those in Fig.~\ref{fig:posteriors}. 
	The final values for the redox potentials in the main text correspond to $\mu = 4000$ iterations in panel A-E, and to $\mu = 8000$ iterations in panel F.
	As can be observed, all six estimates start to converge to a single value when $\mu \simeq 2000$. 
	} 
\label{fig:convergence}
\end{figure}

\newpage

\subsection{Experimental measurement of the redox potentials} \label{subsec:exp}

\begin{figure}[th!]
\centering
\includegraphics[trim={0.0cm 0.0cm 0.0cm 0cm},clip,width=0.95\textwidth]{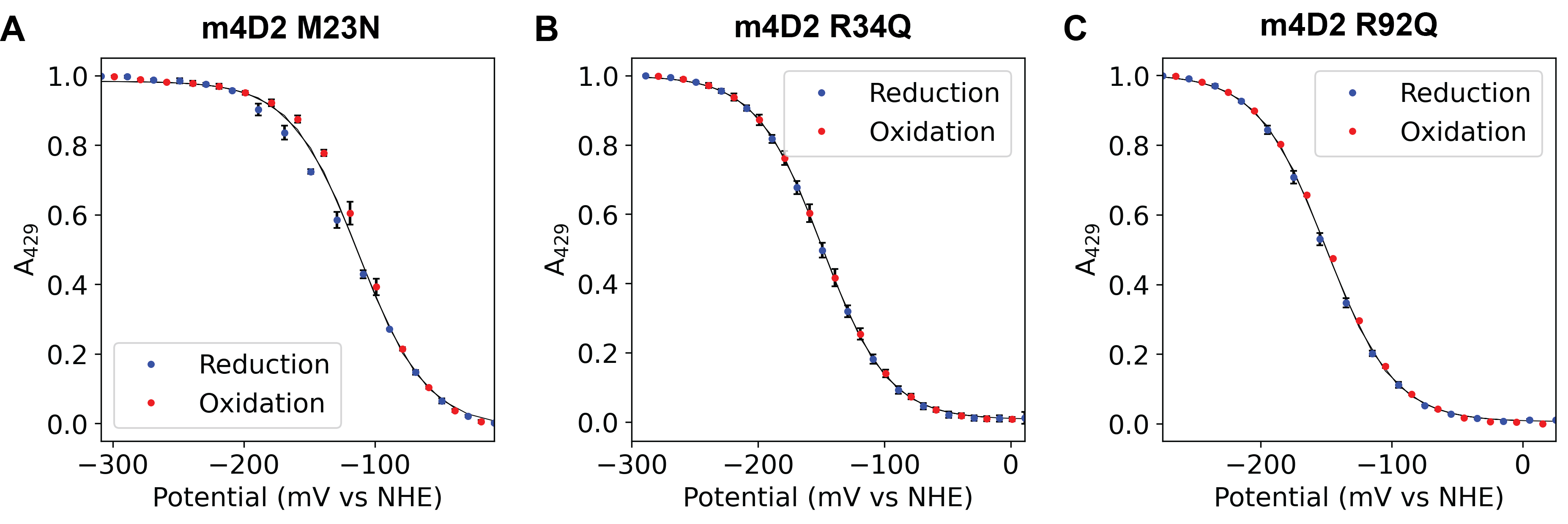}
	\caption{Experimental redox potentiometry of  M23N, R34Q and R92Q recorded in 20 mM CHES, 100 mM KCl, 10\% glycerol, pH 8.6. Data were fitted to a single electron Nernst function.
	}
\label{fig:exp-fig}
\end{figure}

\subsection{Experimental vs predicted redox potentials} \label{subsec:abs}

\begin{table}

\begin{tabular}{|l|r|r|} 
    \hline
    protein &  $ E $ \textbf{experiment} (\si{mV} ) &  $ E $ \textbf{predicted} (\si{mV} ) \\
    \hline
    \hline
    m4D2\cite{Hutchins2023} & -118 (1) & -159 (1) \\
    T19D\cite{Hutchins2023} & -146 (1) &\ -163 (1) \\
    M23N & -119 (1) & -145 (1)  \\
    R34Q & -149 (1) &  -171 (1) \\
    R92Q & -150 (1) &  -173 (1)  \\
    DM\cite{Hutchins2023} & -174 (1) & -171 (1) \\
    \hline
\end{tabular}
\caption{
Experimental (column 2) and predicted (column 3) redox potentials $E$ for m4D2, T19D, M23N, R34Q, R92Q and the double mutant T19D-T77D (DM). These correspond to the methods employed in Fig.~\ref{fig:exp-fig} and Fig.~\ref{fig:convergence}, respectively. 
The predicted redox potentials, in particular, were calculated using the proposed MD+CB method, which post-processes the data generated by the MD simulations via the Crooks-Bayes estimator as described in the main text (Sec.2.4) 
% ~\ref{subsec:FR}
as well as here (Secs.~\ref{subsec:subcrooks} and \ref{subsec:workdata}). The errors associated with $E$ are shown in parentheses. Previously measured redox potentials\cite{Hutchins2023} are referenced in the first column. 
}
\label{table:abs}

\end{table}

\end{document}